\documentclass[review]{elsarticle}
\usepackage{lineno,hyperref}

\modulolinenumbers[5]

\usepackage{amssymb}
\usepackage{latexsym}

\usepackage{url}
\usepackage{xcolor}
\definecolor{newcolor}{rgb}{.8,.349,.1}

\journal{ }

\usepackage{graphicx} 
\usepackage[utf8]{inputenc}
\usepackage{amssymb}
\usepackage{subfigure}
\usepackage{graphicx}
\usepackage{bm}
\usepackage{amsmath}
\usepackage{threeparttable}
\usepackage{CJK}
\usepackage{hyperref}
\usepackage{color}
\usepackage{booktabs}
\usepackage{bookmark}
\usepackage{array}
\usepackage{multirow}
\usepackage{natbib}
\usepackage{setspace}

\biboptions{sort&compress}

\begin{document}




\begin{frontmatter}

\title{Simplified unified wave-particle method for diatomic gases based on Rykov model}%

\author[address2]{Sirui Yang}
\ead{ysr1997@mail.nwpu.edu.cn}
\author[address2,address3]{Sha Liu\corref{cor1}}
\ead{shaliu@nwpu.edu.cn}
\author[address2]{Junzhe Cao}
\ead{caojunzhe@mail.nwpu.edu.cn}
\author[address2,address3]{Chengwen Zhong}
\ead{zhongcw@nwpu.edu.cn}

\address[address2]{National Key Laboratory of Aircraft Configuration Design, Northwestern Polytechnical University, Xi'an, Shaanxi 710072, China}
\address[address3]{Institute of Extreme Mechanics, Northwestern Polytechnical University, Xi'an, Shaanxi 710072, China}

\cortext[cor1]{Corresponding authors,}


\begin{abstract}
During the past decades, the numerical methods based on Navier-Stokes (N-S) equations and direct simulation Monte Carlo (DSMC) methods have been proved effective in simulating flows in the continuum and rarefied regimes, respectively. However, as single-scale methods, they face challenges in addressing common multi-scale problems, which are essential to simulate hypersonic flows around near-space vehicles and the flows in the micro-electro-mechanical systems. Hence, there is an urgent need for a method to predict multi-scale flows. In this work, a quantified model-competition (QMC) mechanism for diatomic multi-scale flows is derived from the integral solution of the Rykov model equations. This mechanism encapsulates both continuum and rarefied behaviors in a cell, weighted according to its local physical scale. By building upon the QMC mechanism, the N-S solver and DSMC solver are directly integrated within a cell to devise a simplified unified wave-particle (SUWP) method for diatomic gases. Specifically, the two-temperature equations considering the rotational energy are introduced into the kinetic inviscid flux (KIF) scheme and the N-S solver. As to the particle part, the collisionless DSMC solver is utilized to describe the non-equilibrium phenomenon. The proposed SUWP method for diatomic gases undergoes validation across a series of cases, including zero-dimensional homogeneous gas relaxation, one-dimensional normal shock structure, two-dimensional flow around the flat and the cylinder, and three-dimensional flows past the sphere and the blunt cone. Additionally, the implementation details of multi-scale wave-particle methods analysis and discussion are also undertaken in this work.
\end{abstract}


\end{frontmatter}

\section{Introduction}

Multi-scale flows refer to the coexistence of continuum and rarefied complex flows within a flow field~\cite{Liu2022_AMS}. The traditional CFD methods Navier-Stokes (N-S) equations and the direct simulation Monte Carlo (DSMC) method based on molecular motion are applicable to continuum flows and rarefied flows respectively~\cite{Bird_Molecular, Blazek2015computational}, and both have high technical maturities in the corresponding industry~\cite{gibbons2023eilmer, candler2015development, bird2013dsmc, white2018dsmcfoam+, zhong2024stochastic}. However, when multi-scale problems are involved, these single-scale methods often fail to provide accurate results or require extensive computational resources~\cite{zhang2019particle}. To calculate mesoscopic scale flows effectively and investigate the flow physical nature at various scales effectively, the development of numerical methods for the multi-scale flows is essential~\cite{Xu_Book_2021}.

At the present stage, multi-scale numerical methods include deterministic method and stochastic particle method. These methods are based on the Boltzmann model equation like the Bhatnagar-Gross-Krook (BGK) equations~\cite{bhatnagar1954model} and the Fokker-Planck (FP) equations~\cite{fokker1914mittlere}. They have gradually expanded their application to other physical domains, offering effective approaches for modeling in various complex scientific fields, like plasma~\cite{liu2017unified, liu2021unified}, phonon~\cite{guo2016discrete} and radiative transport~\cite{liu2023implicit, li2024unified}.

The deterministic unified methods are based on the discrete velocity framework~\cite{oldestDVM}. In these methods, the transport and collision processes are combined by the integral or differential solutions of Boltzmann model equation leading to the multi-scale property. The representative methods in this category include the unified gas kinetic scheme (UGKS)~\cite{xu2010unified, liu2014unified, LiuA2016, Li2018A, Chen2012A}, the discrete unified gas kinetic scheme (DUGKS)~\cite{guo2013discrete, Chen2019Conserved, zhao2023numerical, ZHANG2023107079}, the gas kinetic unified algorithm (GKUA)~\cite{li2009gas, peng2016implicit, Wu2020On}, improved discrete velocity method (IDVM)~\cite{yang2018improved}. And general synthetic iterative scheme (GSIS) is a implicit multi-scale method~\cite{su2020fast}. 
At the present stage, they can deal with multi-scale flows effectively, like micro-nano flows~\cite{wang2022investigation}, jet flows~\cite{chen2020compressible, zhao2024interaction} and radiative transport~\cite{sun2015asymptotic}. In recent years, these methods have been extended through modified BGK-type model equations to facilitate simulations for monoatomic gas mixtures, polyatomic gases and simple chemical reaction flows~\cite{wu2022utility}. Although the cell size and the iteration of the unified method are not limited by the molecular mean free path and mean collision time, such as the traditional DVM, the extra velocity space size of thousands to tens of thousands for every single cell still means long computational time and large memory costs.

On the other hand, the most representative stochastic particle method is the DSMC method~\cite{bird2013dsmc}. The DSMC method separates the transport and collision processes of particles, achieving significant success in simulating high-speed rarefied non-equilibrium flows by modeling molecular collisions through binary collisions. However, due to the decoupling of molecular transport from collisions, molecules undergo separate and successive transport and two-body collision sub steps during every iteration time. Consequently, the cell size must be smaller than the molecular mean free path, and the time step must be smaller than the relaxation time. As the Knudsen number decreases, the required cell size and time step for DSMC simulations also decrease, leading to a substantial increase in computational workload~\cite{Bird_Molecular}. Multi-scale stochastic particle methods also utilize the integral or differential solutions of the Boltzmann model equations to overcome the cell size and time step limitations. The particle-BGK method~\cite{Macrossan2001_nu}, computes the molecular collision process based on the integral solution of the homogenouse BGK model equations, enabling a larger time step or cell size. By discrete solution of full BGK equation, the unified stochastic particle-BGK (USP-BGK) method~\cite{Fei2020A} further coupled the molecular transport and collision to simulate multi-scale flows~\cite{Holway1966New}. By using different BGK equations, the particle-BGK method effectively simulates thermally non-equilibrium gases~\cite{kumar2011modeling, tumuklu2016particle, pfeiffer2018extending}. Although many modified BGK-type model equation are implemented to increase the model accuracy, its simulation results show minor discrepancies compared to DSMC, often manifesting as an ``early rise'' of temperature ahead of shock waves~\cite{xu2021modeling}. Meanwhile, a class of multi-scale particle methods based on the FP equations and the Cubic-FP equations have also been developed~\cite{Gorji2011_fokker}. The particle-FP methods enable the simulation of monoatomic gas mixtures and polyatomic gases flows~\cite{jun2019cubic}. The multi-scale temporal discretization FP method (MTD-FPM)~\cite{fei2017particle} overcomes a key drawback of classical FP-based methods, where transport coefficients gradually deviate from their true values as the time step increases. Besides these methods based on BGK-type and FP-type equations, the asymptotic-preserving Monte Carlo (APMC) method~\cite{ren2014asymptotic} constructed penalty function corrections using the BGK equations, which endows the method with capacity for simulating multi-scale flows. Besides, as well as methods such as moment-guided Monte Carlo method~\cite{degond2011moment} and the expansion of the Boltzmann equation (time relaxed Monte Carlo Methods~\cite{Pareschi2001_TRMC}). In multi-scale particle methods, the number of particles required is smaller than the number of velocity space cells required by the unified methods. However, the fluctuations caused by stochastic processes are inevitable, which means that the method requires higher costs to statistical average when simulating low-speed or unsteady flows to obtain acceptable results.

Additionally, the hybrid methods are also alternative for simulating multi-scale flows. The traditional hybrid methods, like the modular particle-continuum (MPC) method~\cite{wang2003assessment}, combine deterministic or stochastic particle methods to capitalize on their respective advantages by decomposing computational domain~\cite{zhang2019particle}, and the information exchange mechanism at the interface or overlap region between these regions helps to couple the solvers. The information preserving (IP) and other methods to suppress fluctuation~\cite{fan2001statistical} introduced in the hybrid methods to suppress statistical fluctuations in the overlap region. Since rarefied flows are challenging for the N-S, the DSMC method is difficult to calculate continuum flows, due to the huge computational cost the hybrid methods face difficulties when simulating multi-scale flows, such as dividing the suitable computational domain, and the trade-off of computational accuracy and cost~\cite{hash1995hybrid}, and will face more challenges when the scale in the flow field is complex~\cite{glass20013, zhang2019particle}.

In recent years, the unified gas-kinetic wave-particle (UGKWP) method is developed~\cite{LiuUnified2020, zhu2019unified, xu2021unified} based on the BGK model. The UGKWP method can be viewed as a stochastic version of the successful UGKS method, which uses is stochastic particles to represent the non-equilibrium properties of the flows instead of relying on deterministic velocity space. In this method, molecules are categorized into three types: \uppercase\expandafter{\romannumeral1}. free, \uppercase\expandafter{\romannumeral2}. colliding before leaving its origin cell, \uppercase\expandafter{\romannumeral3}. colliding after leaving its origin cell~\cite{ChenY2020A}, which is derived from the time integral solution of the Boltzmann model equation and leads to the multi-scale property. A theoretical analysis of this time integral solution was conducted by Liu et al. and a quantified model-competition (QMC) mechanism was found~\cite{Liu2020Simplified}. This mechanism provides a rational and physically meaningful coupling between continuum and rarefied flow models within the context of multi-scale flow. Based on the QMC mechanism, the simplified unified wave-particle (SUWP) method is proposed~\cite{Liu2020Simplified}. The SUWP method classifies the flow molecules into two categories: colliding molecules and free transport molecules, simulated by the N-S solver and the particle solver respectively. Previous research has constructed that SUWP method for monoatomic gases, and this work focus on extending QMC and SUWP to the more common diatomic gases.

This paper extends the SUWP method to diatomic gases by incorporating the Rykov model~\cite{Rykov1978}. The QMC coupling mechanism for diatomic gases is derived based on the Rykov model and is proved that colliding particles can be described by N-S equations. For the N-S solver, the translational-rotational two-temperature N-S equations based on the 2nd order Chapman-Enskog (C-E) expansion of Rykov model are obtained, and the kinetic inviscid flux (KIF) scheme~\cite{KIF1, KIF2} for two-temperature N-S equations are constructed . For the particle solver, the acceptance-rejection and importance sampling techniques are used to sample particles from the equilibrium distribution of the Rykov equation~\cite{Liu1996Metropolized}, and a suitable particle transport tracking technique is also implemented within the finite volume method framework. Based on the above work, the SUWP method for diatomic gases is established. The accuracy of the method in simulating molecular collision, transport processes and the ability of simulating viscous high speed flow are validated by a series case: relaxation in a diatomic homogeneous gas case, normal shock-structure, flow past the flat plate, cylinder, sphere and blunt cone. 

The rest of this paper is organized as follows. Section \ref{sec::method} presents the SUWP method for diatomic gases. Section \ref{sec::numricalExamples} contains various numerical tests to validate the SUWP-Rykov method. Section \ref{sec::conclusions} is the discussion and concluding remarks. The \ref{sec::appendixA} and \ref{sec::appendixB} provide a brief explanation of the random coordinate sampling and transport treatment of particles in the three-dimensional finite volume method.

\section{\label{sec::method}Simplified Unified Wave-Particle Method with Rykov Model Equation}

\subsection{\label{sec::RykovModel}Rykov model equation for diatomic molecules}

The Boltzmann equation is an kinetic equation to describe the transport and collision of particles, formulated on the scale of particle mean free path and mean collision time. However, solving the collision operator of the Boltzmann equation is complex and challenging. Therefore, previous researchers have developed various models by simplifying the collision operator. Effective model equations are crucial to the multi-scale numerical methods, as the velocity distribution function is evolved in a relaxation way, whose characteristic integral or differential solution contributes the success of multi-scale numerical methods like UGKS, DUGKS, UGKWP, SUWP, et al. 

As a kind of famous Boltzmann model, the BGK-type model has been widely applied in both theoretical and numerical studies, such as the original BGK model~\cite{bhatnagar1954model}, the Ellipsoidal-Statistical BGK (ES-BGK) model~\cite{Holway1966New} and the Shakhov model~\cite{1968Generalization}. These models have the asymptotic preserving and conservation properties. In recent years, there has been a emphasis on developing models that prioritize non-equilibrium performance~\cite{xu2021modeling, yang2021direct, yuan_wu_2022}. 

The Rykov model is a BGK-type model equation constructed using Hermite expansion like the Shakhov model. This model has the added capability of simulating the rotational internal energy evolution of diatomic molecules, providing rotational temperature calculations that align well with DSMC simulation results. In this section, the velocity distribution function of the Rykov model and its conservation equations derived from the C-E expansion are introduced, so that both the wave and the particle part can be designed accurately. 

In the gas-kinetic theory, the states of diatomic gases with rotational mode can be represented by the velocity distribution function $f(\vec{x},\vec{v},\eta ,t)$, where $\vec{x}$ is the physical space coordinates, $\vec{v}$ is the particle transport velocity, $\eta$ is the rotational energy, and $t$ is the time. Macroscopic variables can be calculated by the following integrals (moments)
\begin{equation} \label{eq:macroVariables}
\vec{W}=\int{\vec{\psi }fd\Xi },
\end{equation} 
where $d\Xi =d\vec{v}d{\eta }\text{ }$, $\vec{\psi }={{\left( 1,\vec{v},\frac{1}{2}{{{\vec{v}}}^{{\,}2}}+\eta ,\eta  \right)}^{T}}$ are microscopic variables, and $\vec{W}={{\left( \rho ,\rho \vec{U},\rho E,\rho E_{rot} \right)}^{T}}$ are macroscopic variables. Here, $\rho$ is the density, $\vec{U}$ is the macroscopic velocity, $E$ is the total energy, and ${{E}_{rot}}$ is the rotational energy. The pressure tensor $\mathbf{P}$, the translational heat flux ${{\vec{q}}_{tr}}$ and the rotational heat flux ${{\vec{q}}_{rot}}$ are also defined by the moments as follows:
\begin{equation} \label{eq:pAndQtrAndQrot}
    \begin{split}
      & \mathbf{P}=\int{\vec{c}{\,}\vec{c}}fd\Xi , \\ 
     & {{{\vec{q}}}_{tr}}=\frac{1}{2}\int{{\vec{c}}}{\,}{{{\vec{c}}}^{{\,}2}}fd\Xi , \\ 
     & {{{\vec{q}}}_{rot}}=\int{{\vec{c}}}{\,}\eta fd\Xi , \\ 
    \end{split}
\end{equation} 
where $\vec{c}=\vec{v}-\vec{U}$ is the peculiar velocity. The total heat flux $\vec{q}$ is the sum of ${{\vec{q}}_{tr}}$ and ${{\vec{q}}_{rot}}$.
\begin{equation} \label{eq:qByQtrPlusQrot}
\vec{q}=\int{{\vec{c}}}\left( \frac{1}{2}{{{\vec{c}}}^{{\,}2}}+\eta  \right)fd\Xi ={{\vec{q}}_{tr}}+{{\vec{q}}_{rot}}.
\end{equation} 

The evolution of the diatomic gases distribution function $f$ is governed by the Rykov model equation as follows,
\begin{equation}  \label{eq:RykovModel}
\frac{\partial f}{\partial t}+\vec{v}\cdot \nabla f=\frac{{{g}_{rot}}-{{g}_{tr}}}{{{Z}_{rot}}\tau }+\frac{{{g}_{tr}}-f}{\tau }=\frac{g^{+}-f}{\tau},
\end{equation} 
where $\tau$ is the relaxation time, and ${{Z}_{rot}}$ is the rotational relaxation collision number. The ${{g}_{tr}}$ is the translational equilibrium distribution function and the ${{g}_{rot}}$ is the rotational equilibrium distribution function, whose expression are as follows:
\begin{equation}  \label{eq:gTrAndgRot}
\begin{split}
{{g}_{tr}}&=\rho{{\left( \frac{m}{2\pi {{k}_{B}}{{T}_{tr}}} \right)}^{\frac{3}{2}}}\exp \left( -\frac{m|\vec{c}{\,}{{|}^{2}}}{2{{k}_{B}}{{T}_{tr}}} \right)\left( \frac{1}{{{k}_{B}}{{T}_{rot}}} \right)\exp \left( -\frac{\eta }{{{k}_{B}}{{T}_{rot}}} \right) \\
&\times \left(1-\frac{2m{{{\vec{q}}}_{tr}}\cdot \vec{c}}{15{{k}_{B}}{{T}_{tr}}{{P}_{tr}}}\left( \frac{5}{2}-\frac{m|\vec{c}{\,}{{|}^{2}}}{2k{{T}_{tr}}} \right)+(\sigma -1)\frac{m{{{\vec{q}}}_{rot}}\cdot \vec{c}}{{{k}_{B}}{{T}_{tr}}}\frac{{{k}_{B}}{{T}_{rot}}-\eta }{\theta }\right), \\
{{g}_{rot}}&=\rho{{\left( \frac{m}{2\pi {{k}_{B}}T_{eq}} \right)}^{\frac{3}{2}}}\exp \left( -\frac{m|\vec{c}{\,}{{|}^{2}}}{2{{k}_{B}}T_{eq}} \right)\left( \frac{1}{{{k}_{B}}T_{eq}} \right)\exp \left( -\frac{\eta }{{{k}_{B}}T_{eq}} \right) \\
&\times \left(1-{{\omega }_{0}}\frac{2m{{{\vec{q}}}_{tr}}\cdot \vec{c}}{15{{k}_{B}}T_{eq}P}\left( \frac{5}{2}-\frac{m|\vec{c}{\,}{{|}^{2}}}{2{{k}_{B}}T_{eq}} \right)+{{\omega }_{1}}(\sigma -1)\frac{m{{{\vec{q}}}_{rot}}\cdot \vec{c}}{{{k}_{B}}T_{eq}P}\left( 1-\frac{\eta }{{{k}_{B}}T_{eq}} \right)\right), \\
\end{split}
\end{equation} 
where ${{k}_{B}}$ is the Boltzmann constant, $n$ is the molecular number density, $m$ is the molecular mass, and,
\begin{equation}  \label{eq:thetaAndSigma}
\theta =\iint{{{\left( {{k}_{B}}{{T}_{rot}}-\eta  \right)}^{2}}fd\eta d\vec{v}},\quad \sigma =\frac{{{\mu }}}{mnD},
\end{equation} 
where ${{\mu }}$ is the viscosity, and $D$ is the diffusion coefficient.

The expressions for the translational temperature ${{T}_{tr}}$, rotational temperature ${{T}_{rot}}$, and equilibrium temperature ${T}$ involved in Eq.(\ref{eq:gTrAndgRot}) are as follows
\begin{equation}  \label{eq:tAndTtrAndTrot}
    \begin{split}
 & {{T}_{tr}}=\frac{2{{E}_{tr}}}{3nk_{B}}=\frac{1}{3nk_{B}}\iint{m{{\left| {\vec{c}} \right|}^{2}}f}d\eta d\vec{v}, \\ 
 & {{T}_{rot}}=\frac{2{{E}_{rot}}}{{{K}_{rot}}nk_{B}}=\frac{2}{{{K}_{rot}}nk_{B}}\iint{\eta fd\eta d\vec{v}}, \\ 
 & {{T}_{eq}}=\frac{2\left(E_{tr}+E_{rot}\right)}{\left(3+K_{rot}\right)nk_{B}}=\frac{3{{T}_{tr}}+{{K}_{rot}}{{T}_{rot}}}{3+{{K}_{rot}}}. \\ 
    \end{split}
\end{equation} 
where ${{K}_{rot}}$ is the rotational degrees of freedom. For diatomic gases, ${{K}_{rot}} = 2$.

The terms ${{\omega }_{0}}$ and ${{\omega }_{1}}$ in Eq.(\ref{eq:gTrAndgRot}) are determined by the heat flux relaxation rates~\cite{Rykov1978}. The expressions for the translational heat flux relaxation rate and the rotational heat flux relaxation rate are as follows

\begin{equation}  \label{eq:QtrRateAndQrotRate}
    \begin{split}
  & \frac{\partial {{{\vec{q}}}_{tr}}}{\partial t}=-\frac{1}{\tau }\left( \frac{2}{3}+\frac{1-{{\omega }_{0}}}{3{{Z}_{rot}}} \right){{{\vec{q}}}_{tr}}, \\ 
 & \frac{\partial {{{\vec{q}}}_{rot}}}{\partial t}=-\frac{1}{\tau }\left[ \sigma +\frac{(1-\sigma )\left( 1-{{\omega }_{1}} \right)}{{{Z}_{rot}}} \right]{{{\vec{q}}}_{rot}}. \\
    \end{split}
\end{equation} 
For the nitrogen gas, the typical values used are $\sigma =1/1.55$, ${{\omega }_{0}}=0.2354$, ${{\omega }_{1}}=0.3049$. The relaxation time $\tau$ in Eq.(\ref{eq:QtrRateAndQrotRate}) depends on the translational temperature $T_{tr}$, and is expressed as
\begin{equation}  \label{eq:tauOnlyTtr}
\tau =\frac{{{\mu }}}{n{{k}_{B}}{{T}_{tr}}}.
\end{equation}

The Rykov model is derived in Ref.~\cite{Rykov1978}. For simplicity, the following denotation for integrals in velocity space and substantial derivative are used in the remaining of this paper:
\begin{equation}
    \langle \cdot \rangle=\int \cdot d \vec{v} d \eta, D_t=\frac{D}{D t}=\frac{\partial}{\partial t} + v_{i}\frac{\partial}{\partial x_{i}}.
\end{equation}

The field variables are expressed through the $f$:

\begin{equation} \label{eq:Rykov_macro}
\begin{aligned}
  & \left\langle {f} \right\rangle =\rho ,\left\langle {{v}_{i}} {f} \right\rangle =\rho {{U}_{i}}, \\ 
 & \left\langle \frac{m{{c}^{2}}}{2} {f} \right\rangle =\frac{3}{2}\rho R{{T}_{tr}},\left\langle \eta  {f} \right\rangle =\rho R{{T}_{rot}}, \\ 
 & \langle m c_i c_j {f} \rangle=P_{tr,ij}, \left\langle {{c}_{i}}\frac{m{{c}^{2}}}{2} {f} \right\rangle ={{q}_{tr,i}},\left\langle {{c}_{i}}\eta  {f} \right\rangle ={{q}_{rot,i}}. \\ 
\end{aligned}
\end{equation}

Seek the solution by the method of perturbation to the small parameter $\epsilon$ (In the final results $\epsilon$ will be set to 1) , and the distribution function from the second order C-E expansion of the Rykov can be written as:

\begin{equation}  \label{eq:C-E_Rykov}
    f=g^{+}-\epsilon \cdot \tau \left( \frac{D g}{D t} \right).
\end{equation}

Substitute this perturbation distribution into the Rykov equation Eq.(\ref{eq:RykovModel}), multiply it by the moment vector $\vec{\psi}$, and integrate the resultant equation in the whole velocity space, the macroscopic equation can be found as follows:
\begin{equation}  \label{eq:twoT_Rykov}
\begin{aligned}
& \frac{D\rho }{Dt}+\rho \frac{\partial {{U}_{i}}}{\partial {{x}_{i}}}=0, \\
& \frac{D\rho {{U}_{i}}}{Dt}+\frac{\partial {{P}_{tr,ij}}}{\partial {{x}_{j}}}=0, \\
&  \frac{3}{2}\frac{D\left(\rho R{{T}_{tr}} \right)}{Dt}+\frac{\partial {{q}_{tr,j}}}{\partial {{x}_{j}}}+{{P}_{tr,ij}}\frac{\partial {{U}_{i}}}{\partial {{x}_{j}}}=\frac{1}{\tau {{Z}_{rot}}}\left( \frac{3}{2}\rho R{{T}_{eq}}-\frac{3}{2}\rho R{{T}_{tr}} \right), \\
& \frac{K_{rot}}{2}\frac{D\left( \rho R{{T}_{rot}} \right)}{Dt}+\frac{K_{rot}}{2}\frac{\partial {{q}_{rot,j}}}{\partial {{x}_{j}}}=\frac{1}{\tau {{Z}_{rot}}}\frac{K_{rot}}{2}\left( \rho R{{T}_{eq}} - \rho R{{T}_{rot}} \right), \\
\end{aligned}
\end{equation}
where, 
\begin{equation}  \label{eq:twoT_q}
\begin{aligned}
& {P}_{tr,ij}=p_{tr}\delta_{ij}-\mu_{tr}\left( \frac{\partial U_{i}}{\partial x_{i}} + \frac{\partial U_{j}}{\partial x_{i}} - \frac{2}{3}\delta_{ij} \frac{\partial U_{k}}{\partial x_{k}}\right), \\
& {q}_{tr,i}=-\frac{15}{4}\frac{k_{B}}{m}\mu_{tr}\left(1+0.5\left(1-\omega_0 \right){Z_{rot}}^{-1} \right)^{-1}\frac{\partial T_{tr}}{\partial x_{i}},\\
& {q}_{rot,i}=-\frac{k_{B}}{m}\mu_{tr}\left(\delta+\left(1-\delta \right)\left(1-\omega_1 \right){Z_{rot}}^{-1} \right)^{-1}\frac{\partial T_{rot}}{\partial x_{i}},\\
\end{aligned}
\end{equation}
where $\mu_{tr}$ is the viscosity calculated by using the translational temperature $T_{tr}$~\cite{Rykov1978}.
For accommodation of traditional N-S equation with degenerated translational-rotational temperature, under the condition $Kn \ll 1$, the two-temperature equations can be reduced to an one-temperature form:

\begin{equation}  \label{eq:oneT_Rykov}
\begin{aligned}
& \frac{D\rho }{Dt}+\rho \frac{\partial {{U}_{i}}}{\partial {{x}_{i}}}=0, \\
& \frac{D\rho {{U}_{i}}}{Dt}+\frac{\partial {{P}_{i\text{j}}}}{\partial {{x}_{j}}}=0, \\
& \frac{3+K_{rot}}{2}\frac{D\left( \rho R{{T}_{eq}} \right)}{Dt}+\frac{\partial {{q}_{j}}}{\partial {{x}_{j}}}+{{P}_{ij}}\frac{\partial {{U}_{i}}}{\partial {{x}_{j}}}=0, \\
\end{aligned}
\end{equation}
where, 
\begin{equation}  \label{eq:oneT_q}
\begin{aligned}
& {P}_{ij}=p\delta_{ij}-\mu\left( \frac{\partial U_{i}}{\partial x_{j}} + \frac{\partial U_{j}}{\partial x_{i}} - \frac{2}{3}\delta_{ij} \frac{\partial U_{k}}{\partial x_{k}}\right) - \frac{4}{15}Z_{rot} \mu\delta_{ij} \frac{\partial U_{k}}{\partial x_{k}}, \\
& {q}_{i}=-\frac{3}{2}\frac{k_{B}}{m}\mu\left(\frac{5}{2+\left(1-\omega_0 \right){Z_{rot}}^{-1}} + \frac{2/3}{\delta+\left(1-\delta \right)\left(1-\omega_1\right){Z_{rot}}^{-1}}\right)\frac{\partial T_{eq}}{\partial x_{i}}.\\
\end{aligned}
\end{equation}
The second coefficient of viscosity (bulk viscosity) $\zeta=(4/15)Z_{rot}\mu$ appears in Eq.(\ref{eq:oneT_q}), and the bulk stress is defined from the relations of $p$ and $p_{tr}$:
\begin{equation}  \label{eq:bulkViscosity}
p_{tr}-p=nk_B(T_{tr}-T_{eq})=-\zeta\frac{\partial U_{k}}{\partial x_{k}}.
\end{equation}

It can be seen that the bulk viscosity stems from the relaxation of translational and rotational temperature, whose process is expressed in the source term of energy equation of the two temperature formula (Eq.(\ref{eq:twoT_Rykov})), and is expressed in the constitution relation of the one temperature formula (Eq.(\ref{eq:oneT_q})). In two temperature formulation, the difference between translational and rotational temperatures are faithful recorded, while in one temperature formulation, it is approximated by the small perturbation for Kn<<1. In this work, the two temperature formulation is adopted. More details. proof process can be referred to Ref~\cite{Rykov1978, Brunofluids7110356}.

\subsection{\label{sec::SUWP-Rykov}Simplified unified wave-particle method}

\subsubsection{\label{sec::QMC}Quantified model-competition mechanism}

\begin{figure}[!h]
    \centering
    \includegraphics[width=0.4\textwidth]{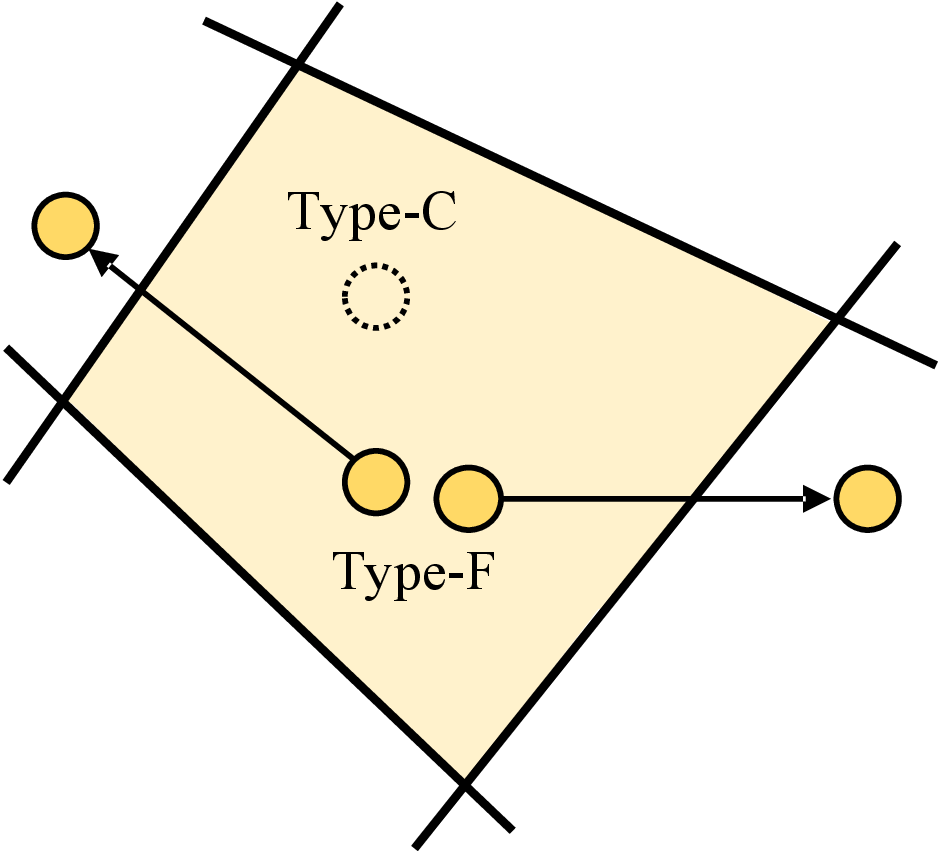}
    \caption{\label{Fig:particle_Categorization} Categorization of molecules in a cell of the SUWP method.}
\end{figure}

As illustrated in Fig.\ref{Fig:particle_Categorization}, the SUWP method classifies gas molecules into two categories: free transport molecules (Type-F) and colliding molecules (Type-C). The Type-F molecules are simulated by the particle solver, while the Type-C molecules are modeled by wave and simulated by the N-S solver. For the accurate updating of particle and macroscopic information within each cell, a reliable mechanism for determining the weights of both types of molecules is essential, which is derived as QMC mechanism as follows.

The time integral solution for BGK-type equations can be represented as follows
\begin{equation}  \label{eq:timeIntegralSolution}
f(\vec{x},\vec{v}, \eta,\Delta t)={{e}^{-\frac{\Delta t}{\tau }}}f({{\vec{x}}_{0}},\vec{v}, \eta, 0)+\frac{1}{\tau }\int_{0}^{\Delta t}{g\left( \vec{x}{\,}',\vec{v}, \eta, t' \right){{e}^{\frac{t'}{\tau }}}dt'},\text{ }
\end{equation} 
where the $\vec{x}_{0}$ is the original coordinate of particle at time $0$, and $\Delta t$ is the time step, which is physically the observation time.

The initial velocity of a group of molecules with velocity $\vec{v}$ is represented by the velocity distribution function $f({{\vec{x}}_{0}},\vec{v}, \eta,0)$, where the subscript $0$ indicates the variable at the initial time. If this group of molecules undergoes free transport, its transport path can be represented as $\vec{x}{\,}'={{\vec{x}}_{0}}+\vec{v}t'$, with $t'$ ranging from $0$ to $\Delta t$.

In the first term on the right hand side (RHS) of Eq.(\ref{eq:timeIntegralSolution}), ${{e}^{-\frac{\Delta t}{\tau }}}$ represents the ratio of molecules that do not experience collisions, maintain their velocity distribution from the initial time $f({{\vec{x}}_{0}},\vec{v}, \eta, 0)$, and move along the path defined by ${{\vec{x}}_{0}}+\vec{v}\Delta t$. 
Correspondingly, $1-{{e}^{-\frac{\Delta t}{\tau }}}$ portion of molecules experience at least one collision, who changes their velocities, and then follows the equilibrium velocity distribution $g\left( \vec{x}',\vec{v}, \eta,t' \right)$. By using the Taylor expansion to expand the second term of the RHS $g\left( \vec{x}',\vec{v}, \eta,t' \right)$ to the second order in both space and time, it denoted by
\begin{equation}  \label{eq:timeIntegralTerm}
h(\vec{x},\vec{v}, \eta,\Delta t)=\left\{ 
\begin{aligned}
\left( 1-{{e}^{-\frac{\Delta t}{\tau }}} \right)\left[ g-\tau \left( \vec{v}\cdot \frac{\partial g}{\partial \vec{x}}+\frac{\partial g}{\partial t} \right) \right]
+{{e}^{-\frac{\Delta t}{\tau }}}\Delta t\left( \vec{v}\cdot \frac{\partial g}{\partial \vec{x}}+\frac{\partial g}{\partial t} \right) \\
+\left( \Delta t-{{e}^{-\frac{\Delta t}{\tau }}}\Delta t \right)\frac{\partial g}{\partial t} 
\end{aligned}
\right\}
_{(\vec{x},\vec{v}, \eta,0)}.\text{ }
\end{equation} 
If the distribution function $g$ is the Rykov distribution function $g^{+}$ shown in Eq.(\ref{eq:gTrAndgRot}) and drop the high-order term, $h(\vec{x},\vec{v}, \eta, \Delta t)$ can be expressed as:
\begin{equation}  \label{eq:hydroEqn01}
h(\vec{x},\vec{v}, \eta,\Delta t)=\left( 
1-{{e}^{-\frac{\Delta t}{\tau }}} \right){{\left\{
\begin{aligned}
 \left[ g^{+} -\tau \left( \vec{v}\cdot \frac{\partial g}{\partial \vec{x}}+\frac{\partial g}{\partial t} \right) \right] 
 + \frac{\Delta t{{e}^{-\frac{\Delta t}{\tau }}}}{1-{{e}^{-\frac{\Delta t}{\tau }}}}\left( \vec{v}\cdot \frac{\partial g}{\partial \vec{x}}+\frac{\partial g}{\partial t} \right)
\end{aligned}
 \right\}}_{(\vec{x},\vec{v}, \eta,0)}}.
\end{equation} 
The term in the square bracket of Eq.(\ref{eq:hydroEqn01}) corresponds to the second-order C-E expansion of the Rykov distribution function, and the second term in the curly bracket is a scale-dependent anti-dissipative term with the opposite sign of the dissipative term.

In the macroscopic solver, the viscous flows are caused by the dissipation term. Therefore, for multi-scale flows, the viscous coefficient should be modified by the following scale-dependent adjustment parameter:
\begin{equation}  \label{eq:coeffVis}
{{c}_{vis}}=1-\left( \frac{\Delta t}{\tau } \right)\frac{{{e}^{-\frac{\Delta t}{\tau }}}}{1-{{e}^{-\frac{\Delta t}{\tau }}}}.
\end{equation} 
And $h(\vec{x},\vec{v}, \eta,\Delta t)$ is expressed as follows
\begin{equation}  \label{eq:newTypeHf}
h(\vec{x},\vec{v}, \eta,\Delta t)={{\left\{ \left( 1-{{e}^{-\frac{\Delta t}{\tau }}} \right)\left[ {{g}^{+}}-{{c}_{vis}} \tau {{D}_{t}}\left( {{g}} \right) \right] \right\}}_{(\vec{x},\vec{v}, \eta,0)}}.
\end{equation} 

The velocity distribution function at the present time step can be expressed as:
\begin{equation}  \label{eq:fOfTheTimestep}
f(\vec{x},\vec{v}, \eta,\Delta t)={{e}^{-\frac{\Delta t}{\tau }}}f(\vec{x},\vec{v}, \eta,0)+\left( 1-{{e}^{-\frac{\Delta t}{\tau }}} \right) {{\left[ {{g}^{+}}-{{c}_{vis}} \tau {{D}_{t}}\left( {{g}} \right) \right]}_{(\vec{x},\vec{v}, \eta,0)}},
\end{equation} 
where the terms in the square bracket are just a modified 2nd order C-E distribution for Rykov equations (Eq.(\ref{eq:C-E_Rykov})), and they correspond to wave part (N-S equations as Eq.(\ref{eq:twoT_Rykov})). Therefore, numerical solvers can be developed by the coupling of the N-S and DSMC solver.

The corresponding QMC weights for Type-F molecules, denoted ${{w}_{free}}$, and Type-C molecules, denoted ${{w}_{hydro}}$, can be expressed as 
\begin{equation}  \label{eq:weightOfCAndF}
{{w}_{free}}={{e}^{-\frac{\Delta t}{\tau }}},\quad {{w}_{hydro}}=1-{{e}^{-\frac{\Delta t}{\tau }}}.
\end{equation} 
If we use the difference approximation of the Boltzmann equation along the properties as the QMC weights, the weights for Type-F molecules ${{w}_{free}}$ and Type-C molecules ${{w}_{hydro}}$ are
\begin{equation}  \label{eq:weightOfCAndFButDiff}
{{w}_{free}}=\frac{\tau }{\tau +\Delta t},\quad {{w}_{hydro}}=\frac{\Delta t}{\tau +\Delta t}.
\end{equation} 
In the SUWP-Rykov, the relaxation time is calculated by using the translational temperature according to Eq.(\ref{eq:tauOnlyTtr}).

\subsubsection{\label{sec::NumericalScheme}Hybrid framework}
In this study, the mass, momentum, energy, and rotational energy are defined as $\vec{Q}$
\begin{equation}  \label{eq:consVar}
\vec{Q}=\left( \begin{matrix}
  \rho   \\ 
  \rho \vec{U}  \\ 
  \rho E  \\ 
  \rho {{E}_{rot}}  \\ 
\end{matrix} \right) \Omega,
\end{equation} 
where $\Omega $ is the volume of a cell. And the total variables in the cell are denoted as ${{\vec{Q}}_{total}}$, and its evolution equation can be written as:
\begin{equation}  \label{eq:macroVar}
\vec{Q}_{total\text{ }}^{n+1}=\vec{Q}_{total\text{ }}^{n}-\Delta t \sum{{{{\vec{F}}}_{hydro\text{ }}}A+\vec{S} +\vec{Q}_{micro}^{n},}
\end{equation} 
where $A$ is the area of the cell interface, and ${{\vec{F}}_{hydro}}$ is the macroscopic flux contributed by Type-C molecules. The contribution of the Type-F molecules transport through the cell interfaces is denoted by $\vec{Q}_{micro}^{n}$. The total variables are updated by these two types of molecules. In Eq.(\ref{eq:macroVar}), $\vec{S}=\left( 0, \vec{0}, -\dot{\omega}, \dot{\omega} \right)^{T}$ is source term caused by internal energy relaxation, where the internal energy of the inelastic collision is assigned according to the internal degrees of freedom: 
\begin{equation}  \label{eq:sourceTerm}
\dot{\omega}=\frac{w_{hydro}}{Z_{rot}}\left( \frac{K_{rot}}{3+K_{rot}} \rho E - \rho E_{rot} \right).
\end{equation}

The macroscopic N-S solver calculates the macroscopic flux ${{\vec{F}}_{hydro}}$ based on the present total variables $\vec{Q}_{total}^{n}$ in cell. Macroscopic flux is represented as follows by Eq.(\ref{eq:fOfTheTimestep}):
\begin{equation}  \label{eq:Fhydro}
{{\vec{F}}_{hydro}}={{w}_{hydro}}\left( {{{\vec{F}}}_{inv}}+{{c}_{vis}}{{{\vec{F}}}_{vis}} \right),
\end{equation} 
where ${{\vec{F}}_{inv}}$ and ${{\vec{F}}_{vis}}$ are the inviscid and viscous fluxes obtained by the macroscopic solver, respectively. 

\begin{figure}[!h]
    \centering
    \includegraphics[trim=0 10 0 0, clip, width=0.95\textwidth]{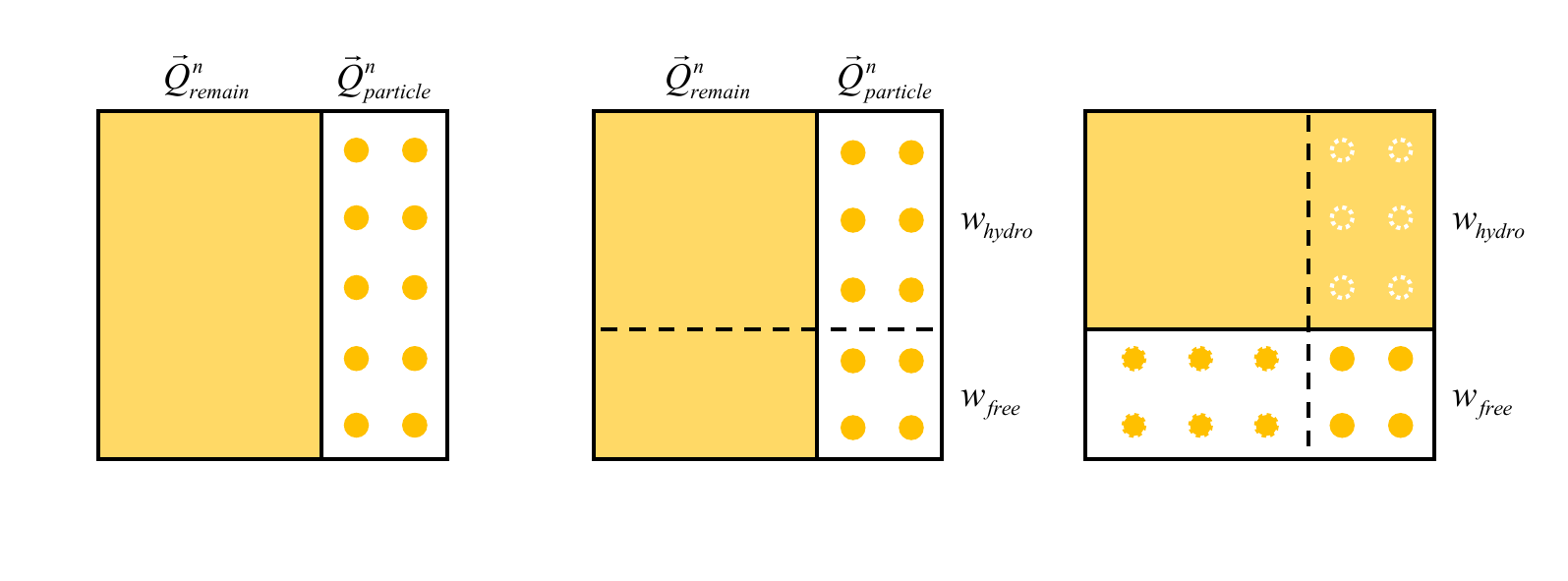}
    \caption{\label{Fig:suwpProcess} Particle remove and reconstruct process of SUWP method.}
\end{figure}

For each time step, there are the present total variables $\vec{Q}_{total}^{n}$ and particle information within the cell. By statistically counting the particles in the cell, their represented macroscopic variables are obtained as $\vec{Q}_{particle}^{n}$. Then, by subtracting the particle variables $\vec{Q}_{particle}^{n}$ from the total variables $\vec{Q}_{total}^{n}$, the remaining non-particle variables are calculated as $\vec{Q}_{remain}^{n}$:
\begin{equation}  \label{eq:qRemain}
\vec{Q}_{remain}^{n}=\vec{Q}_{total}^{n}-\vec{Q}_{particle}^{n}.
\end{equation} 

The particle solver is then used to remove and reconstruct particles in the target cell using QMC weights and remaining non-particle variables $\vec{Q}_{remain}^{n}$. The process is illustrated in Fig.\ref{Fig:suwpProcess}. The particles representing the Type-C molecules are removed from the original Type-F ones, and then, based on the weights for Type-F molecules, the corresponding particles are reconstructed from the remaining non-particle variables $w^{*}$. The new reconstructed particles are then randomly distributed in cell. This removing and reconstruction process actually realized the information exchange between N-S solver and DSMC solver. 

Upon completion in the above process, the particles in cell undergo a time step transport, and $\vec{Q}_{micro}^{n}$ of cell is obtained from the particle in cell:
\begin{equation}  \label{eq:qMicro}
    \vec{Q}_{micro}^{n}=
    \left( 
    \begin{matrix}
         & \sum\limits_{i=1}^{N_{a}}{m_{i}} \\ 
         & \sum\limits_{i=1}^{N_{a}}{m_{i}\vec{v}_{i}} \\ 
         & \sum\limits_{i=1}^{N_{a}}{m_{i}\left( \frac{1}{2}{{\vec{v}_{i}{\,}}^{2}}+{\eta}_{i} \right)} \\ 
         & \sum\limits_{i=1}^{N_{a}}{m_{i}{\eta}_{i}} \\ 
    \end{matrix}
    \right)_{a}
    -
    \left( 
    \begin{matrix}
         & \sum\limits_{i=1}^{N_{b}}{m_{i}} \\ 
         & \sum\limits_{i=1}^{N_{b}}{m_{i}\vec{v}_{i}} \\ 
         & \sum\limits_{i=1}^{N_{b}}{m_{i}\left( \frac{1}{2}{{\vec{v}_{i}{\,}}^{2}}+{\eta}_{i} \right)} \\ 
         & \sum\limits_{i=1}^{N_{b}}{m_{i}{\eta}_{i}} \\ 
    \end{matrix}
    \right)_{b},
\end{equation} 
where subscripts ``b'' and ``a'' respectively represent the variables of the particles before and after the transport process in the target cell. The $N$ is the total number of particles in this cell before or after the transport process. 

The following subsections describe the detailed of particle solver and the macroscopic solver used in SUWP-Rykov.

\subsubsection{\label{sec::ParticleSolver}Particle solver}

In the SUWP method, the particle solver is responsible for particle removal, reconstruction, transport, and counting the flux of particles across cell interfaces.

~\\

\textbf{The removal and reconstruction of particles}:

The process of particle removal and reconstruction are performed synchronously. First, all particles in a cell are traversed to statistically determine variables $\vec{Q}_{particle}^{n}$. Then, the remaining non-particle variables $\vec{Q}_{remain}^{n}$ in cell are calculated using Eq.(\ref{eq:qRemain}).

Based on $w_{hydro}$, new Type-C molecules are removed from the Type-F molecules. A random number $R_n$ that is uniformly distributed in the interval [0,1] is used to test all the particles in cell. If the random number of the particle is less than the weight of the Type-C molecules, i.e. $R_n<{{w}_{hydro}}$, then this particle is removed. 

For reconstructive particles, using the remaining non-particle variables $\vec{Q}_{remain}^{n}$ and the weight of the Type-F molecules are multiplied to give the macroscopic variables ${{w}_{free}}\cdot \vec{Q}_{remain}^{n}$ for Type-F molecules , which is new emerged from macroscopic variables for Type-C particles. The velocity and rotational energy of the sampled particle follow the Rykov velocity distribution function given by Eq.(\ref{eq:RykovModel}). A random number $R_n<{1/{Z}_{rot}}$ ($R_n$ uniformly distributed in the interval$[0,1]$) is used to determine whether the reconstructed particle follows the translational equilibrium distribution ($R_n<1/Z_{rot}$) or the rotational equilibrium distribution ($R_n\geq1/Z_{rot}$) of Eq.(\ref{eq:gTrAndgRot}). Based on a chosen distribution, the particle mass, velocity, and rotational energy are assigned to every generated particle. 

After the reconstruction process, the cell gets its particles representing the Type-F particles a the present time step.

~\\

\textbf{The transport of the particle}:
\begin{figure}[!h]
    \centering
    \includegraphics[width=0.95\textwidth]{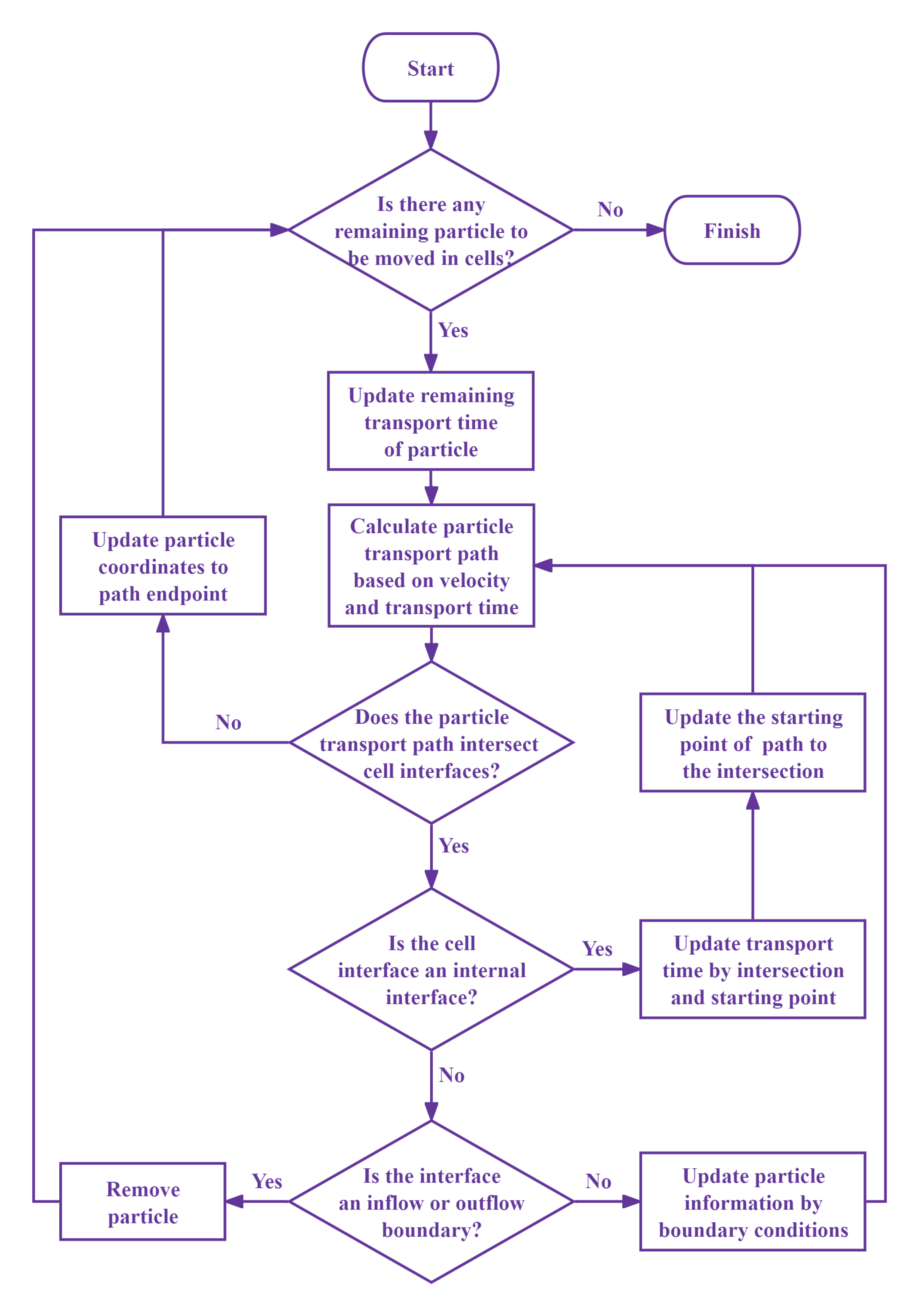}
    \caption{\label{Fig:suwpTransport} Particle transport process of SUWP method.}
\end{figure}

During the transport process, particles move to their respective coordinates in the physical space based on their velocities and the time step. Let the starting point coordinate of the particle transport path be denoted as ${{\vec{P}}_{start}}$ and the end point coordinate as ${{\vec{P}}_{end}}$. The transport process can be expressed as follows:

\begin{equation}  \label{eq:transportPoint}
{{\vec{P}}_{end}}={{\vec{P}}_{start}}+\vec{v}\cdot \Delta t,
\end{equation} 

In a structured mesh, there is a direct correspondence between the coordinates of cells and their indices, which allows particles to be moved directly to the endpoint of their transport path, then looks for the cell where the endpoint is located. However, in an unstructured mesh, the cost of searching for cells by coordinates is relatively high. Therefore, it is more efficient to determine the destination cell for a particle by intersections of its transport path with cell interfaces.

In the solver implemented in this paper, unstructured meshes are used for better coupling with the common macroscopic solvers. The process of particle transport is illustrated in Fig.\ref{Fig:suwpTransport} and can be expressed as the following five steps: 

\textbf{{Step 1:}} At each time step, first check if there are any particles need to move in the cell. If there are, assign the time step $\Delta_t$ as the remaining transport time for a particle.

\textbf{{Step 2:}} For a moving particle, calculate its transport path based on the remaining transport time and particle velocity. Check the transport path for intersections with any cell interface.

\textbf{{Step 3:}} If there is no intersection, it means the particle has not moved out of the starting cell, so update the particle coordinates directly to the endpoint coordinates. If there is an intersection, determine the type of interface. 

\textbf{{Step 4:}} For internal field interfaces, move the particle to the corresponding adjacent cell before continuing transporting. For boundary interfaces, update the particle's information according to the boundary conditions, and then allow the particle to continue its transport. 

\textbf{{Step 5:}} For particles that cross the cell interface, calculate their effect on the variables according to Eq.(\ref{eq:qMicro}) and use this to update the flow field. 

\textbf{{Step 6:}} Repeat the above steps until all particles in the field have completed their transport.

\subsubsection{\label{sec::MacroscopicSolver}Macroscopic solver}

In the SUWP method, the macroscopic solver is responsible for calculating the macroscopic fluxes at the cell interfaces and updating the macroscopic variables in the cells. The present SUWP-Rykov method uses the KIF~\cite{KIF1, KIF2} as the inviscid flux scheme in the macroscopic solver. This scheme combines the kinetic flux-vector splitting (KFVS) method and the totally thermalized transport (TTT) method through a weighted coupling~\cite{Xu1998GaskineticSF}. The SUWP-Rykov method adds a rotational energy equation and rotational energy flux to the basic KIF method. It is in a similar to the way of adding the vibrational energy equation to the traditional two-temperature N-S solver~\cite{anderson1989hypersonic}.

The rotational energy equation can be obtained by integrating the rotational energy with Eq.\ref{eq:twoT_Rykov}. It should be noted that the pressure in the momentum equation of the two-temperature KIF is the $p_{tr}$.

The KFVS flux in this work can be written as:

\begin{equation}  \label{eq:momOfKFVS}
\vec{K}=\left\langle(\vec{v} \cdot \vec{n}) {\vec{\psi}} g_L\right\rangle_R+\left\langle(\vec{v} \cdot \vec{n}) {\vec{\psi}} g_R\right\rangle_L,
\end{equation} 
where the left half moment $\left\langle\cdot\right\rangle_L$ and right half moment $\left\langle\cdot\right\rangle_R$ are defined as

\begin{equation}  \label{eq:momOfKFVS_LR}
\begin{aligned}\langle\cdot\rangle_L & =\int_{-\infty}^0 d v_1 \int_{-\infty}^{+\infty} d v_2 \int_{-\infty}^{+\infty} d v_3 \int_{0}^{+\infty}(\cdot) d {\eta}, \\ \langle\cdot\rangle_R & =\int_0^{+\infty} d v_1 \int_{-\infty}^{+\infty} d v_2 \int_{-  \infty}^{+\infty} d v_3 \int_{0}^{+\infty}(\cdot) d {\eta}.\end{aligned}
\end{equation} 

According to the Eq.(\ref{eq:momOfKFVS}), the KFVS flux $\vec{K}$ can be represented in an explicit form:

\begin{equation}  \label{eq:kfvsFlux}
\begin{aligned}
{{K}_{mass}} & =\frac{1}{2}\left( {{\rho }_{L}}{{u}_{L}}+{{\rho }_{R}}{{u}_{R}} \right)+\frac{1}{2}\left( {{\rho }_{L}}{{u}_{L}}{{\delta }_{L}}-{{\rho }_{R}}{{u}_{R}}{{\delta }_{R}} \right)+\frac{1}{2}\left( {{\rho }_{L}}{{\theta }_{L}}-{{\rho }_{R}}{{\theta }_{R}} \right), \\ 
{{K}_{xmon}} & =\frac{1}{2}\left[ \left( {{\rho }_{L}}u_{L}^{2}+{{p}_{tr,L}} \right)+\left( {{\rho }_{R}}u_{R}^{2}+{{p}_{tr,R}} \right) \right]+\frac{1}{2}\left[ \left( {{\rho }_{L}}u_{L}^{2}+{{p}_{tr,L}} \right){{\delta }_{L}} \right.\left. -\left( {{\rho }_{R}}u_{R}^{2}+{{p}_{tr,R}} \right){{\delta }_{R}} \right] \\
 & +\frac{1}{2}\left( {{\rho }_{L}}{{u}_{L}}{{\theta }_{L}}-{{\rho }_{R}}{{u}_{R}}{{\theta }_{R}} \right), \\ 
{{K}_{ymon}} & =\frac{1}{2}\left( {{\rho }_{L}}{{u}_{L}}{{v}_{L}}+{{\rho }_{R}}{{u}_{R}}{{v}_{R}} \right)+\frac{1}{2}\left( {{\rho }_{L}}{{u}_{L}}{{v}_{L}}{{\delta }_{L}}-{{\rho }_{R}}{{u}_{R}}{{v}_{R}}{{\delta }_{R}} \right)+\frac{1}{2}\left( {{\rho }_{L}}{{v}_{L}}{{\theta }_{L}}-{{\rho }_{R}}{{v}_{R}}{{\theta }_{R}} \right),\\ 
{{K}_{zmon}} & =\frac{1}{2}\left( {{\rho }_{L}}{{u}_{L}}{{w}_{L}}+{{\rho }_{R}}{{u}_{R}}{{w}_{R}} \right)+\frac{1}{2}\left( {{\rho }_{L}}{{u}_{L}}{{w}_{L}}{{\delta }_{L}}-{{\rho }_{R}}{{u}_{R}}{{w}_{R}}{{\delta }_{R}} \right)+\frac{1}{2}\left( {{\rho }_{L}}{{w}_{L}}{{\theta }_{L}}-{{\rho }_{R}}{{w}_{R}}{{\theta }_{R}} \right), \\ 
{{K}_{energy}} & ={{K}_{tr}} + {{K}_{rot}}, \\ 
{{K}_{tr}} & =\frac{1}{2}\left( {{\rho }_{L}}{{u}_{L}}{{h}_{L}}+{{\rho }_{R}}{{u}_{R}}{{h}_{R}} \right)+\frac{1}{2}\left( {{\rho }_{L}}{{u}_{L}}{{h}_{L}}{{\delta }_{L}}-{{\rho }_{R}}{{u}_{R}}{{h}_{R}}{{\delta }_{R}} \right) \\
 & +\frac{1}{2}\left[ \left( {{\rho }_{L}}{{h}_{L}}-\frac{{{p}_{tr,L}}}{2} \right){{\theta }_{L}}-\left( {{\rho }_{R}}{{h}_{R}}-\frac{{{p}_{tr,R}}}{2} \right){{\theta }_{R}} \right], \\ 
{{K}_{rot}} & =\frac{1}{2}\left( {{\rho }_{L}}{{E}_{rot,L}}{{u}_{L}}+{{\rho }_{R}}{{E}_{rot,R}}{{u}_{R}} \right)+\frac{1}{2}\left( {{\rho }_{L}}{{E}_{rot,L}}{{u}_{L}}{{\delta }_{L}}-{{\rho }_{R}}{{E}_{rot,R}}{{u}_{R}}{{\delta }_{R}} \right) \\
 & +\frac{1}{2}\left( {{\rho }_{L}}{{E}_{rot,L}}{{\theta }_{L}}-{{\rho }_{R}}{{E}_{rot,R}}{{\theta }_{R}} \right), \\ 
\end{aligned}
\end{equation} 
where the expression of $\delta$ and $\theta$ is
\begin{equation}  \label{eq:deltaAndTheta}
\begin{split}
& {{\delta }_{\alpha }}=\operatorname{erf}\left( \frac{{{u}_{\alpha }}}{\sqrt{2R{{T}_{tr,\alpha }}}} \right), \\ 
 & {{\theta }_{\alpha }}=\sqrt{\frac{2R{{T}_{tr,\alpha }}}{\pi }}\exp \left( -\frac{u_{\alpha }^{2}}{2R{{T}_{tr,\alpha }}} \right), \\
\end{split}
\end{equation} 
where the subscript $\alpha$ can be $L$ or $R$.

The TTT flux $\vec{G}$ in the present KIF is a simple Euler flux using the kinetic averaged values
\begin{equation}  \label{eq:tttFlux}
\begin{split}
& {{G}_{mass}}=\overline{\rho }\overline{u}, \\ 
 & {{G}_{xmon}}=\overline{\rho }{{{\overline{u}}}^{2}}+\overline{p_{tr}}, \\ 
 & {{G}_{ymon}}=\overline{\rho }\overline{u}\overline{v}, \\ 
 & {{G}_{zmon}}=\overline{\rho }\overline{u}\overline{w}, \\
 & {{G}_{energy}}={{G}_{tr}} + {{G}_{rot}}, \\ 
 & {{G}_{tr}}=\overline{\rho }{\overline{u}}\overline{h_{tr}}, \\ 
 & {{G}_{rot}}=\overline{\rho }{\overline{u}}\overline{{{E}_{rot}}},
\end{split}
\end{equation} 
where the kinetic averaged values can be obtained from the following TTT process:
\begin{equation}  \label{eq:tttProcess}
\begin{split}
 & \overline{\rho }=\frac{1}{2}\left( {{\rho }_{L}}+{{\rho }_{R}} \right)+\frac{1}{2}\left( {{\rho }_{L}}{{\delta }_{L}}-{{\rho }_{R}}{{\delta }_{R}} \right), \\ 
 & \overline{\rho u}=\frac{1}{2}\left( {{\rho }_{L}}{{u}_{L}}+{{\rho }_{R}}{{u}_{R}} \right)+\frac{1}{2}\left( {{\rho }_{L}}{{\theta }_{L}}-{{\rho }_{R}}{{\theta }_{R}} \right)+\frac{1}{2}\left( {{\rho }_{L}}{{u}_{L}}{{\delta }_{L}}-{{\rho }_{R}}{{u}_{R}}{{\delta }_{R}} \right), \\ 
 & \overline{\rho v}=\frac{1}{2}\left( {{\rho }_{L}}{{v}_{L}}+{{\rho }_{R}}{{v}_{R}} \right)+\frac{1}{2}\left( {{\rho }_{L}}{{v}_{L}}{{\delta }_{L}}-{{\rho }_{R}}{{v}_{R}}{{\delta }_{R}} \right), \\ 
 & \overline{\rho w}=\frac{1}{2}\left( {{\rho }_{L}}{{w}_{L}}+{{\rho }_{R}}{{w}_{R}} \right)+\frac{1}{2}\left( {{\rho }_{L}}{{w}_{L}}{{\delta }_{L}}-{{\rho }_{R}}{{w}_{R}}{{\delta }_{R}} \right), \\ 
 & \overline{\rho E_{tr}}=\frac{1}{2}\left( {{\rho }_{L}}{{E}_{tr,L}}+{{\rho }_{R}}{{E}_{tr,R}} \right)+\frac{1}{4}\left( {{\rho }_{L}}{{u}_{L}}{{\theta }_{L}}-{{\rho }_{R}}{{u}_{R}}{{\theta }_{R}} \right)+\frac{1}{2}\left( {{\rho }_{L}}{{E}_{L}}{{\delta }_{L}}-{{\rho }_{R}}{{E}_{R}}{{\delta }_{R}} \right), \\ 
 & \overline{\rho {{E}_{rot}}}=\frac{1}{2}\left( {{\rho }_{L}}{{E}_{rot,L}}+{{\rho }_{R}}{{E}_{rot,R}} \right)+\frac{1}{2}\left( {{\rho }_{L}}{{E}_{rot,L}}{{\delta }_{L}}-{{\rho }_{R}}{{E}_{rot,R}}{{\delta }_{R}} \right), \\
\end{split}
\end{equation} 
where $\overline{u}=\overline{\rho u}/\overline{\rho }$, $\overline{v}=\overline{\rho v}/\overline{\rho }$, $\overline{w}=\overline{\rho w}/\overline{\rho }$,
$\overline{e_{tr}}=\overline{\rho E_{tr}}/\overline{\rho} - 1/2(\overline{u}^2+\overline{v}^2+\overline{w}^2)$,
$\overline{p_{tr}}={(\gamma-1)\overline{\rho}\overline{e_{tr}}}$ and $\overline{h_{tr}}={\overline{\rho E_{tr}} /\overline{\rho} +\overline{p} /\overline{\rho}}$.

In KIF, the KFVS fluxes and TTT fluxes are coupled as
\begin{equation}  \label{eq:coupleKFVSandTTT}
{{\vec{F}}_{inv}}=\beta \vec{K}+\left( 1-\beta  \right)\vec{G},
\end{equation} 
where $\beta$ is a kind of weight (the weight of the free transport mechanism at a cell interface):
\begin{equation}  \label{eq:kifbeta}
\beta=F_{s}\left(\Delta P,0,1\right)F_{s}(Ma,0.5Ma_{back},Ma_{back}),
\end{equation} 
where $F_{s}$ is a smooth function consisting of sine function. The expressions of $F_{s}$, $\Delta P$ and $Ma_{back}$ are in Ref.~\cite{KIF2}.
The coupling approach and proper chosen of $\beta$ ensure that most continuum regions in the flow field adopt the TTT scheme. Meanwhile, the KFVS scheme possesses the capability to capture discontinuities. By using the expression of $\beta$, the KFVS scheme dominates near shock waves, while the TTT scheme dominates near the boundary layer.

In this paper, the gradients of macroscopic variables are computed using the least squares method with the application of the Venkatakrishnan gradient limiter~\cite{venkatakrishnan1993implicit}.
The viscous flux ${{\vec{F}}_{vis}}$ is calculated by the central scheme. The physical variables and their gradients at the cell interfaces are computed by taking the weighted average of the center values of two adjacent cells. These values are then utilized in calculating the viscous flux. The computational framework for the flux is the same with that for traditional N-S equations, except the viscous flux is slight is modified by Eq.(\ref{eq:Fhydro}) to match the multi-scale physics. By treating the translational energy and rotational energy separately, the flux of the N-S solver is in good consistency with the evolution of colliding particles based on the Rykov model.

\subsubsection{\label{sec::BoundaryCondition}Boundary condition}
The boundary condition of the present SUWP-Rykov method are also a coupling one. For the inflow boundary, the particle solver and the macroscopic solver compute the detailed information of inflow particles and macroscopic net flux, respectively, according to weights ${{w}_{free}}$ and ${{w}_{hydro}}$. For the wall boundary, the particle solver updates the particle velocity and energy of the particles hitting the wall according to wall speed and temperature, while the macroscopic solver calculates the flux by directly multiplying the N-S one by weight ${{w}_{hydro}}$.
It can be seem that the handling of the macroscopic boundary condition is similar to the classical N-S solver. Therefore, the following discussion focuses on the boundary condition of the particle solver.

~\\

\textbf{Inflow and outflow boundary conditions}:

The inflow boundary treatment of the particle solver is consistent with the DSMC method, except the inflow mass flux is set according to the scale weights as follows:

\begin{equation}  \label{eq:boundaryInflow}
\begin{split}
  & {{{\vec{F}}}_{DSMC}}=\left\{ \rho \exp \left( -\frac{u_{n}^{2}}{2RT} \right)\sqrt{\frac{RT}{2\pi }} \right.\left. +\frac{{{u}_{n}}}{2}\left[ 1+erf\left( \frac{{{u}_{n}}}{\sqrt{2RT}} \right) \right] \right\}, \\ 
 & {{{\vec{F}}}_{inflow}}={{w}_{free}}\cdot {{{\vec{F}}}_{DSMC}}. \\ 
\end{split}
\end{equation} 

The final mass of particles entering from the inflow boundary is ${{\vec{F}}_{inflow}}\cdot \Delta t \cdot 
A$, where $A$ is the area of the inflow interface. The remaining transport time for particles entering from the interface is $R_n\cdot \Delta t$ ($R_n$ is a real random number in $[0,1]$), and their coordinates are randomly distributed on the inflow boundary (The method can be referred to in the \ref{sec::appendixA}). 

~\\

\textbf{Wall boundary condition}:

Particles should update their velocities and rotational energies after hitting the solid wall according to the Maxwell distribution at the wall. Therefore, the bounce back velocity in face coordinate system of wall can be sampled from:

\begin{equation}  \label{eq:boundaryWall}
\vec{v}={{\vec{U}}_{wall}}+{}\left( \begin{aligned}
  & \sqrt{2R{{T}_{wall\text{ }}}}\sqrt{-\log \left( R_{n1} \right)} \\ 
 & \sqrt{2R{{T}_{wall\text{ }}}}\sqrt{-\log \left( R_{n2} \right)}\cos \left( 2\pi \cdot R_{n3} \right) \\ 
 & \sqrt{2R{{T}_{wall\text{ }}}}\sqrt{-\log \left( R_{n2} \right)}\sin \left( 2\pi \cdot R_{n3} \right) \\ 
\end{aligned} \right),
\end{equation} 
where ${{\vec{U}}_{wall}}$ is the wall movement speed. $T_{wall}$ is the wall temperature. The expression for the rotational energy of particles that hit the wall is
\begin{equation}  \label{eq:boundaryRotE}
\eta =-\log \left( R_n \right)R{{T}_{wall}},
\end{equation} 

\subsubsection{\label{sec::FlowChart}Process of SUWP-Rykov}

\begin{figure}
    \centering
    \includegraphics[width=0.95\textwidth]{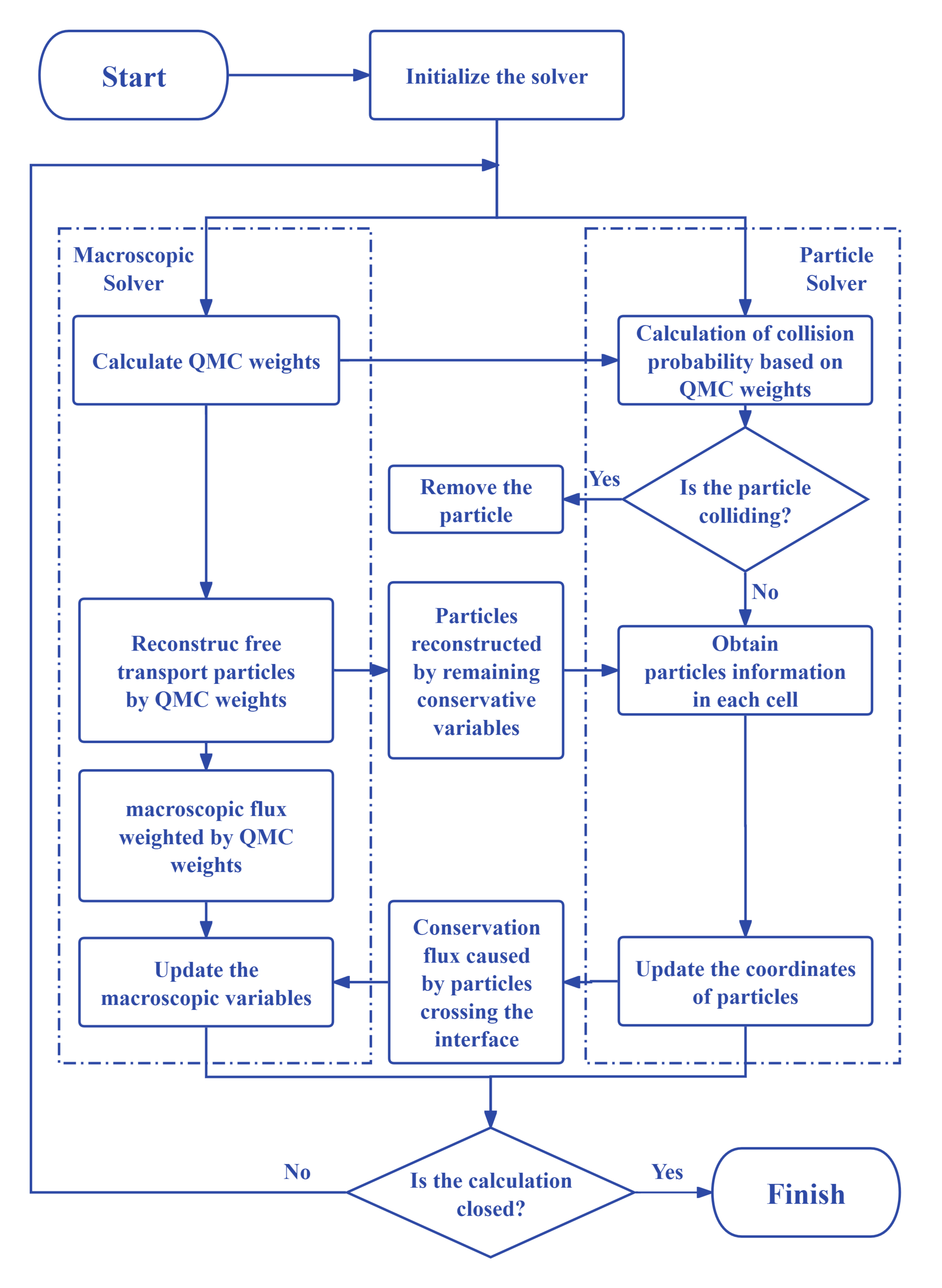}
    \caption{\label{Fig:solverProcess} The route of SUWP-Rykov method.}
\end{figure}

The computational process of the SUWP-Rykov method is illustrated in Fig.\ref{Fig:solverProcess}.
\begin{itemize}
\item[$\bullet$]At the initial of a time step, each cell holds both particle and macroscopic information. Then, calculate the weights for Type-C molecules and Type-F molecules at the present time step by the QMC mechanism.
\item[$\bullet$]New Type-C particles is removed based on QMC weights (absorb them back into the macroscopic information). New Type-F particles denotes reconstructed particles also based on QMC weights by sampling from the Rykov equilibrium distribution.
\item[$\bullet$]The particle solver moves the existing particles to their new cells, updates their coordinates, and records the flux caused by particles crossing cell interfaces.
\item[$\bullet$]The macroscopic solver calculates the macroscopic flux at cell interfaces by the multiplying the N-S flux by the weights for Type-C particles.
\item[$\bullet$]The macroscopic variables are updated by their changes obtained from the statistics in particle solver and the flux calculated by the macroscopic solver (Eq.(\ref{eq:macroVar})).
\item[$\bullet$]The updating of particle information has already been completed during the particle transport step and requires no additional processing.
\end{itemize}

\section{\label{sec::numricalExamples}Numerical results and Discussions}
\subsection{\label{sec::relax0D}Rotational relaxation in a homogeneous gas}
The test case of the rotational relaxation in a homogeneous gas is used to verify the accuracy of the SUWP-Rykov method in simulating molecular collision processes. The conditions are from Bird's monograph~\cite{Bird_Molecular}. At the initial time, the equilibrium temperature ${{T}_{eq}}$ is 300$K$, the rotational ${{T}_{rot,0}}$ is 0$K$, the translational temperature $T_{tr,0}$ is determined to be 500$K$ from Eq.(\ref{eq:tAndTtrAndTrot}). The initial particle velocities are sampled from a Maxwell distribution function.

The equilibrium temperature ${{T}_{eq}}$ remains constant, and the predicted values for the translational temperature ${{T}_{tr}}$ and the rotational temperature ${{T}_{rot}}$ are as follows:
\begin{equation}  \label{eq:case0Drelax_predictT}
\begin{split}
  & {{T}_{rot}}={{T}_{eq}}+({{T}_{rot,0}}-{{T}_{eq}})\cdot {{e}^{\frac{-t}{{{Z}_{rot}}\tau }}}, \\ 
 & {{T}_{tr}}={{T}_{eq}}+({{T}_{tr,0}}-{{T}_{eq}})\cdot {{e}^{\frac{-t}{{{Z}_{rot}}\tau }}}. \\ 
\end{split}
\end{equation} 


Over time, molecules within the flow field collide with each other, causing the energy transfer from translational motion to rotational one. 

\begin{figure}[!h]
    \centering
    \includegraphics[width=0.6\textwidth]{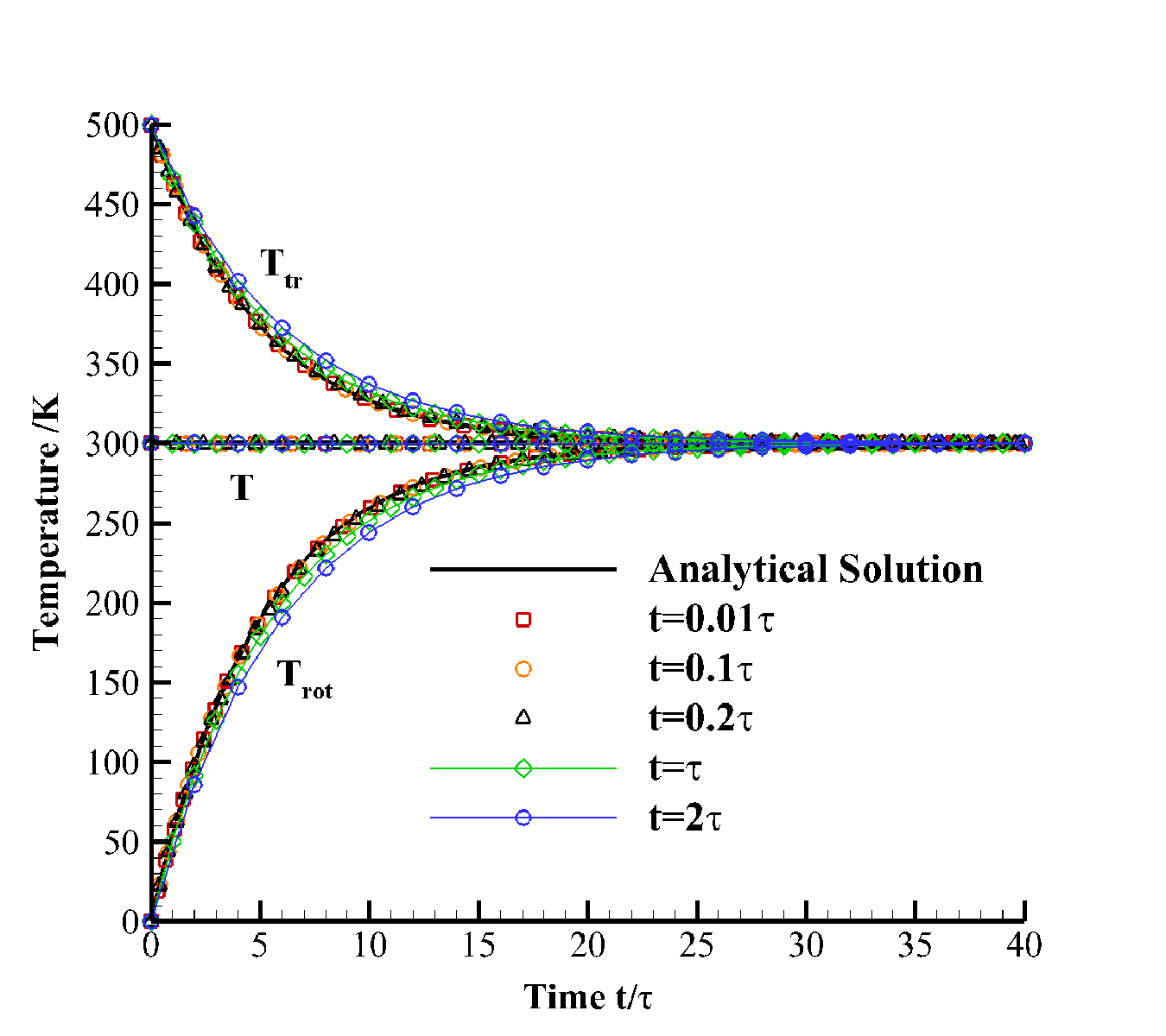}
    \caption{\label{Fig:case01_relax0D}Rotational relaxation in a homogeneous gas.}
\end{figure}

In calculations, the nitrogen gas is chosen as the working gas. The temperature exponent $\omega$ of the VHS model is set to 0.75 and a rotational collision number ${{Z}_{rot}}$ is set to 5. The simulation domain is a cubic region with dimensions of $1\times1\times1$, and the number of simulated particles in the flow field is 4000 to reduce fluctuations. Simulations are performed with five different time steps, $t=0.01\tau$, $0.1\tau$, $0.2\tau$, $\tau$, $2\tau$, and the simulation results are shown in Fig.\ref{Fig:case01_relax0D}.

It can be observed that if the time step is too large, the relaxation rate of the rotational temperature is smaller than the predicted value, in a slightly slower relaxation to the equilibrium state. It is because the relaxation of temperature in the numerical scheme (Eq.(\ref{eq:macroVar})) is approximated by a trapezoid rule, a too large iteration time will leads to extra and nonphysical truncation errors. When the time step $t\le 0.2\tau $, the relaxation curves for both translational temperature and rotational temperature predicted by the SUWP-Rykov coincide with the analytical solution.

\subsection{\label{sec::shock1D}Normal shock structure}

As a standard benchmark, shock structures are used to verify the ability of numerical methods to simulate strong non-equilibrium flows. In classical N-S methods, shock waves are typically treated as discontinuities. However, shock waves are actually smooth continuum structures, and within this region, the flows exhibit strong non-equilibrium. In this paper, the SUWP-Rykov method is used to simulate the nitrogen shock structures for upstream Mach numbers, $Ma = 4$ and $Ma = 7$. The shock structures simulated by the SUWP method are compared with the DSMC~\cite{liu2014unified} and the DUGKS solver mentioned in Ref.\cite{Chen2019Conserved}.

In this case, the inlet and outlet boundary conditions are determined based on the Rankine-Hugoniot relations.
 \begin{equation}  \label{eq:case1D_RH}
\begin{split}
 & \frac{{{\rho }_{2}}}{{{\rho }_{1}}}=\frac{(\gamma +1)M{{a_1}^{2}}}{(\gamma -1)M{{a_1}^{2}}+2}, \\ 
 & \frac{{{T}_{2}}}{{{T}_{1}}}=\frac{\left( 1+\frac{\gamma -1}{2}M{{a_1}^{2}} \right)\left( \frac{2\gamma }{\gamma -1}M{{a_1}^{2}}-1 \right)}{M{{a_1}^{2}}\left( \frac{2\gamma }{\gamma -1}+\frac{\gamma -1}{2} \right)}, \\ 
\end{split}
\end{equation} 
where subscripts ``1'' and ``2'' respectively represent the flow parameters ahead and behind of the shock, corresponding to the inlet and outlet, respectively. The specific heat ratio of the diatomic gas $\gamma =1.4$. The heat index of the VHS model is $\omega =0.74$, the rotational collision number is ${{Z}_{rot}}=2.4$. The flow field is uniformly divided into $200\times1\times1$ cells, with the length of the computational domain being 80 times the mean free path, and the total number of particles in the flow field is 253, 000. For VHS model, the mean free path is defined by the following equation:
 \begin{equation}  \label{eq:case1Dshock_meanFreePath}
{{\lambda }_{\text{mfp}}}=\frac{2(7-2\omega )(5-2\omega )}{15}\frac{\mu }{{{\rho }_{1}}}\sqrt{\frac{1}{2\pi R{{T}_{1}}}}.
\end{equation} 

To facilitate comparison with the DUGKS and DSMC data from the reference literature, the temperature is non-dimensionalized using the following equation:
 \begin{equation}  \label{eq:case1Dshock_dimensionless}
\hat{T}=\frac{T}{{{T}_{1}}},{{\hat{T}}_{tr}}=\frac{{{T}_{tr}}}{{{T}_{1}}},{{\hat{T}}_{rot}}=\frac{{{T}_{rot}}}{{{T}_{1}}}.
\end{equation} 

\begin{figure}[!h]
\centering
\subfigure[\label{Fig:case1D_shockMa4}]{
\includegraphics[width=0.45\textwidth]{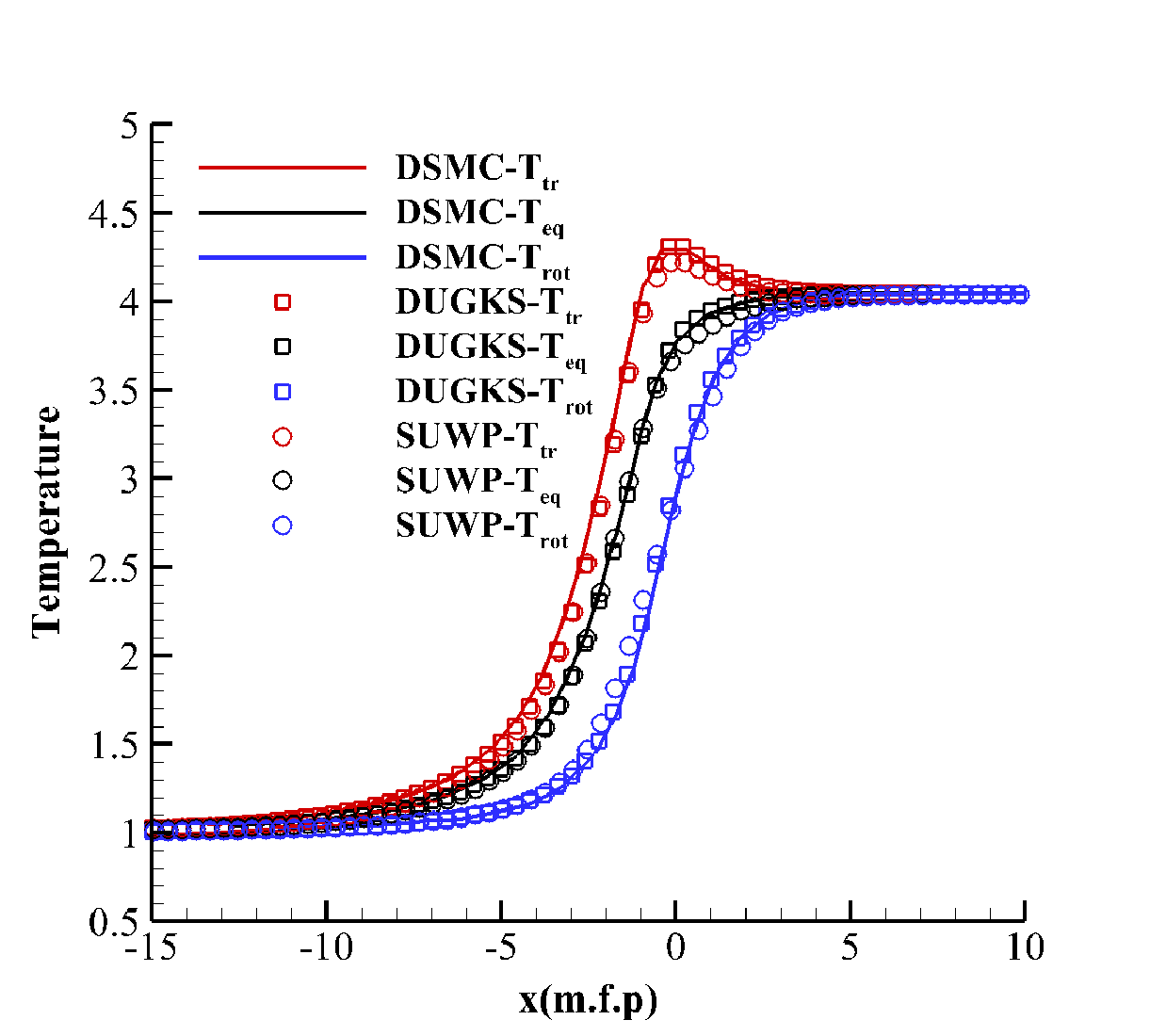}
}\hspace{0.05\textwidth}%
\subfigure[\label{Fig:case1D_shockMa7}]{
\includegraphics[width=0.45\textwidth]{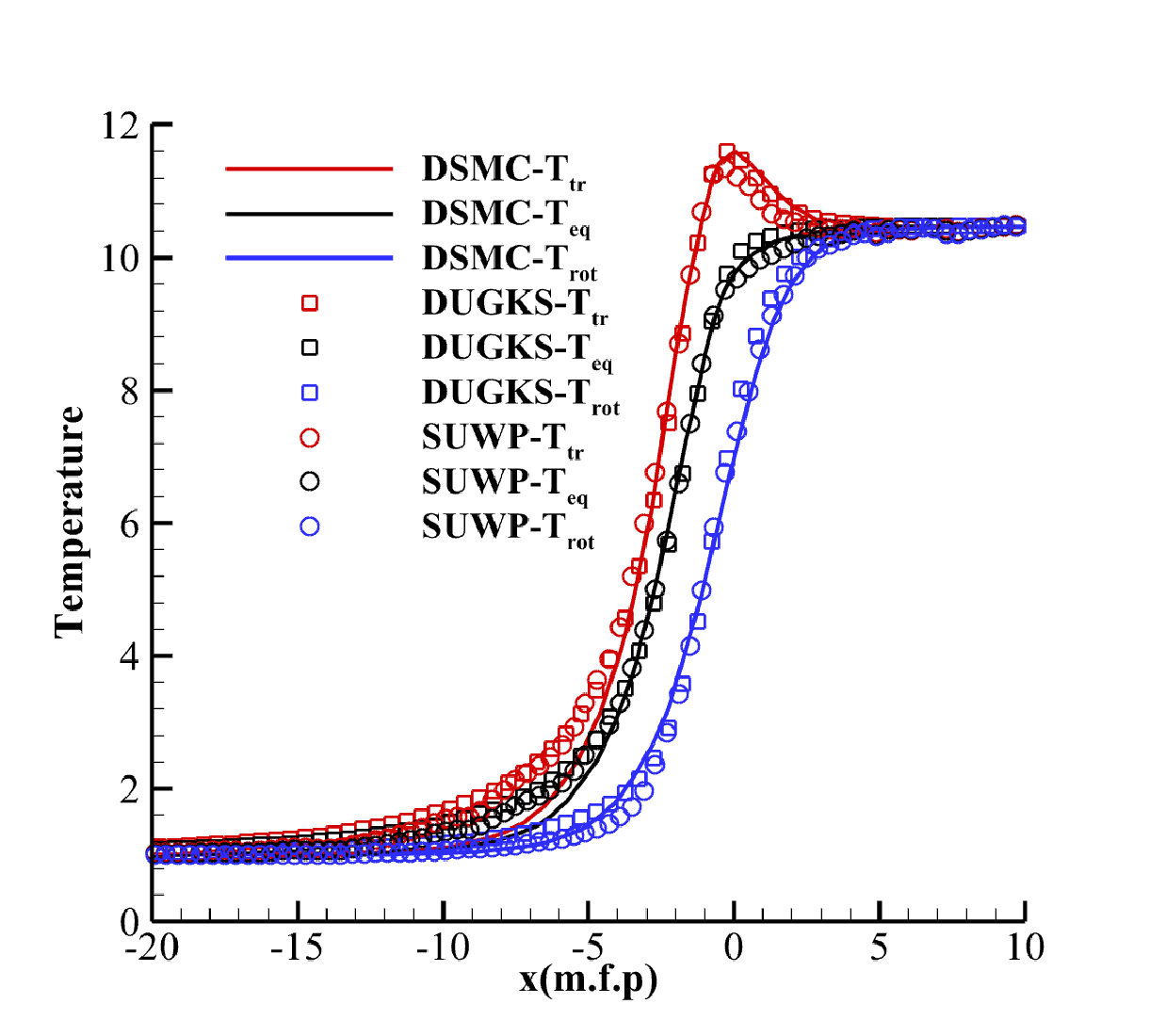}
}
\caption{\label{Fig:case1D_shock} Nitrogen shock structure, (a) Ma=4, (b) Ma=7.}
\end{figure}

The temperature profiles are shown in Fig. \ref{Fig:case1D_shock}. Under the Ma=4 condition, a good agreement between the SUWP-Rykov and the diatomic DUGKS and DSMC methods is achieved. Under the Ma=7 condition, the equilibrium temperature and the translational temperature of SUWP-Rykov and DUGKS match well with each other, but have a slight ``early rise'' in front of the shock. This is influenced by the BGK-type equation used in the present SUWP and DUGKS, and can be improved by using a velocity dependent relaxation rate~\cite{xu2021modeling, yuan_wu_2022}.

\subsection{\label{sec::flat2D}Flow past a flat plate}

\begin{figure}[!h]
    \centering
    \includegraphics[width=0.65\textwidth]{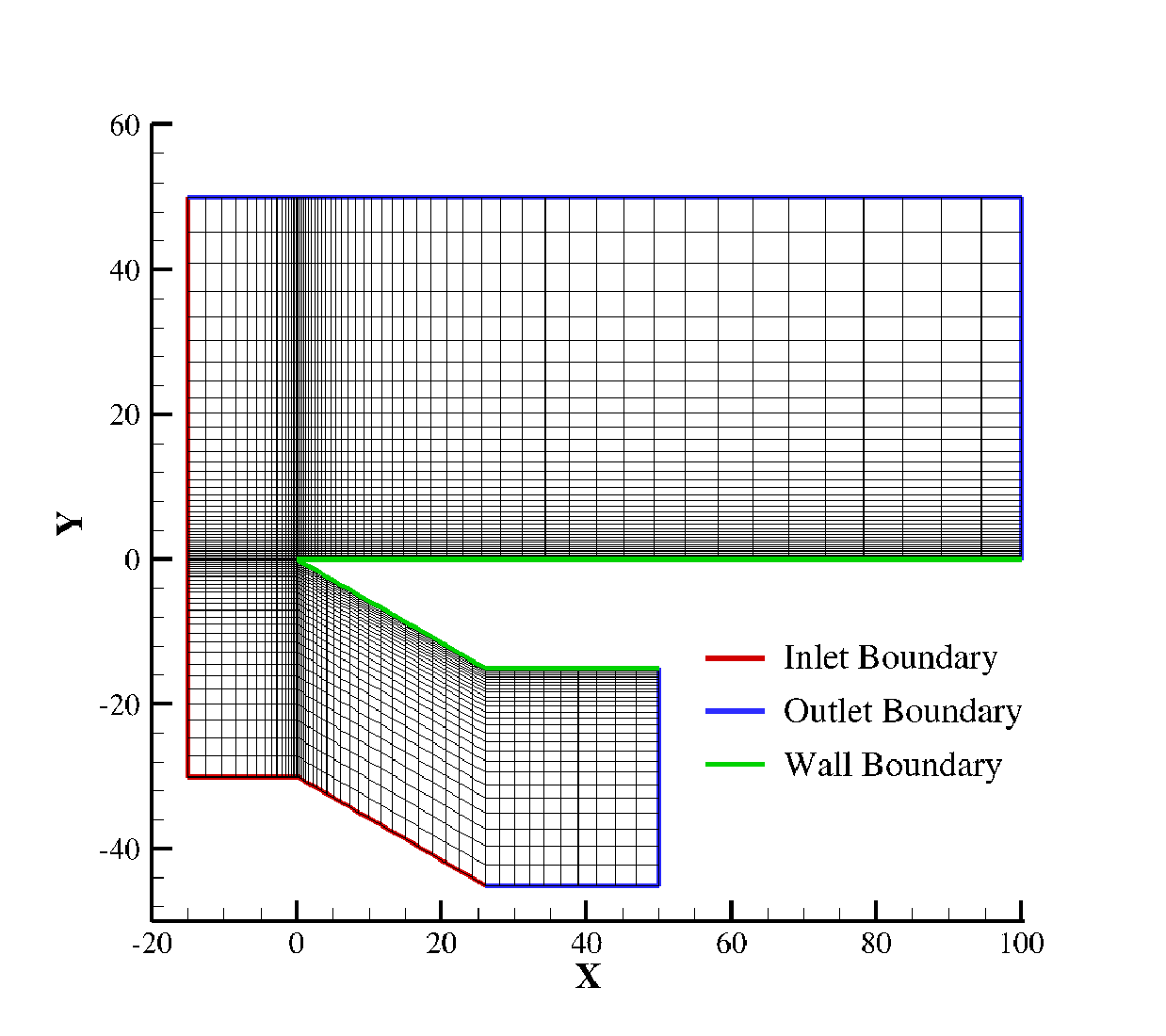}
    \caption{\label{Fig:case03_flatMesh}Computational mesh for the flat plate simulation.}
\end{figure}

The case of flow past a flat plate is based on the wind tunnel experiment run34~\cite{tsuboi2005experimental} conducted by Tsuboi and Matsumotod. It is often chosen to test the ability of a method for simulating non-equilibrium phenomena in the boundary layer of multi-scale flow. The computational domain and physical mesh are shown in Fig.\ref{Fig:case03_flatMesh}. The flow field is divided into a total of 3644 cells. The wall temperature of the flat plate $T_{wall}$ is 290K, and the thickness of the flat plate is 15mm. The angle of the front-end sharp corner is 30 degrees. In this case, the working gas is nitrogen, and the specific heat ratio is $\gamma =1.4$. The temperature exponent of the VHS model $\omega =0.75$. The rotational collision number is ${{Z}_{rot}}=3.5$. The free-stream density is ${{\rho }_{\infty }}=6.15\times {{10}^{-5}}kg\cdot {{m}^{-3}}$, the velocity is ${{U}_{\infty }}=1073.94m/s$, the Mach number is $Ma=4.89$ and temperature is ${{T}_{\infty }}=116K$. With a reference length of 1mm, $Kn=0.78$.

\begin{figure}[!h]
\centering
\subfigure[\label{Fig:case2D_flat_Con_Rho}]{
\includegraphics[width=0.45\textwidth]{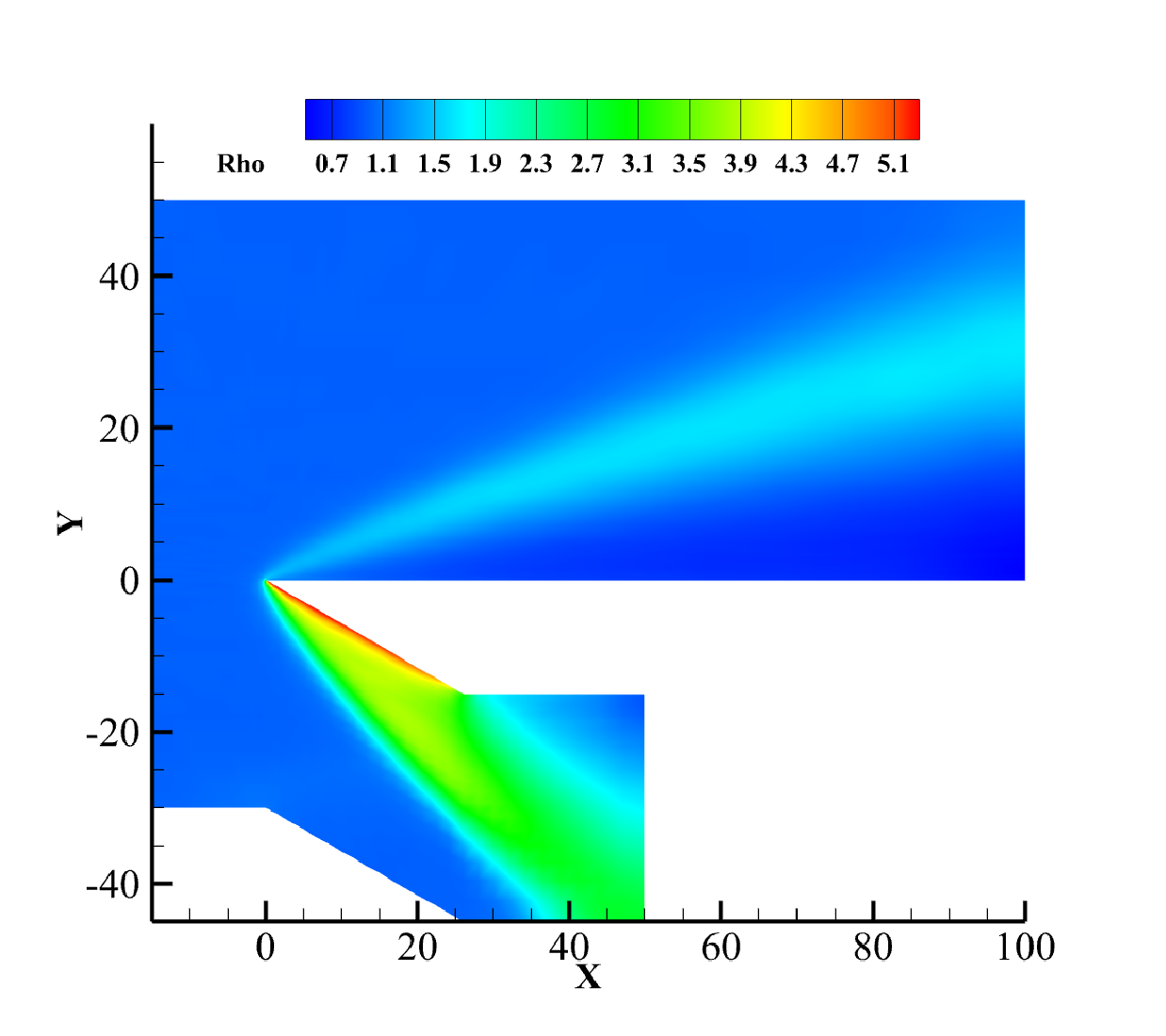}
}\hspace{0.05\textwidth}%
\subfigure[\label{Fig:case2D_flat_Con_U}]{
\includegraphics[width=0.45\textwidth]{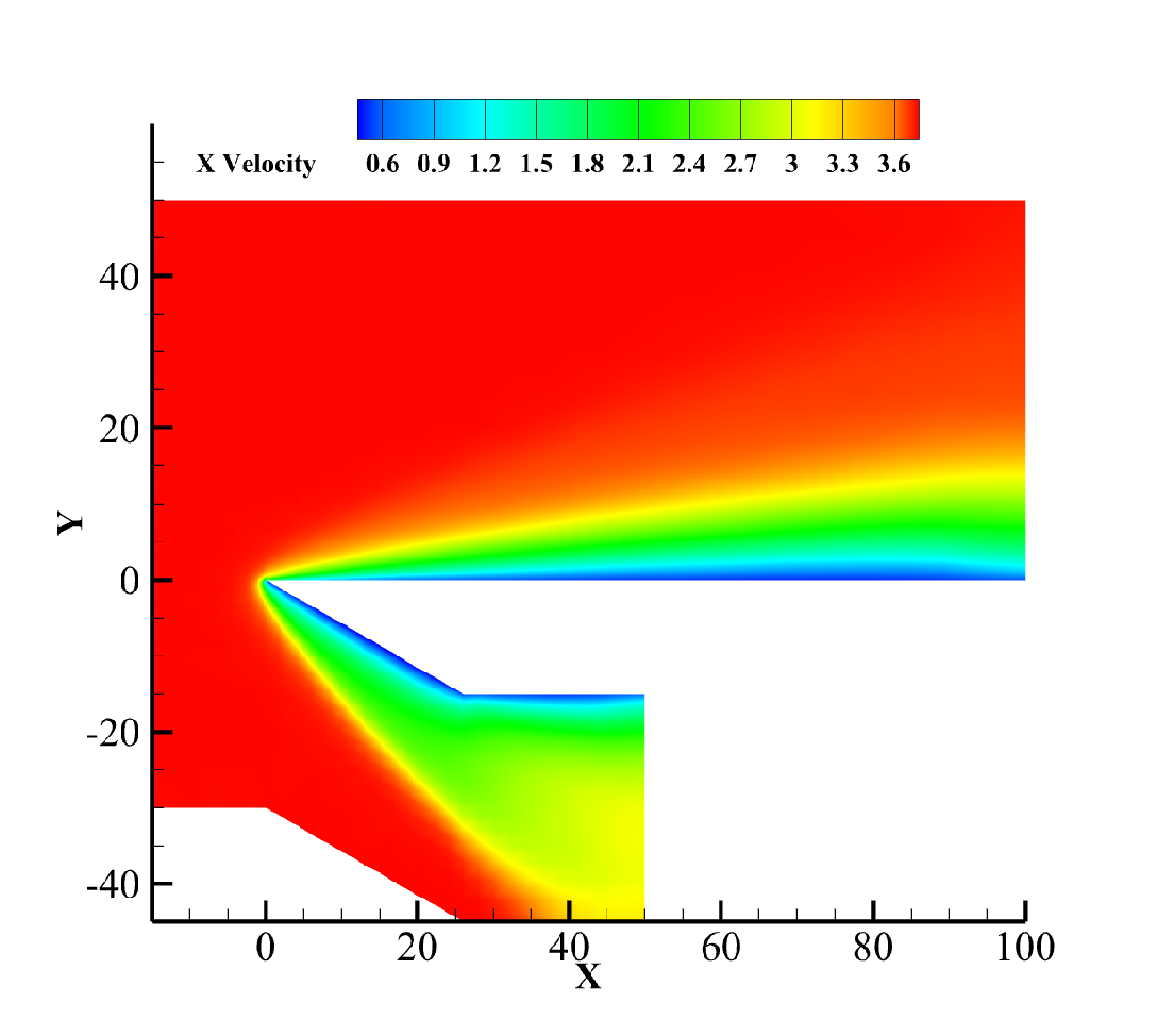}
}\\
\subfigure[\label{Fig:case2D_flat_Con_Teq}]{
\includegraphics[width=0.45\textwidth]{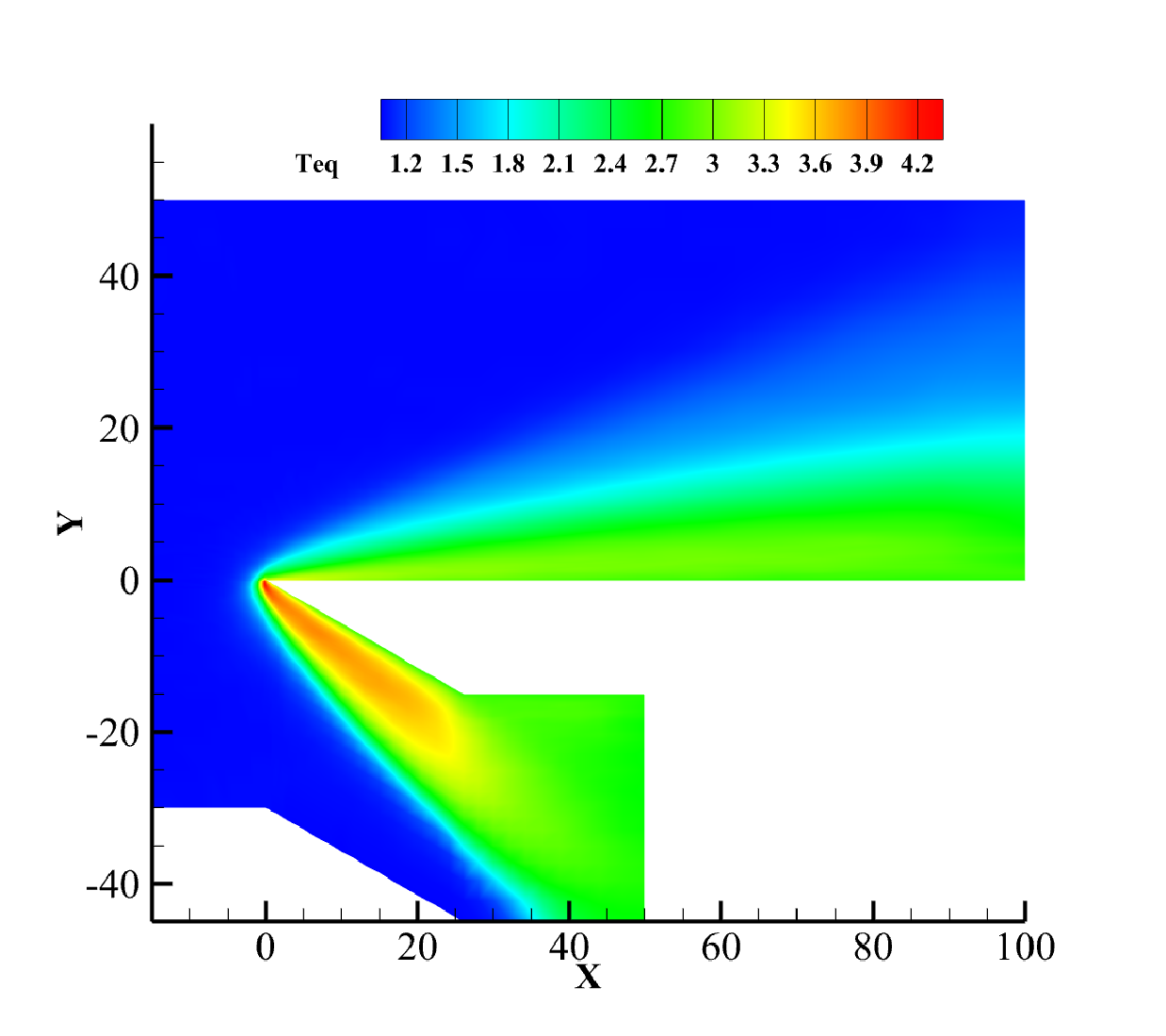}
}\hspace{0.05\textwidth}%
\subfigure[\label{Fig:case2D_flat_Con_Trot}]{
\includegraphics[width=0.45\textwidth]{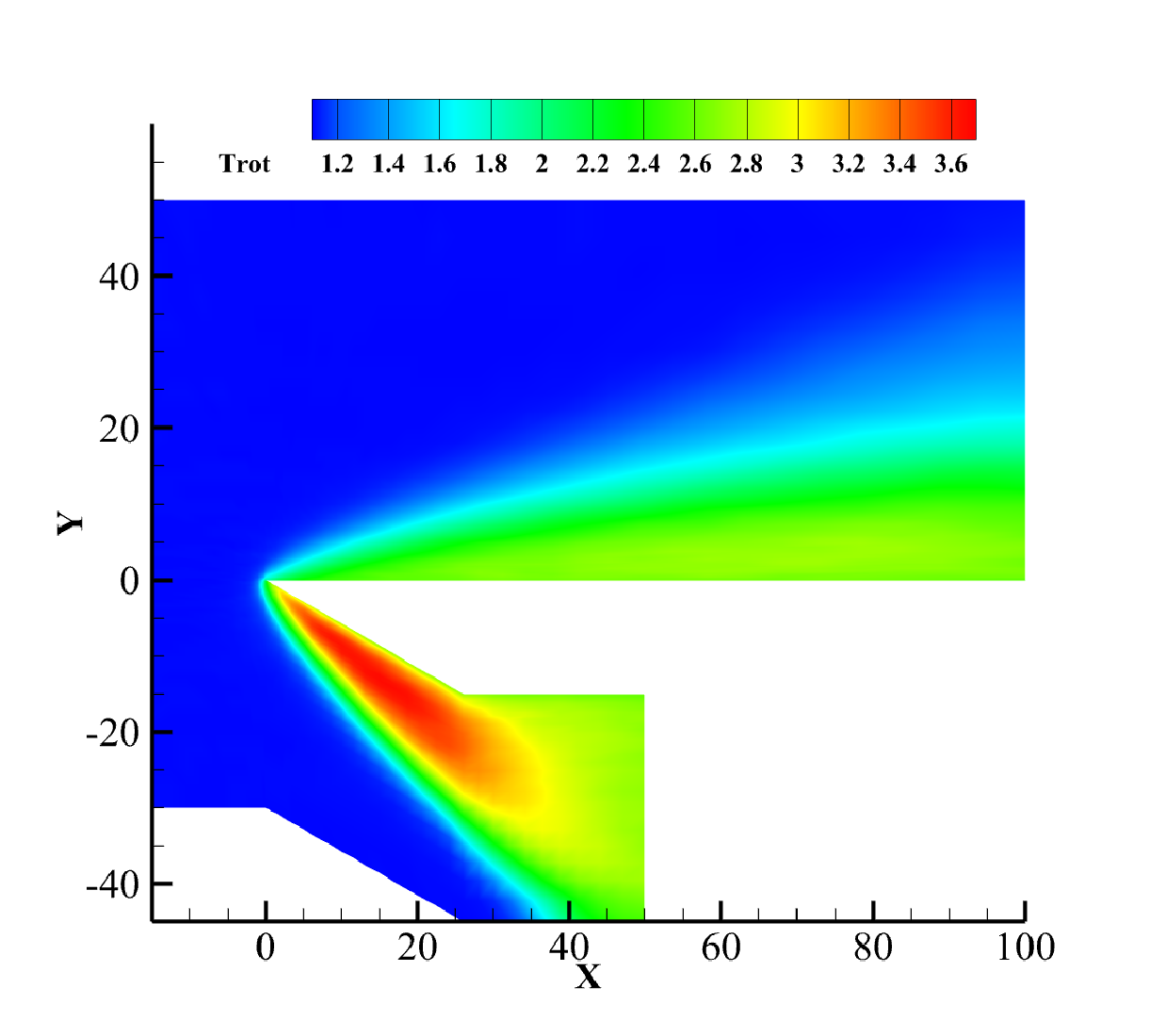}
}\\
\caption{\label{Fig:case2D_flatCon} Contours of flow around the flat plate, (a) density, (b) X-Velocity, (c) equilibrium temperature, (d) rotational temperature.}
\end{figure}

Figure \ref{Fig:case2D_flatCon} shows the contour flood of density, X-direction velocity, equilibrium temperature, and rotational temperature for the supersonic flow past the flat plate problem. In Fig.\ref{Fig:case2D_flat_Con_Rho}, a clear shock boundary layer interaction phenomenon can be observed, for this rarefied flow, both the shock wave and boundary lay are thick, and they merged into each other.

\begin{figure}[!h]
\centering
\subfigure[\label{Fig:case2D_flat_X05}]{
\includegraphics[width=0.45\textwidth]{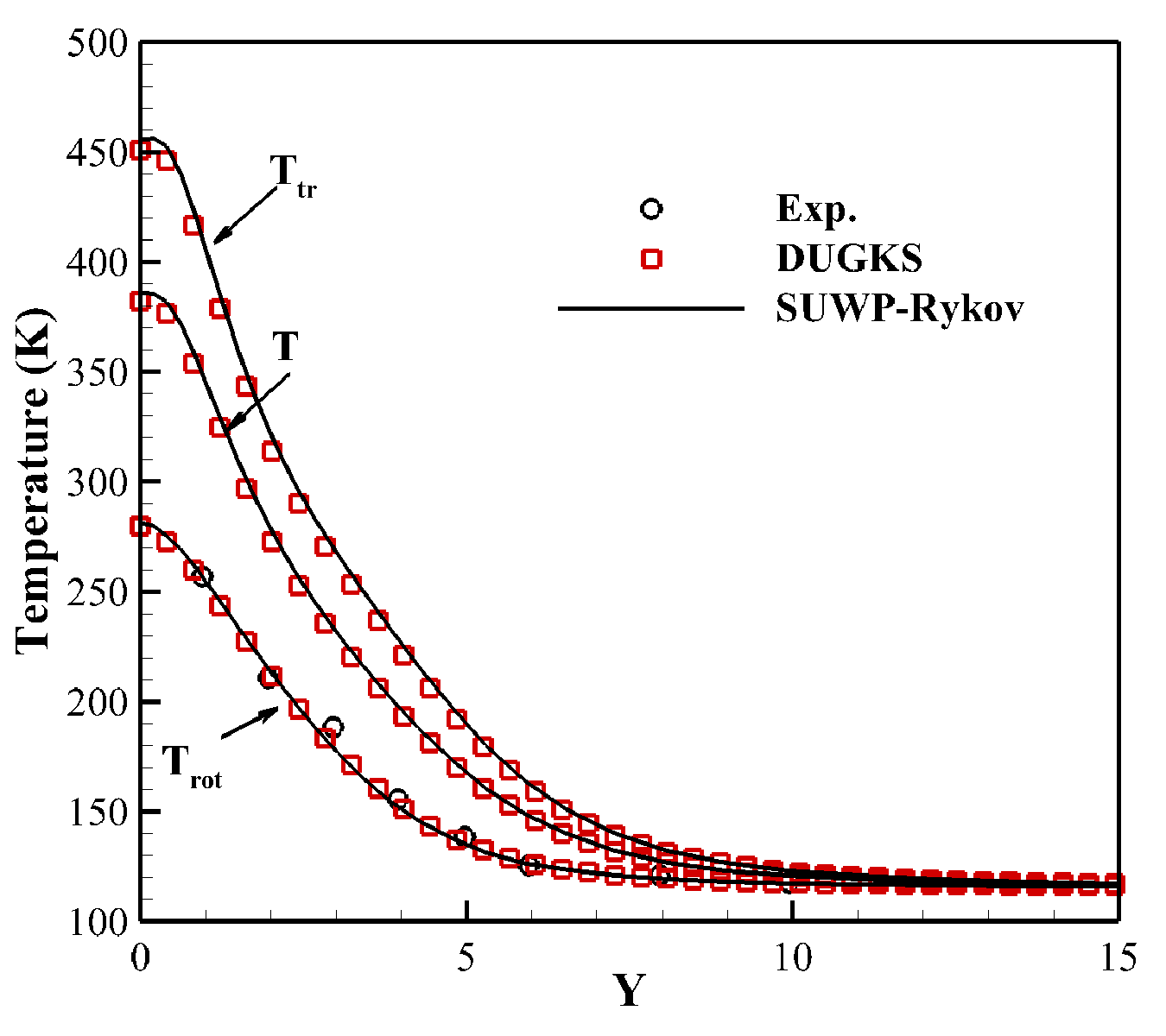}
}\hspace{0.05\textwidth}%
\subfigure[\label{Fig:case2D_flat_X10}]{
\includegraphics[width=0.45\textwidth]{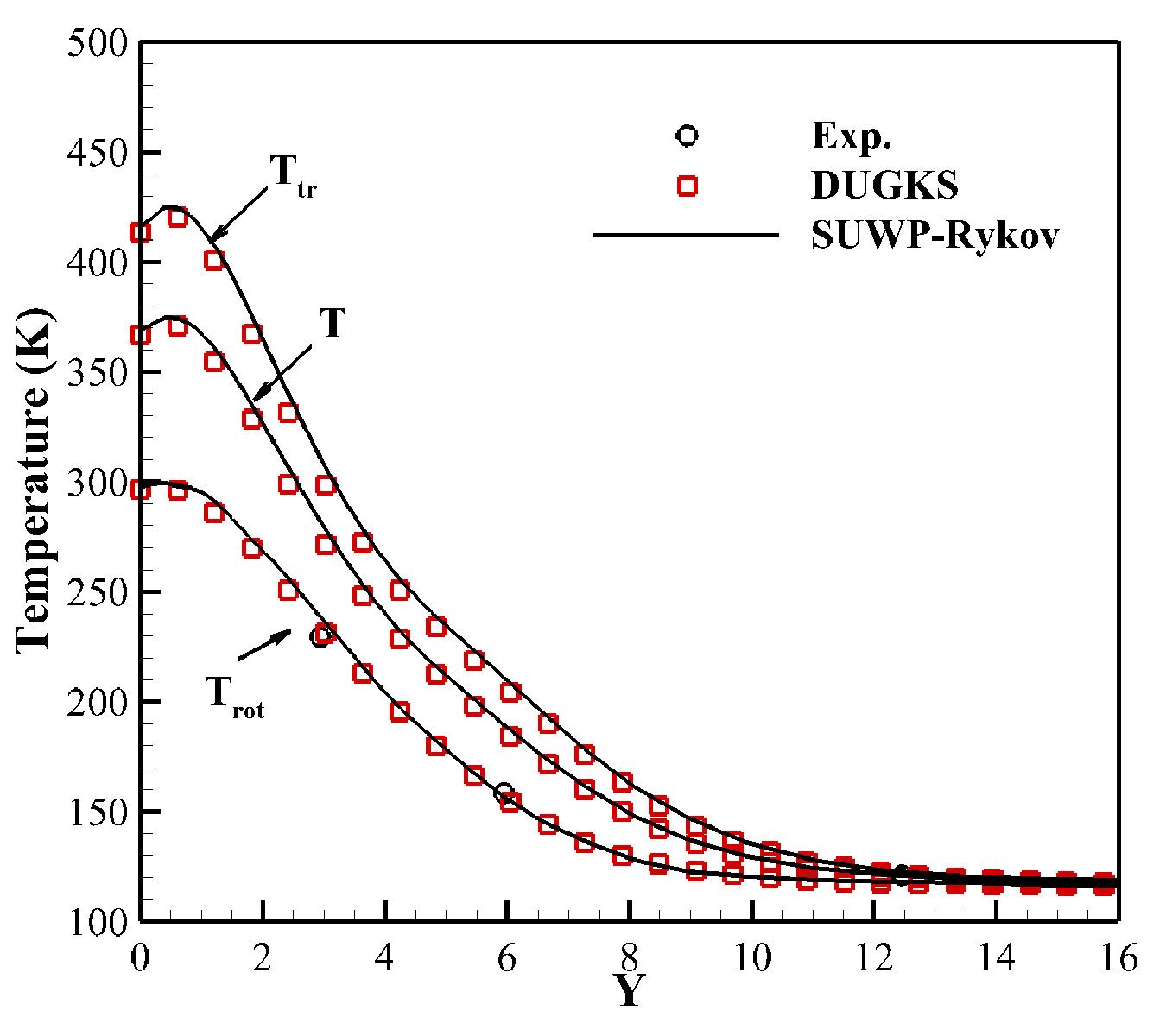}
}\\
\subfigure[\label{Fig:case2D_flat_X20}]{
\includegraphics[width=0.45\textwidth]{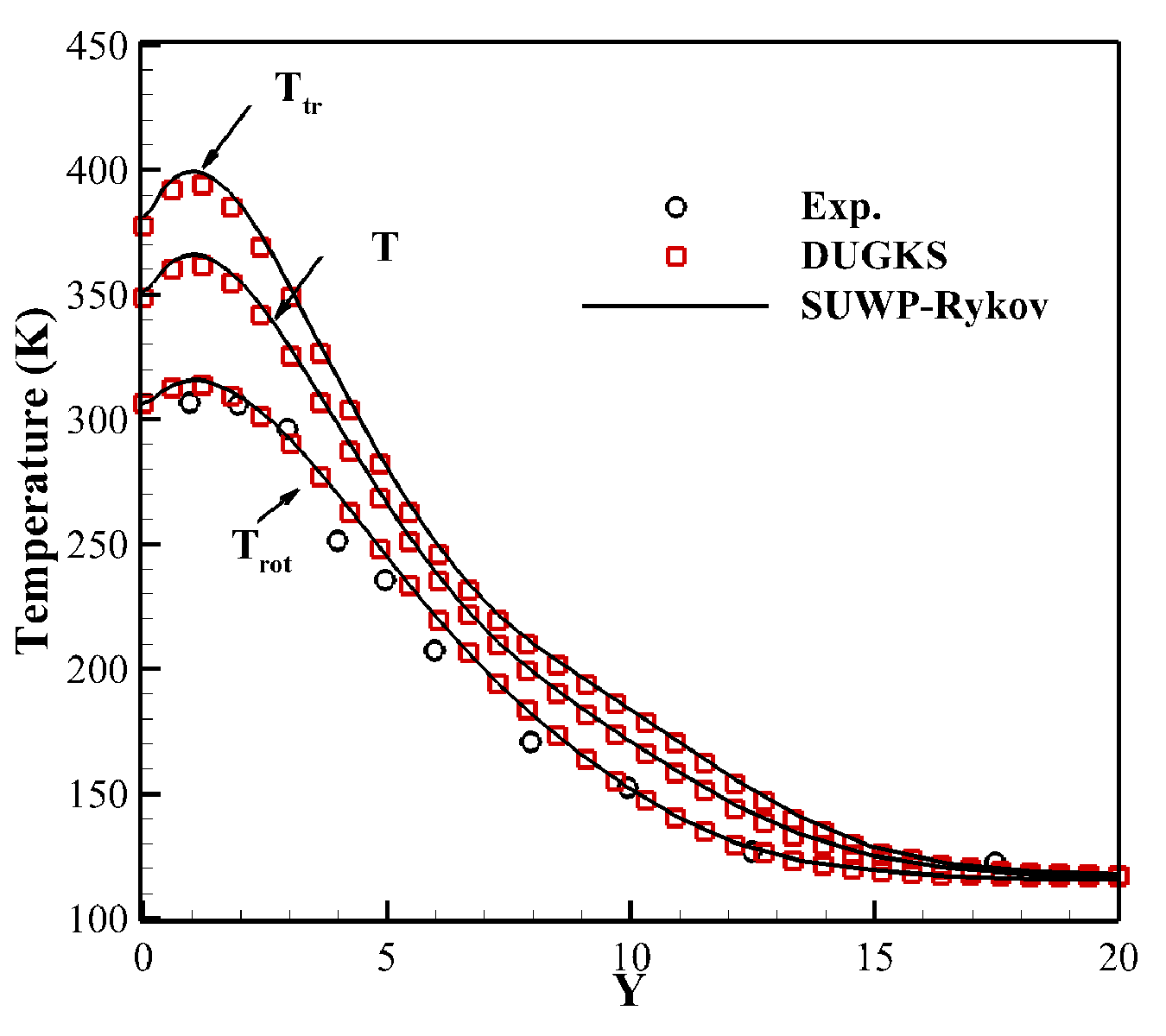}
}\\
\caption{\label{Fig:case2D_flatX}Temperature profiles along vertical lines at (a) X=5mm, (b) X=10mm, (c) X=20mm.}
\end{figure}

Figure \ref{Fig:case2D_flatX} compares the results predicted by SUWP-Rykov with the UGKS-Rykov method~\cite{liu2014unified, YUAN2020106972, ZHANG2023107079} and the experiments of Tsuboi et al~\cite{tsuboi2005experimental}. It can be observed that on the temperature profiles at along the vertical lines X = 5mm, X = 10mm and X = 20mm up the plate, the results from SUWP-Rykov agrees well with those from both UGKS and the experimental data. This demonstrates that the SUWP-Rykov method is capable of simulating non-equilibrium phenomena within the boundary layer of multi-scale flow in this benchmark case. 

\subsection{\label{sec::cyl2D}Hypersonic flow past a cylinder}

The hypersonic flow around a cylinder is a benchmark case for testing the ability of a numerical method for capture the high non-equilibrium in flow field and on solid surface. The Knudsen number chosen for this case is 0.01, where both continuum and rarefied flows are in the flow fields, and both N-S and DSMC methods have difficulties in simulation of this flow. Therefore, this case can also be used to test the multi-scale properties.

\begin{figure}[!h]
    \centering
    \includegraphics[width=0.45\textwidth]{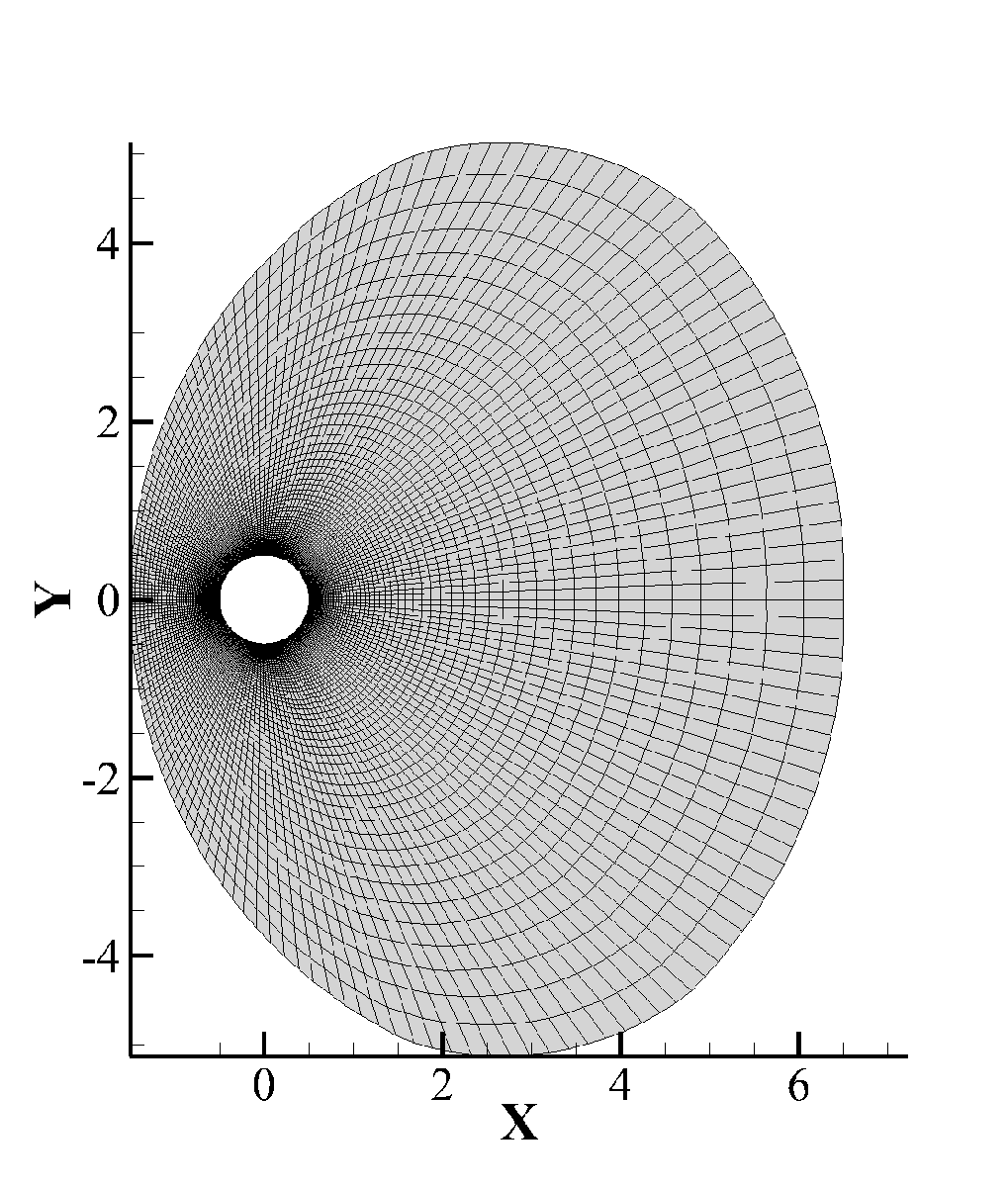}
    \caption{\label{Fig:case2D_cyl_Mesh}Computational mesh for the flow past a cylinder simulation.}
\end{figure}

The free-stream Mach number is set to 5.0. The working gas is nitrogen. The VHS model is used with a temperature exponent $\omega =0.74$, and the rotational collision number $Z_{rot}$ is set to 3.5. The non-dimensional density, velocity, and temperature are ${{\rho }_{\infty }}=1.0$, ${{U}_{\infty }}=4.1833$, and ${{T}_{\infty }}=1.0$, respectively, and the viscosity coefficient is 0.0684. The wall temperature ${{T}_{wall}}=1.0$. The radius of the cylinder is 0.5, and the domain is approximately elliptical, the distance from the inlet to the center of the cylinder is about 1.5, and the distance from the outlet to the center of the cylinder is about 6.5. The physical mesh used for the calculation is shown in Fig.\ref{Fig:case2D_cyl_Mesh} with 88 cells in the axial direction and 180 cells in the radial direction. The minimum height of the cell is 0.001.

\begin{figure}
\centering
\subfigure[\label{Fig:case2D_cyl_Con_Rho}]{
\includegraphics[trim={5} {5} {5} {5}, clip, width=0.45\textwidth]{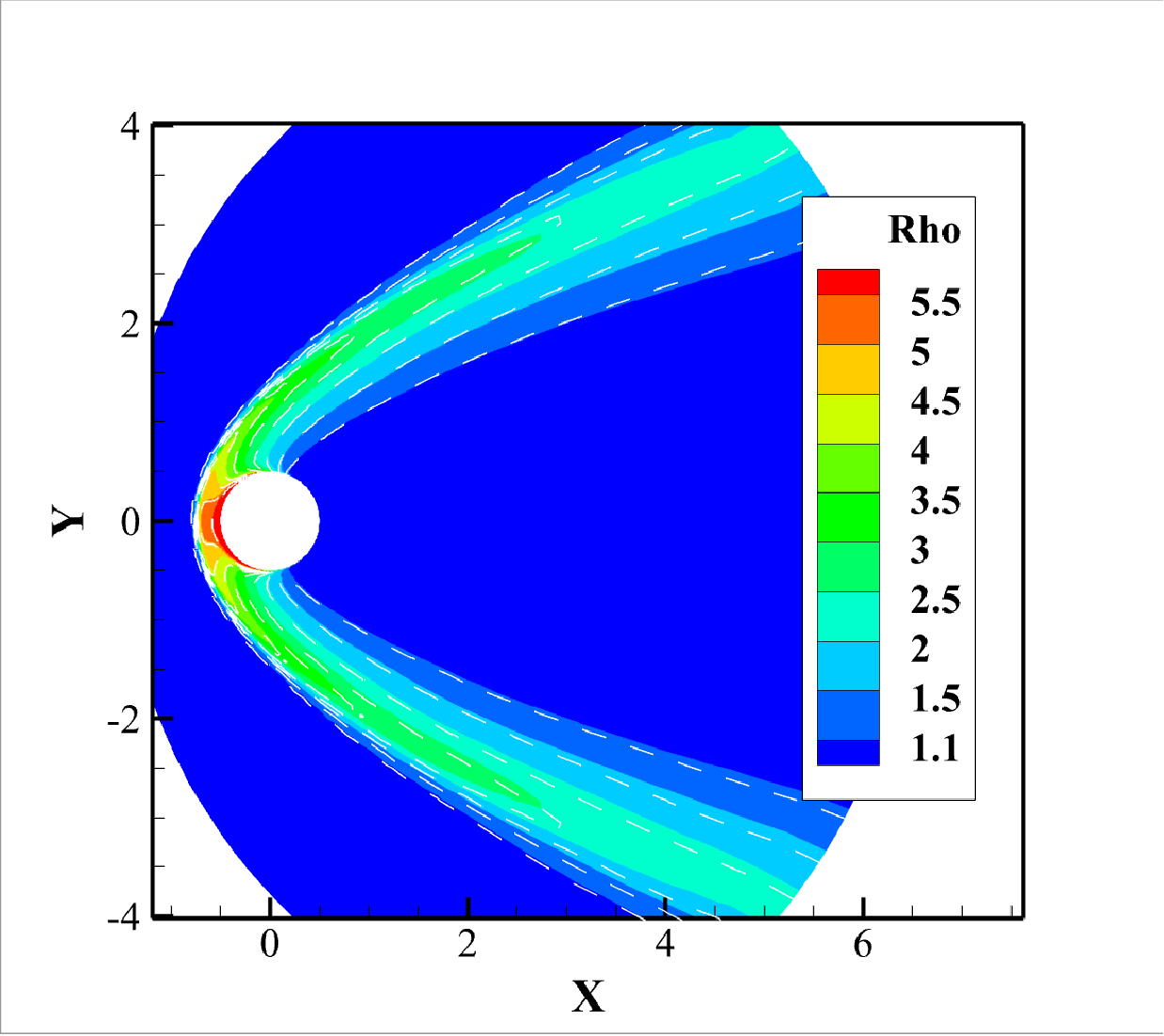}
}\hspace{0.05\textwidth}%
\subfigure[\label{Fig:case2D_cyl_Con_U}]{
\includegraphics[trim={5} {5} {5} {5}, clip,width=0.45\textwidth]{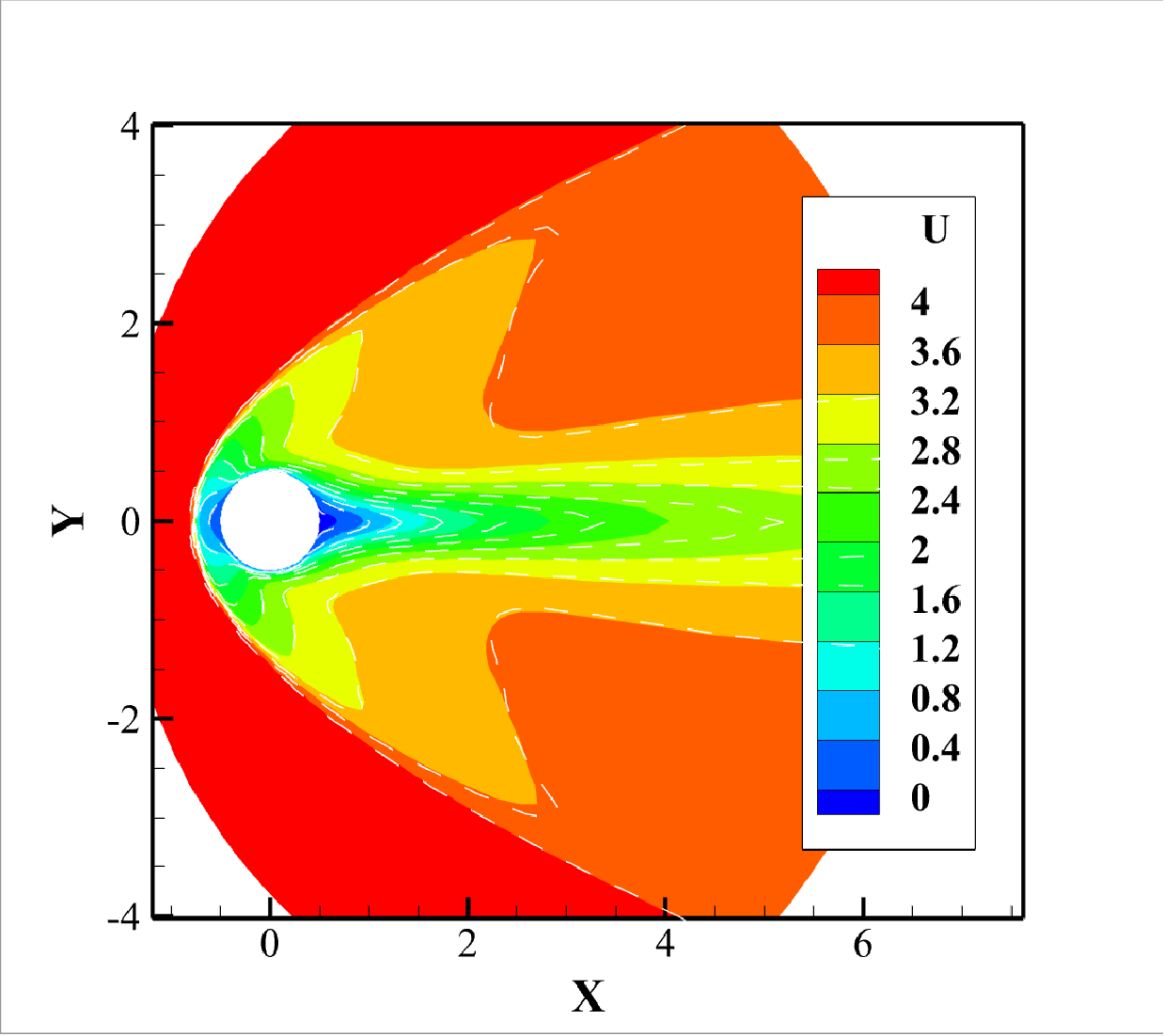}
}\\
\subfigure[\label{Fig:case2D_cyl_Con_V}]{
\includegraphics[trim={5} {5} {5} {5}, clip,width=0.45\textwidth]{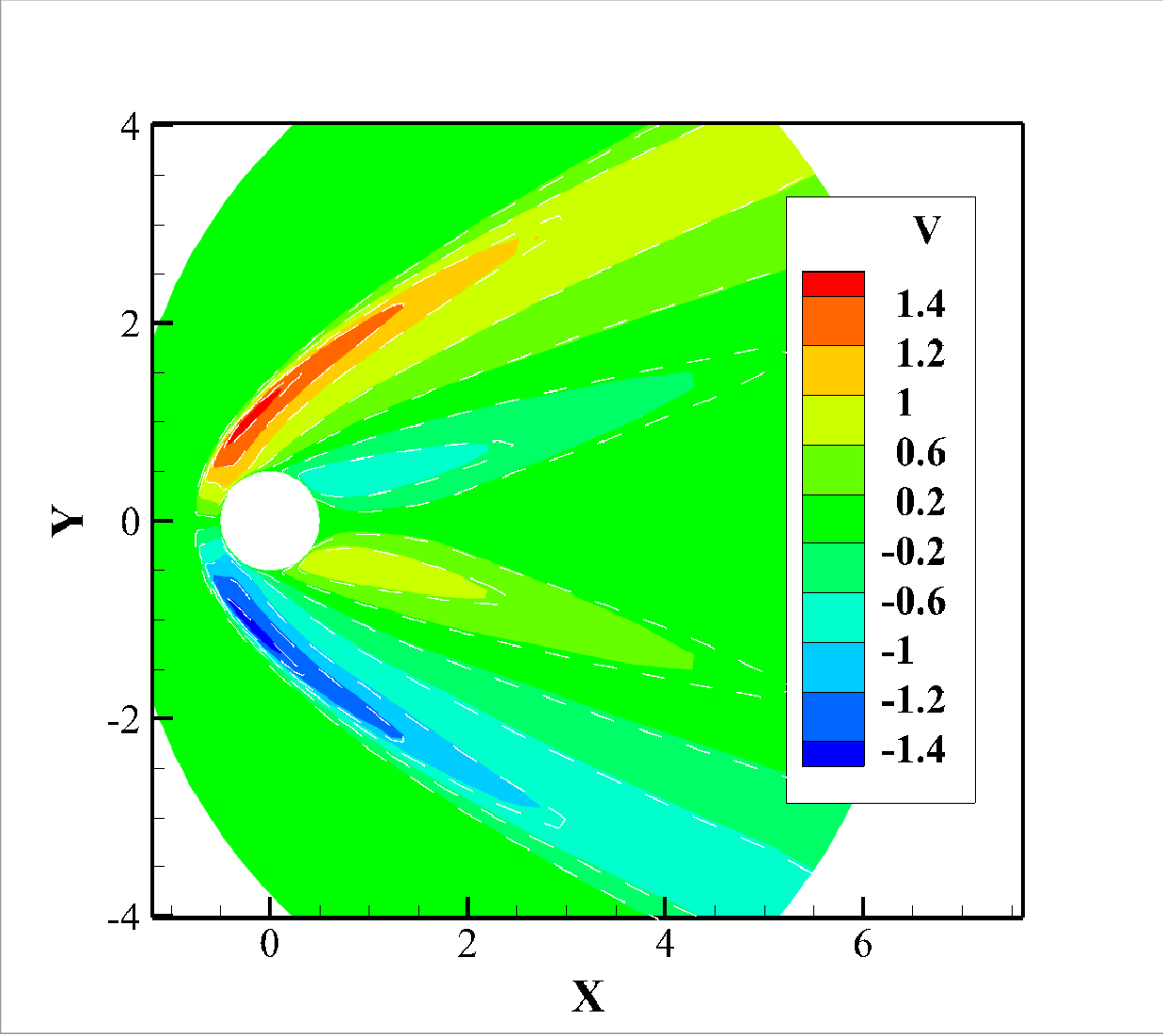}
}\hspace{0.05\textwidth}%
\subfigure[\label{Fig:case2D_cyl_Con_Teq}]{
\includegraphics[trim={5} {5} {5} {5}, clip,width=0.45\textwidth]{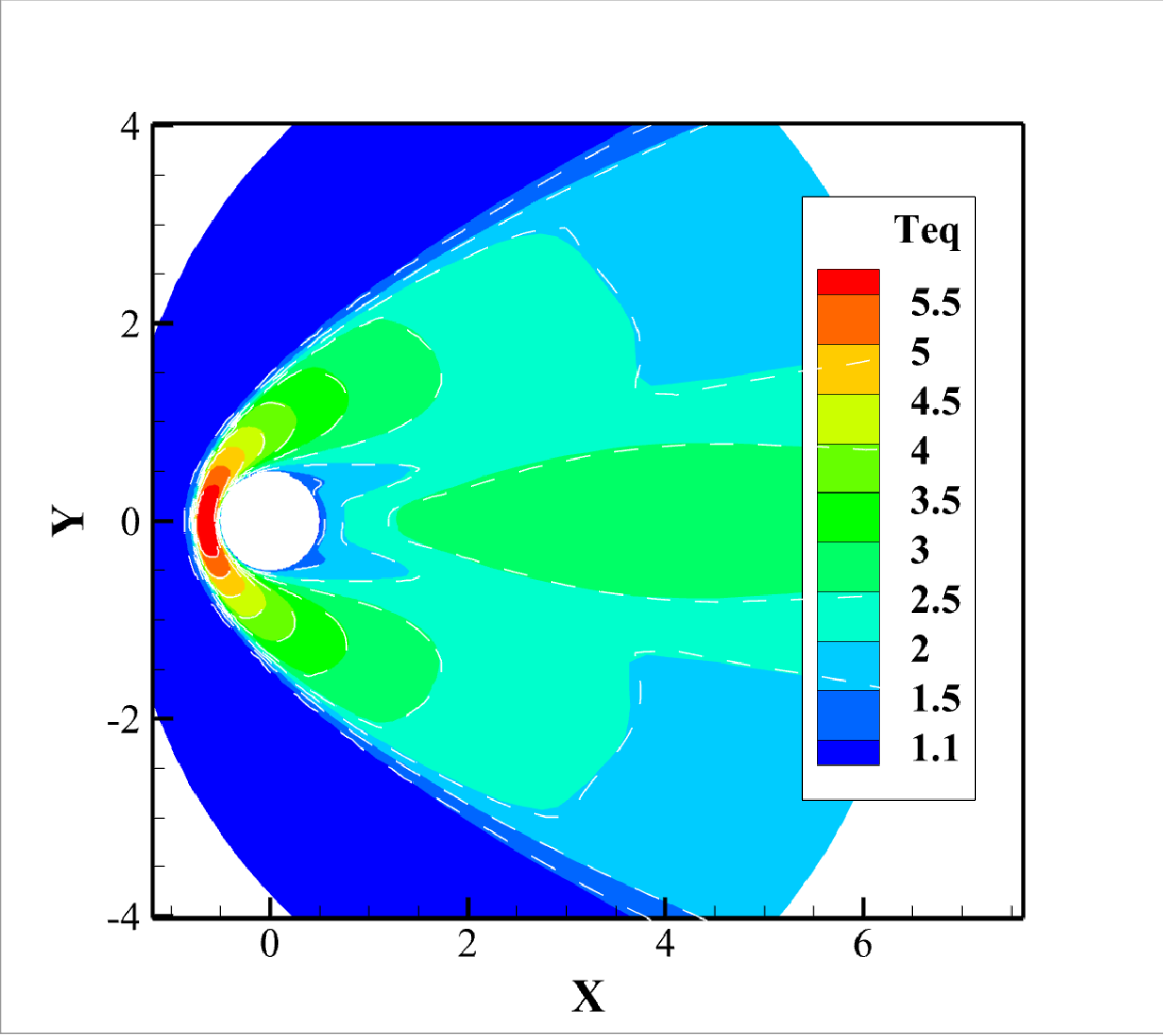}
}\\
\subfigure[\label{Fig:case2D_cyl_Con_Ttr}]{
\includegraphics[trim={5} {5} {5} {5}, clip,width=0.45\textwidth]{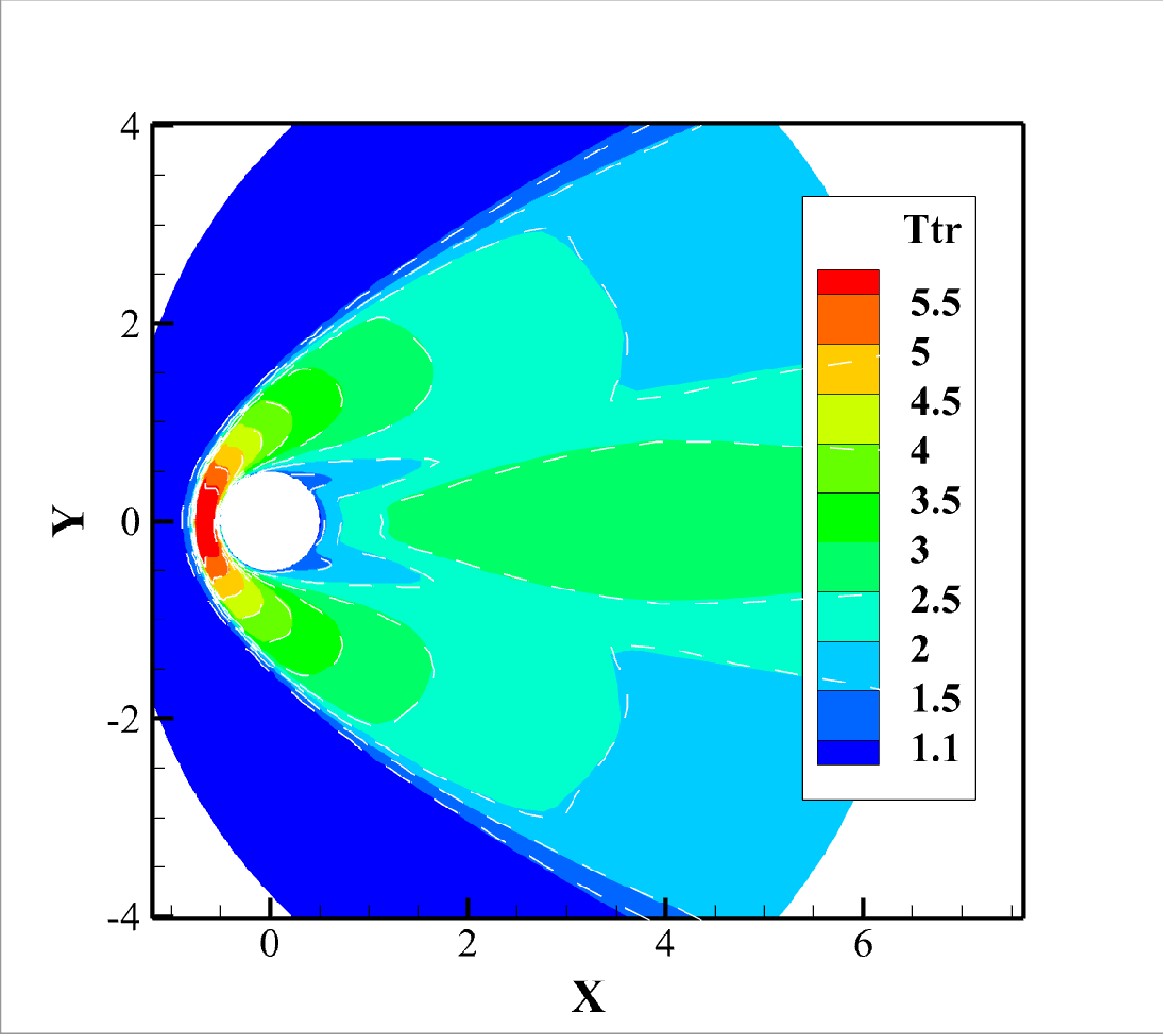}
}\hspace{0.05\textwidth}%
\subfigure[\label{Fig:case2D_cyl_Con_Trot}]{
\includegraphics[trim={5} {5} {5} {5}, clip,width=0.45\textwidth]{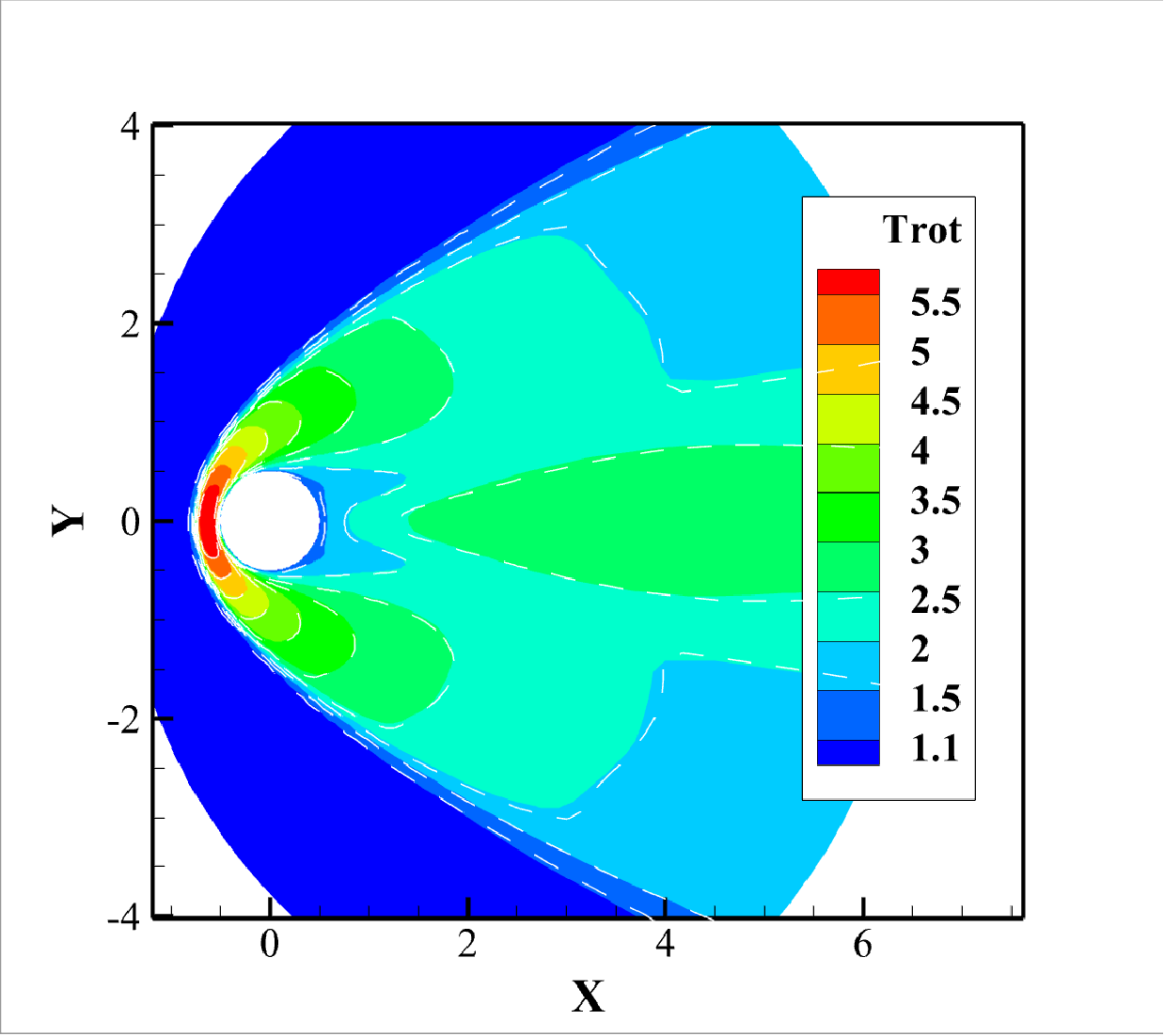}
}\\
\caption{\label{Fig:case2D_cylCon}Macroscopic variable contours for flow around the cylinder, (a) density, (b) X-Velocity, (c) Y-Velocity, (d) equilibrium temperature, (e) translational temperature, (f) rotational temperature. White dashed lines: DUGKS, flood contour: SUWP-Rykov.}
\end{figure}

\begin{figure}[!h]
\centering
\subfigure[\label{Fig:case2D_cyl_Slice_Rho}]{
\includegraphics[width=0.45\textwidth]{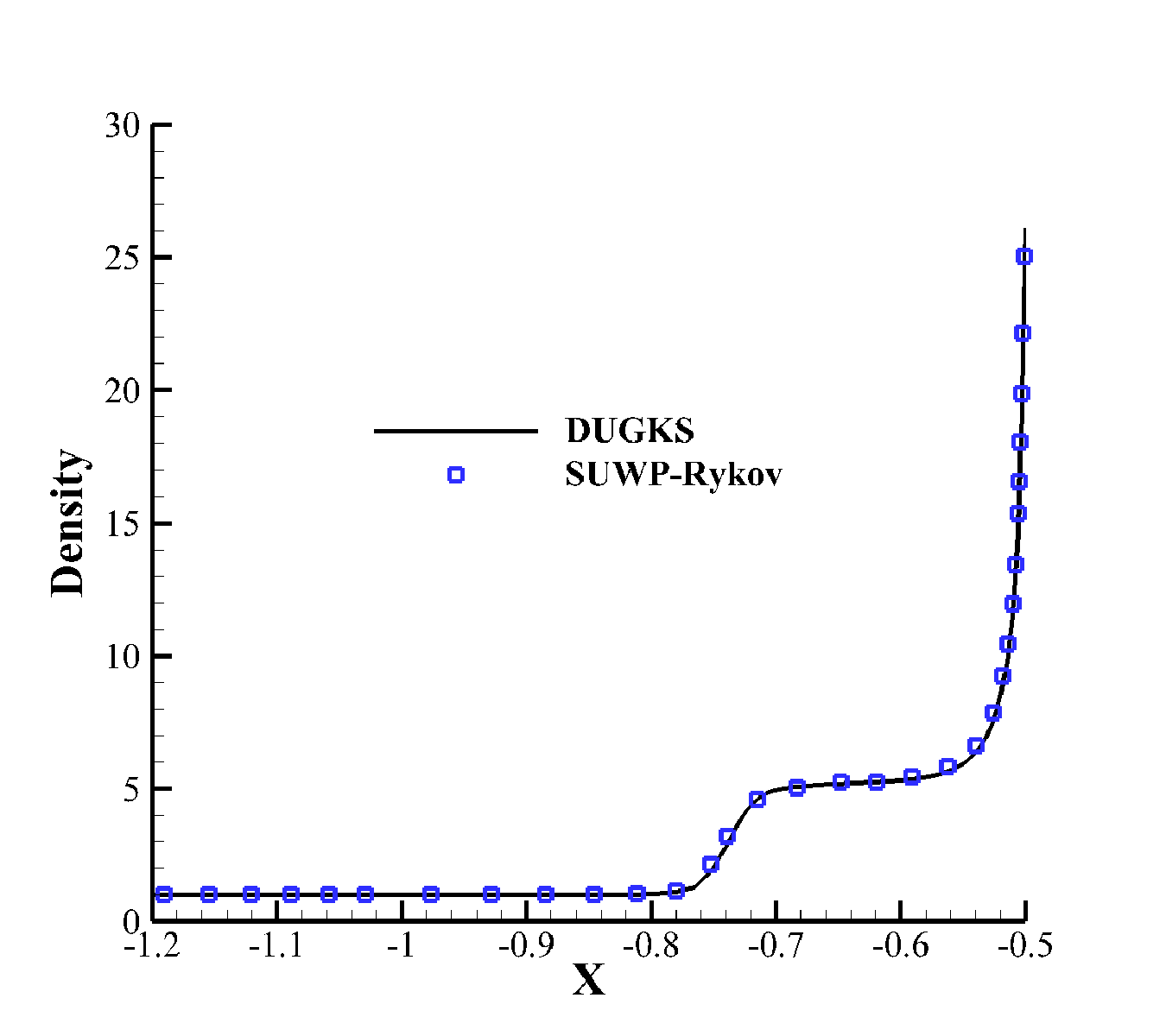}
}\hspace{0.05\textwidth}%
\subfigure[\label{Fig:case2D_cyl_Slice_U}]{
\includegraphics[width=0.45\textwidth]{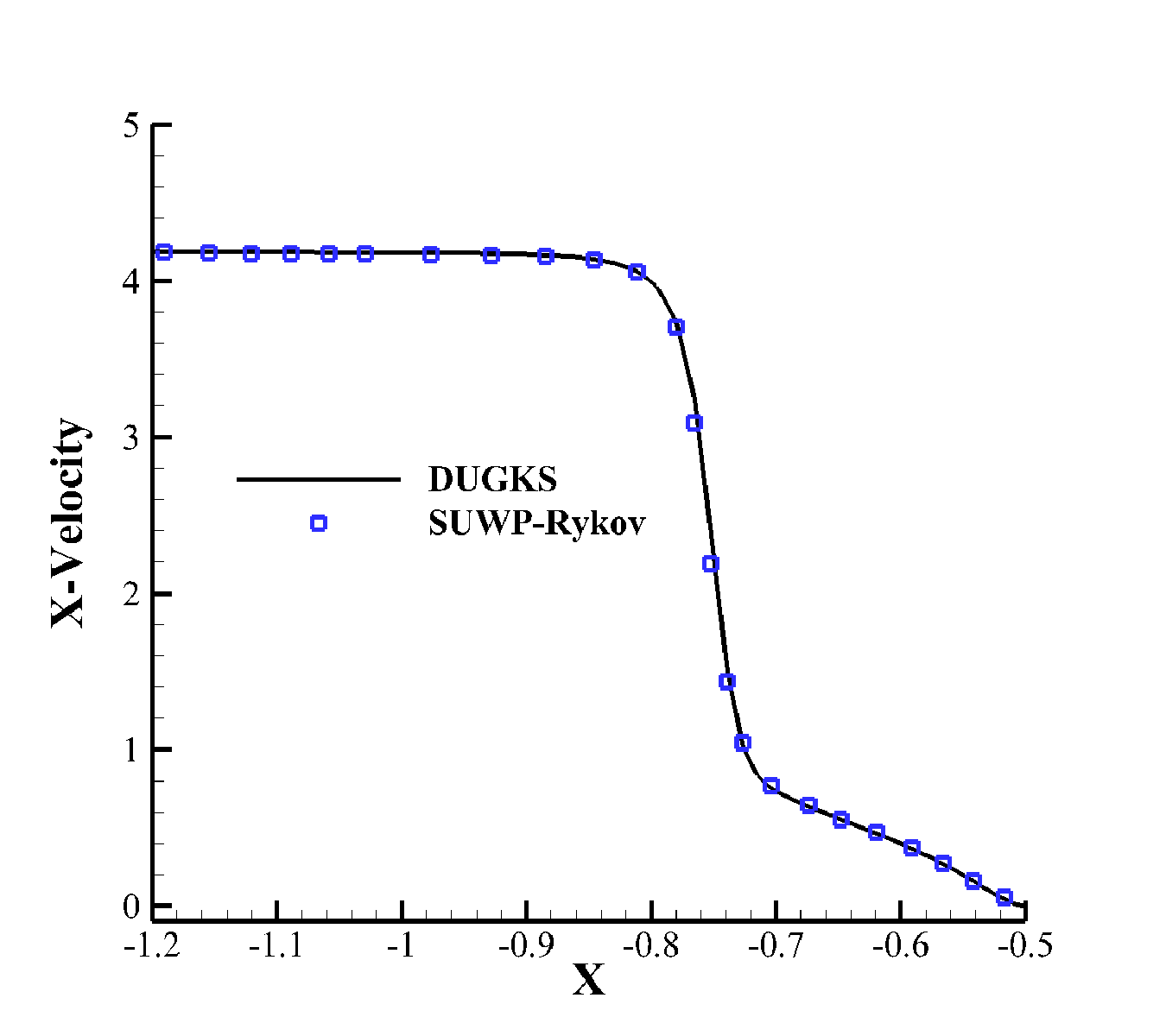}
}\\
\subfigure[\label{Fig:case2D_cyl_Slice_T}]{
\includegraphics[width=0.45\textwidth]{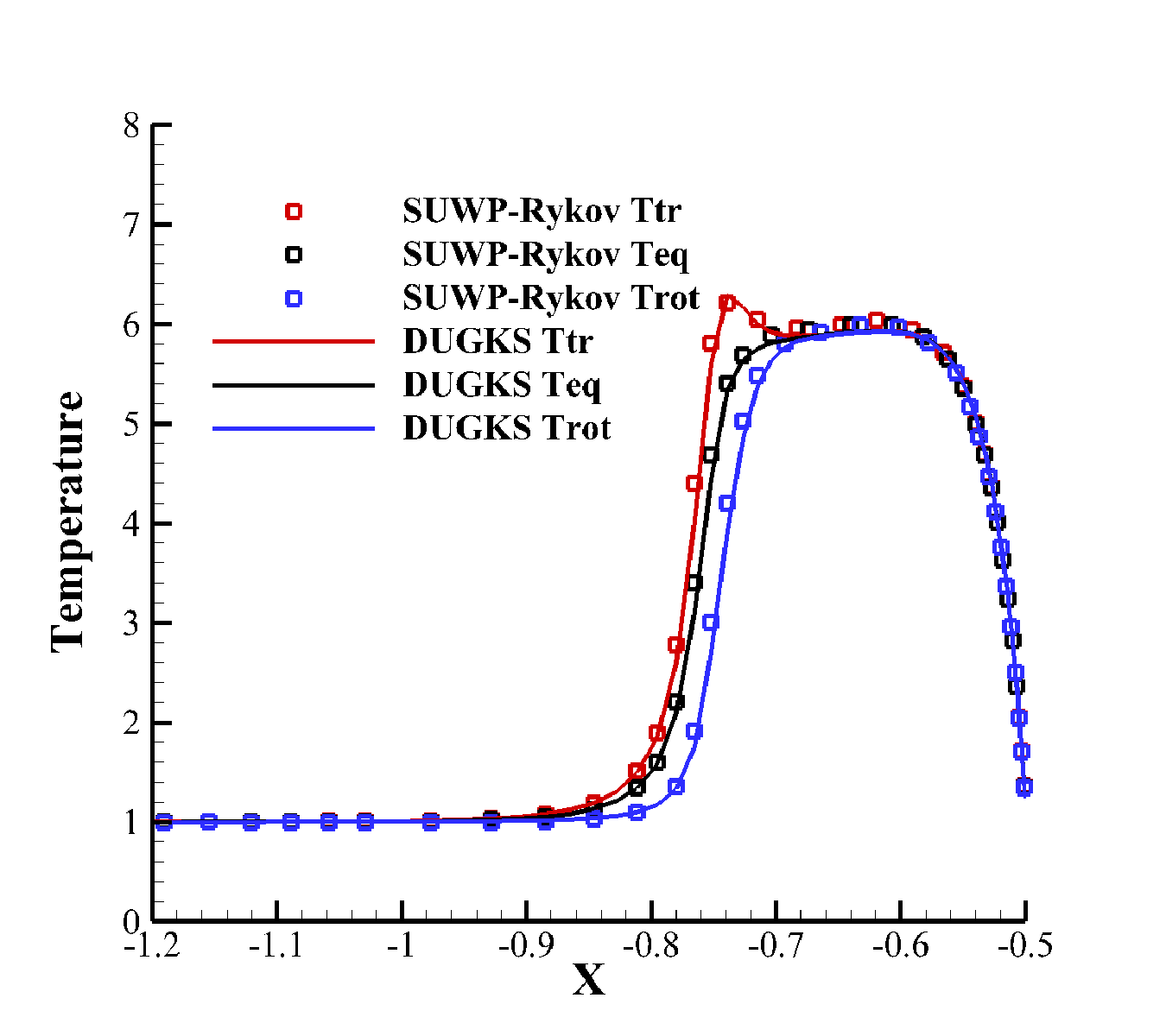}
}
\subfigure[\label{Fig:case2D_cyl_Wall_CpCfCh}]{
\includegraphics[width=0.45\textwidth]{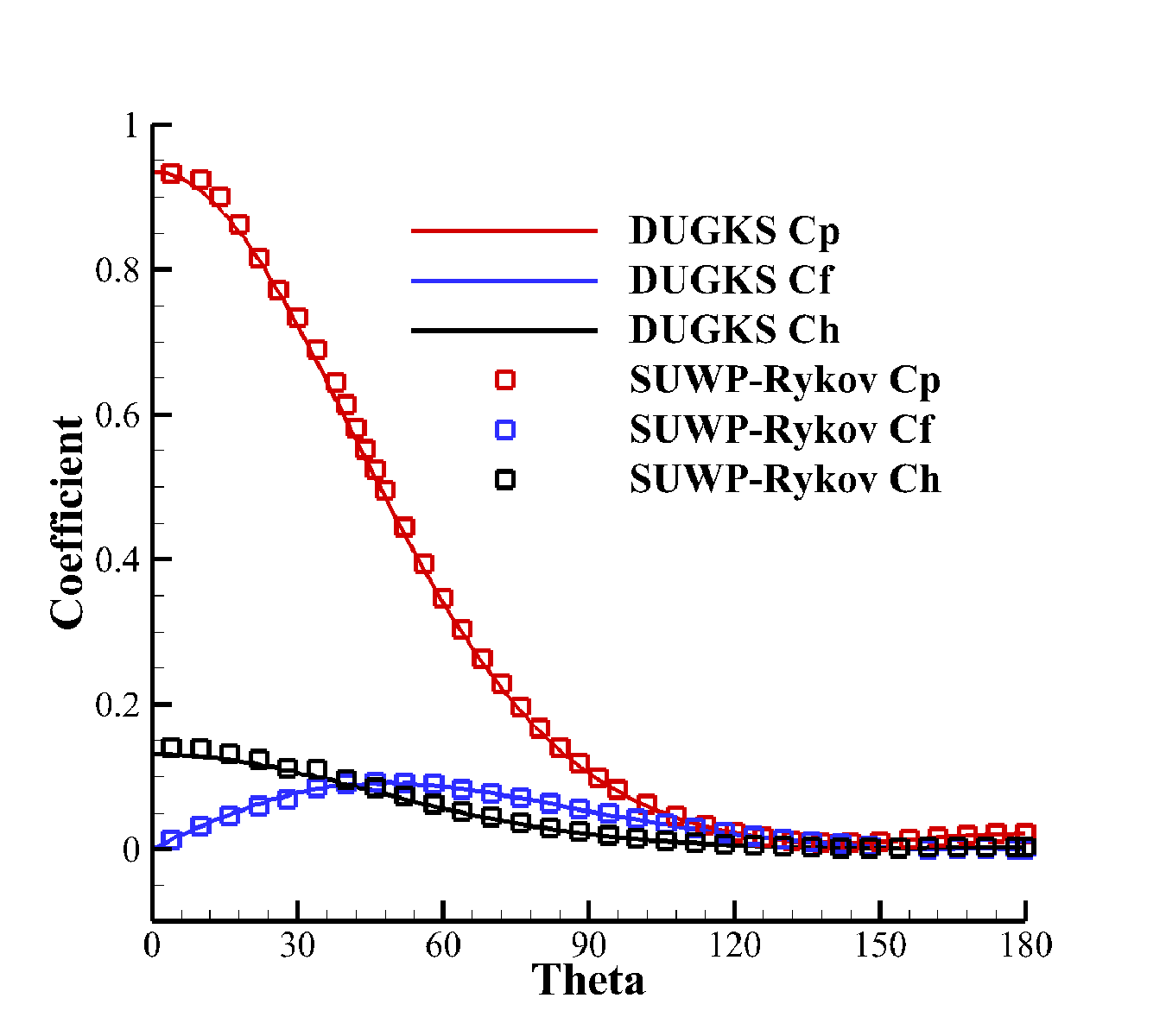}
}\\
\caption{\label{Fig:case2D_cyl_slice} The (a) density, (b) X-velocity, and (c) temperature distributions of the cylinder stagnation line and (d) pressure stress and heat flux coefficient of the cylinder wall.}
\end{figure}

Figure \ref{Fig:case2D_cylCon} shows the contours of pressure, X-direction velocity, Y-direction velocity, equilibrium temperature, translational temperature, and rotational temperature for the hypersonic flow around the cylinder obtained from the calculation. The comparison between the results of the SUWP-Rykov and the DUGKS is shown in the Fig.\ref{Fig:case2D_cyl_slice}. It can be observed that SUWP-Rykov and DUGKS give good agreement in their calculated results. The comparison between the wall pressure coefficient, friction coefficient, and heat transfer coefficient are presented. The wall pressure coefficient and friction coefficient show good agreement with DUGKS, while the peak value of the heat conduction coefficient is slightly higher.

\subsection{\label{sec::sphere3D}Hypersonic flow past a sphere in transitional regime}

The supersonic flow around a sphere is conducted to validate the ability of SUWP-Rykov for simulating three-dimensional multi-scale flow. The radius of the sphere is 1 mm and its surface is divided into a $16\times16$ mesh. The height of the cells adjacent to the wall is $5\times {{10}^{-5}}m$, and the total number of cells in the flow field is 29,700. The working gas for this case is nitrogen, and the specific heat ratio for the diatomic gas $\gamma =1.4$. The VHS model is used with temperature exponent $\omega =0.74$. The rotational collision number $Z_{rot}$ is set to 3.5. The free stream density ${{\rho }_{\infty }}=6.87\times {{10}^{-4}}kg\cdot {{m}^{-3}}$, the velocity ${{U}_{\infty }}=698.46m/s$, the free stream temperature ${{T}_{\infty }}=65K$, and the wall temperature ${{T}_{wall}}=302K$. The Mach number is set to 4.25, and the Knudsen number is set to 0.031 with a reference length of 1mm. 

\begin{figure}[!h]
\centering
\subfigure[\label{Fig:case3D_sphere_Con_Rho}]{
\includegraphics[width=0.45\textwidth]{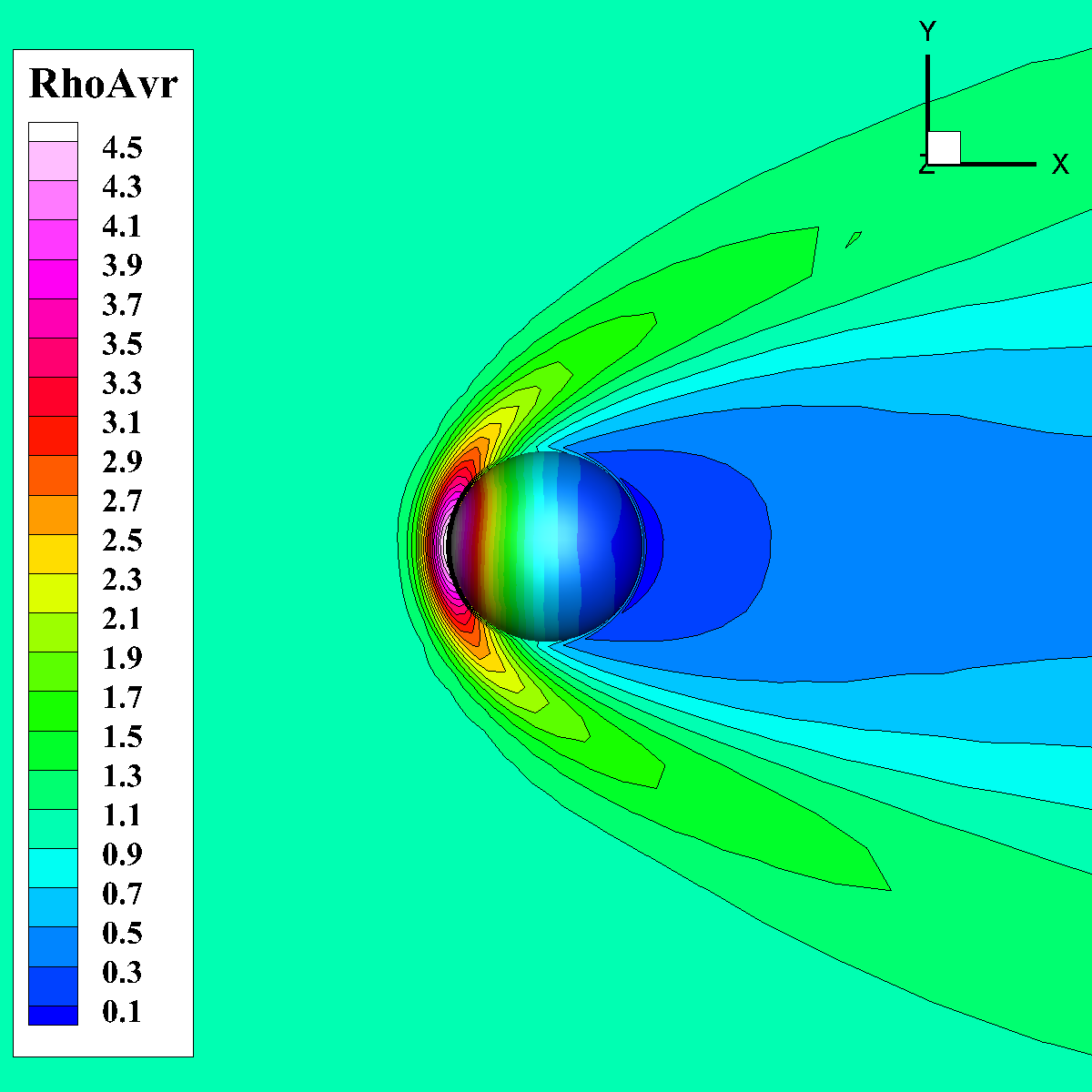}
}\hspace{0.05\textwidth}%
\subfigure[\label{Fig:case3D_sphere_Con_U}]{
\includegraphics[width=0.45\textwidth]{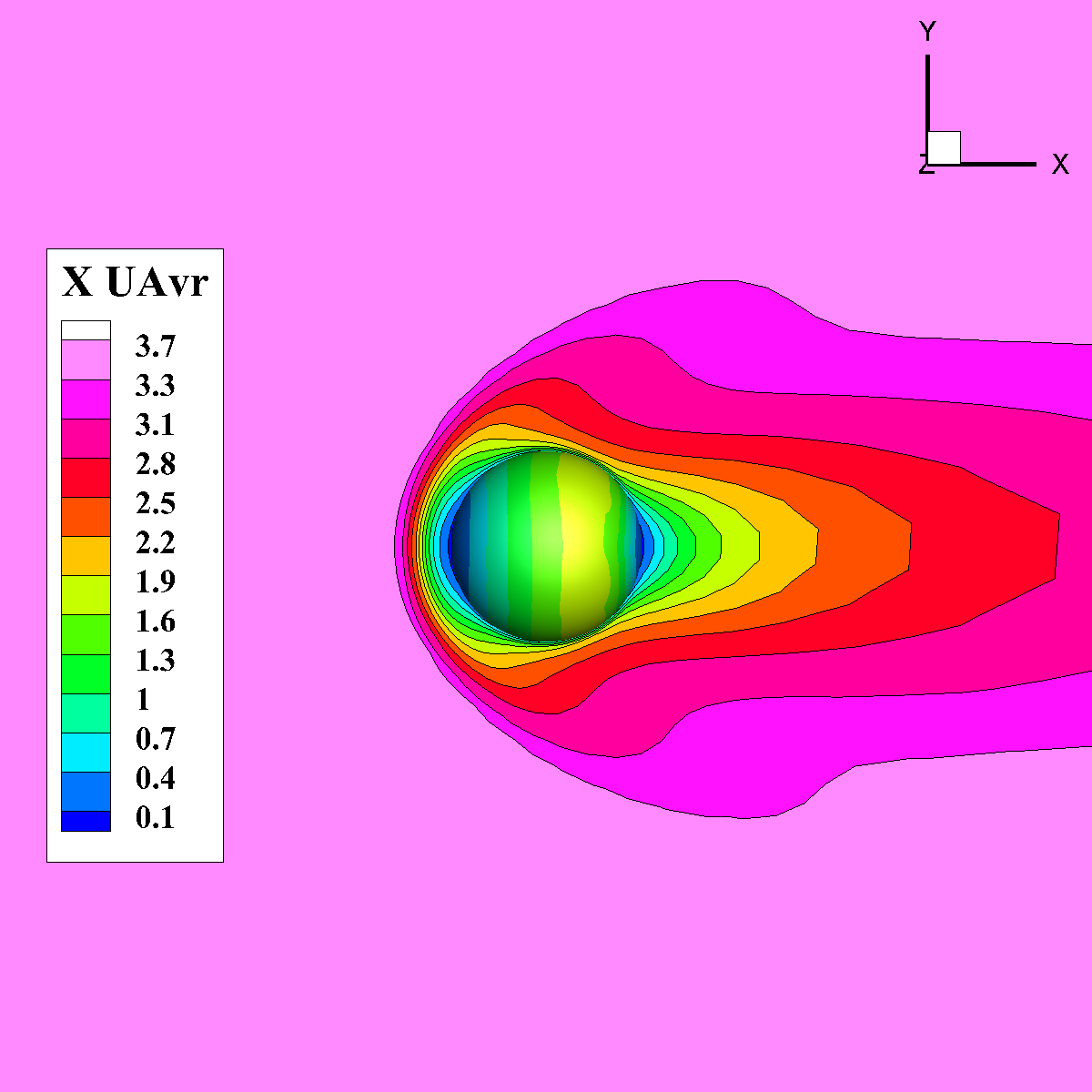}
}\\
\subfigure[\label{Fig:case3D_sphere_Con_Teq}]{
\includegraphics[width=0.45\textwidth]{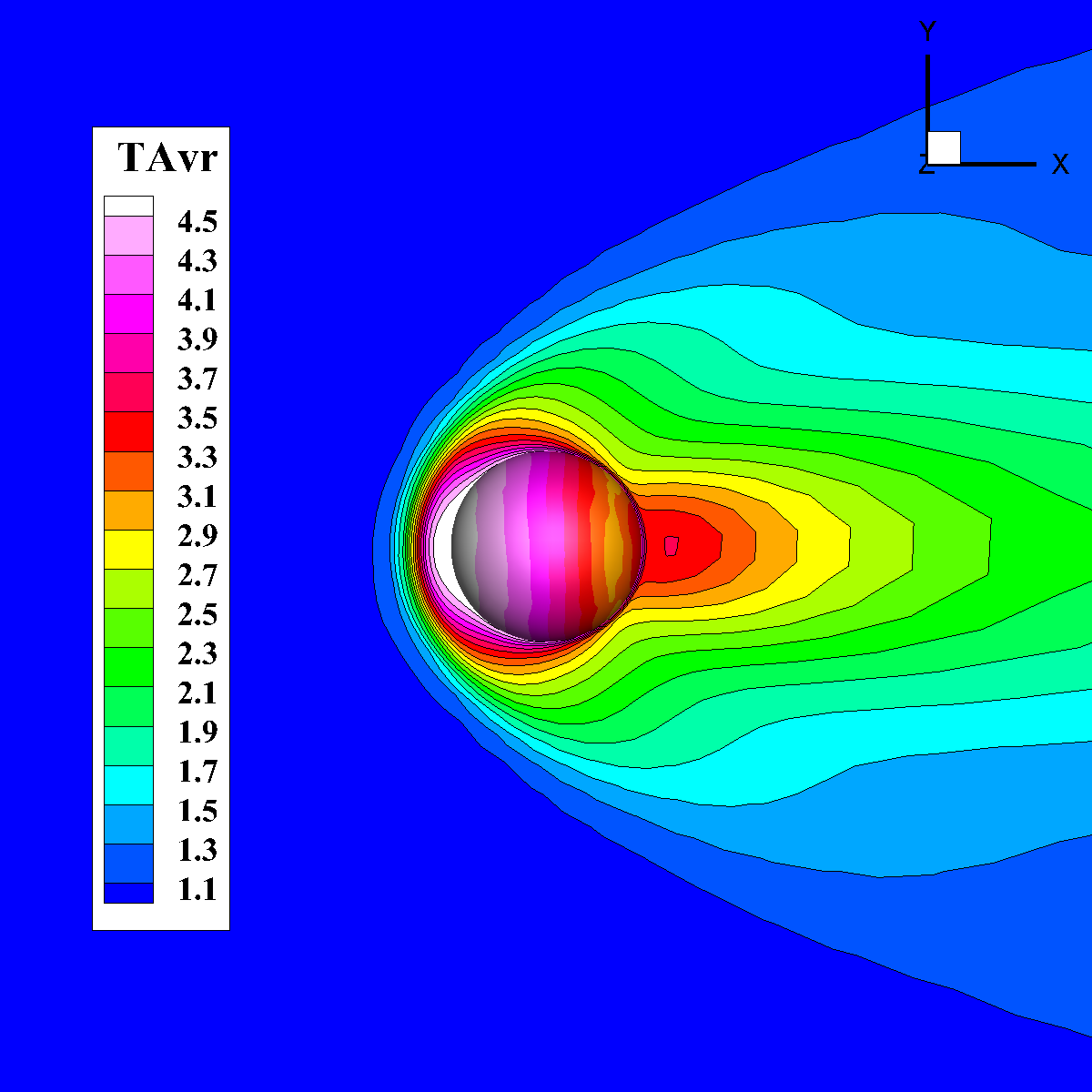}
}\hspace{0.05\textwidth}%
\subfigure[\label{Fig:case3D_sphere_Con_Trot}]{
\includegraphics[width=0.45\textwidth]{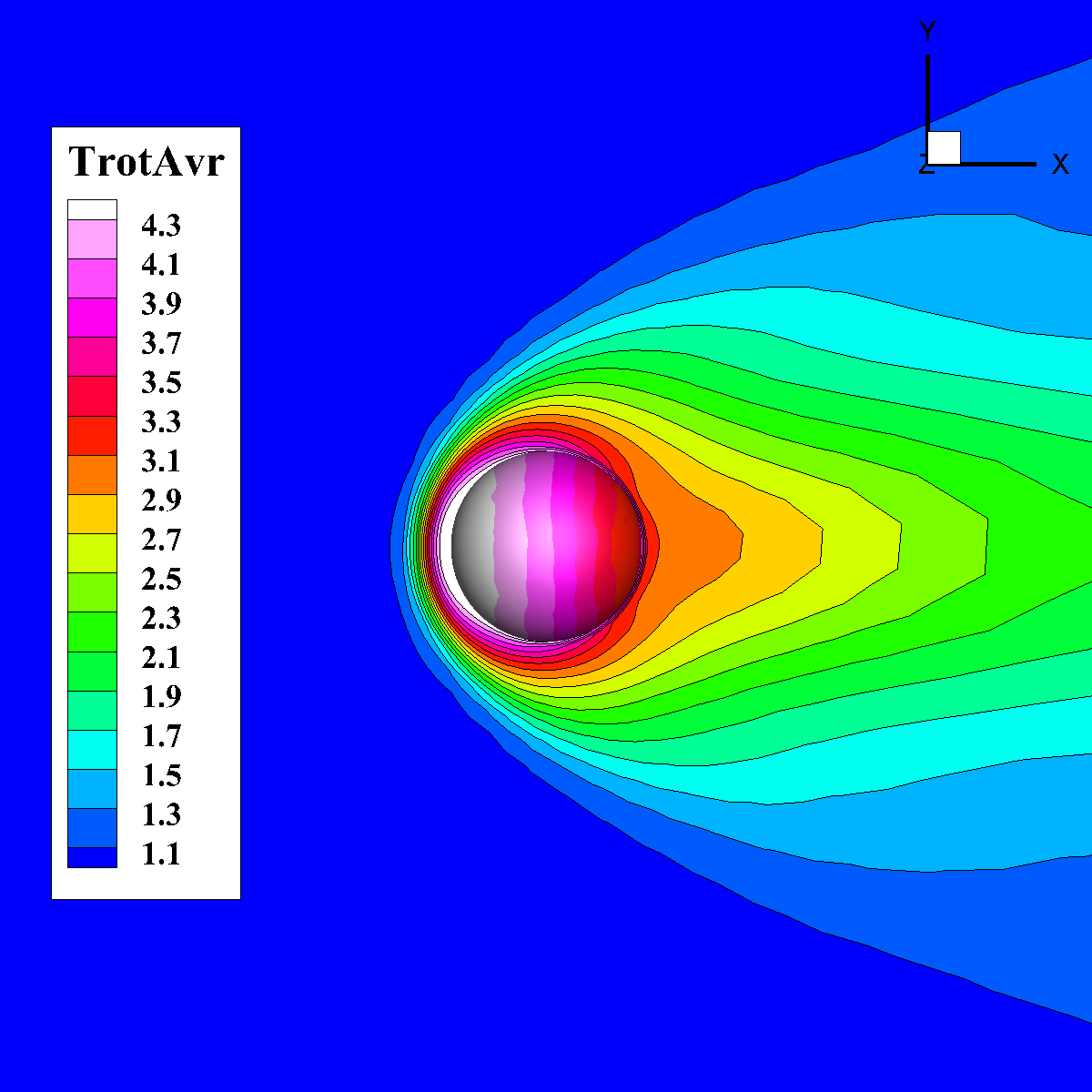}
}\\
\caption{\label{Fig:case3D_sphereCon}Macroscopic variable contours around the sphere, (a) density, (b) X-Velocity, (c) equilibrium temperature, (d) rotational temperature.}
\end{figure}

Figure \ref{Fig:case3D_sphereCon} shows the contours of the density, X-direction velocity, temperature, and rotational temperature. Wendt et al.~\cite{wendt1971drag} conducted wind tunnel experiments on supersonic flow around a spherical body (the working gas of Wendt is air, so it was used for simulation by equivalent parameters). Jiang et al.~\cite{Jiang2019} performed simulations of supersonic flow around a spherical body using the UGKS and GKS. In this case, the drag coefficients of the sphere obtained from SUWP-Rykov simulations are compared with Wendt's experimental results and Jiang's simulation results, as shown in Table \ref{table:case3D_sphere_drag}. The drag coefficients obtained from SUWP-Rykov simulations show a deviation of only 0.96\% from the experimental values.
\begin{table}[htb]   
\begin{center}   
\caption{Comparison of the drag coefficients for the sphere.}  
\label{table:case3D_sphere_drag} 
\begin{tabular}{cccc}
\toprule 
 & Experiment   &  SUWP & UGKS \\ 
\midrule
Experiment  & 1.35 & 1.363 & 1.355\\
Relative error & - & 0.96\% & 0.39\%\\
\bottomrule 
\end{tabular}
\end{center}   
\end{table}

For the case, the time-averaging starts from 2500 steps and continues for 13000 steps with an initial ﬁeld computed by 1000 steps from KIF. The total computation takes 13500 time steps, and runs on a workstation with AMD EPYC 7763 at 2.45GHz with 48 cores. Compared with the UGKS case~\cite{xu2021unified} with the same steps (runs on a workstation with Intel Xeon Platinum 8168 at 2.70 GHz with 48 cores), the results are shown in Table \ref{table:case3D_sphere_efficient}.

\begin{table}[htb]   
\begin{center}   
\caption{Comparison of the computational cost in $Ma=4.25$ and $Kn=0.031$ sphere case.}  
\label{table:case3D_sphere_efficient} 
\begin{tabular}{ccc}
\toprule 
& Time (hour) & Clock speed-hours (GHz $\cdot$ hour) \\ 
\midrule
Implicit UGKS & 265.6 & 717.1\\
SUWP & 2.8 & 6.9\\
Speedup Ratio & 94.9 & 103.9\\
\bottomrule 
\end{tabular}
\end{center}   
\end{table}

Here is a discussion about why SUWP is more efficient than DSMC in the same mesh. Denote the number of DSMC particles in a cell as $N$. The number of collision pairs of DSMC in each time step can be expressed as follows:
\begin{equation}  \label{eq:dsmcNTC}
{{N}_{collisionParis}}=\frac{1}{2}N(N-1){{F}_{N}}{{({{\sigma }_{T}}c{}_{r})}_{\max }}\Delta t/{\Omega},
\end{equation} 
where ${{F}_{N}}=n{\Omega}/N$ represents the number of real molecules represented by each simulated molecule, $n$ is the number density of the gas, and ${\Omega}$ is the volume of the present cell. According to HS model, the value of ${({{\sigma }_{T}}c{}_{r})}_{\max }$ is set to $5\pi {{d}^{2}} \sqrt{RT}$. The number of collision pairs of DSMC can be expressed as~\cite{Bird_Molecular}:
\begin{equation}  \label{eq:dsmcNumOfCollision}
{{N}_{collisionParis}}=\frac{25}{32}(N-1)\cdot \sqrt{\pi } \frac{\Delta t}{\tau },
\end{equation} 

For the SUWP method, we assume that particle transport has little influence on the proportion of particles in the cell, then the number of remaining particles in the cell at the beginning of the time step can be regarded as ${{N}_{remain}}\approx N\cdot {{w}_{free}}$, and the number of sampled and deleted particles can be expressed as ${{N}_{sample}}=N\cdot (1-{{w}_{free}})\cdot {{w}_{free}}$ and ${{N}_{remove}}=N\cdot (1-{{w}_{hydro}})\cdot {{w}_{hydro}}$. The number of particles to be transported and counted can be expressed as ${{N}_{transport, SUWP}}=N\cdot {{w}_{free}}$ and ${{N}_{statistic, SUWP}}=N\cdot {{w}_{free}}$.

The QMC weights are set to Eq.(\ref{eq:weightOfCAndF}), the run time of DSMC and SUWP can be represented as:
\begin{equation}  \label{eq:runTimeOfDSMCAndSUWP}
\begin{aligned}
&{{t}_{DSMC}}=\frac{25}{32}\sqrt{\pi }(N-1)\cdot \frac{\Delta t}{\tau }\cdot {{P}_{collision}}+N\cdot {{P}_{transport}}+N\cdot {{P}_{statistic}},&\\
&{{t}_{SUWP}}=N\cdot \frac{\tau }{\tau +\Delta t}\cdot [\frac{\Delta t}{\tau +\Delta t}\cdot {{P}_{sample}}+{{P}_{transport}}+{{P}_{statistic}}]+{{t}_{N-S}},&\\
\end{aligned}
\end{equation} 
where the $t_{N-S} << t_{SUWP}$, and the $P_{X}$ represents the time required to perform sequential $X$ operations on a particle. The $P_{sample}$ is slightly smaller than $P_{collision}$. 

It's easy to know that $\frac{\tau }{\tau +\Delta t}$ is less than 1 in both continuum and rarefied regimes. Therefore, in the same situation, SUWP method is more efficient.

\subsection{\label{sec::cone3D}Hypersonic flow past a blunt cone}

\begin{figure}[!h]
    \centering
    \includegraphics[trim=50 150 50 150, clip, width=0.8\textwidth]{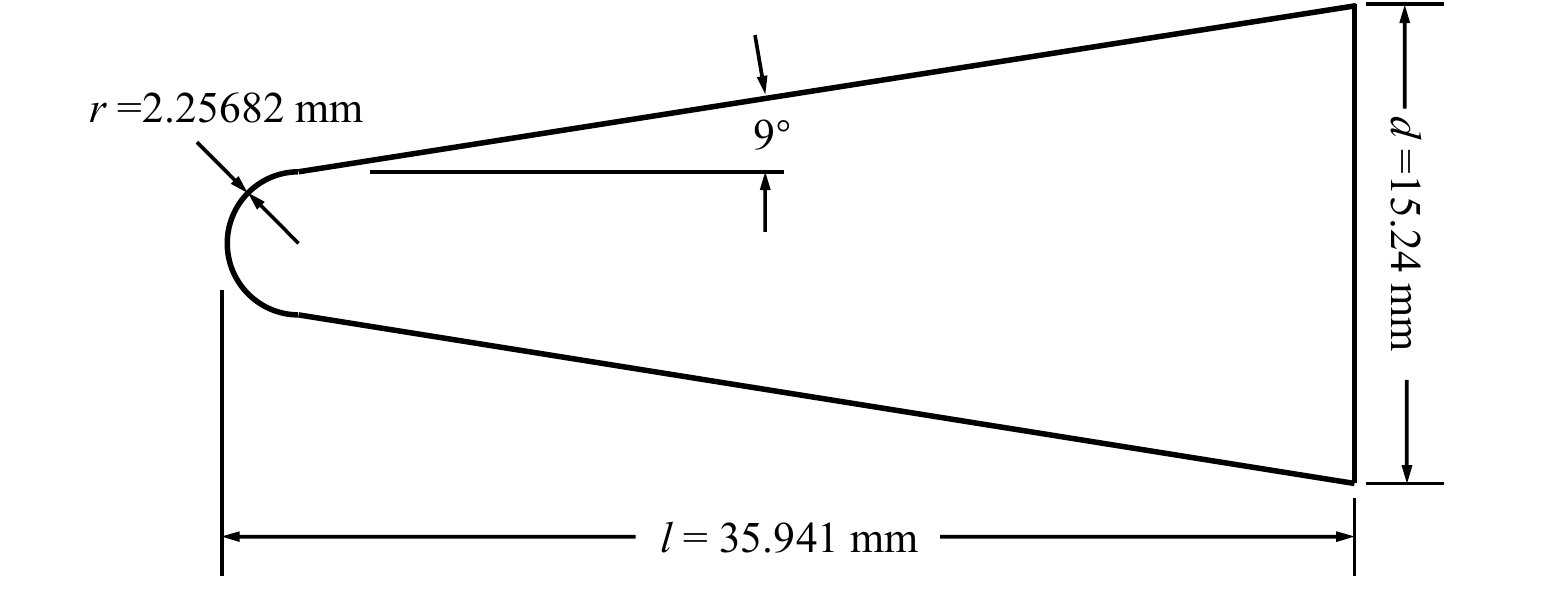}
    \caption{\label{Fig:cone9_shape}Geometric of blunt cone~\cite{rui_zhang_conservative_2024}.}
\end{figure}

The flow around a blunt cone includes both the compressible flow around the nose and the boundary layer along the wall, as well as the expansion flow behind the body. The geometric shape of the blunt cone is shown in the Fig.\ref{Fig:cone9_shape}, with a half-cone angle of 9 degrees. The free stream gas is nitrogen, and the specific heat ratio for the diatomic gas is $\gamma =1.4$. The VHS model is used with a temperature exponent $\omega =0.74$. The free stream temperature is 143.5K, and the wall temperature is 600K. The Mach number is 10.15, and the Knudsen number is 0.052. The Reynolds number is 232.8 and the angle of attack is 25 degrees. The rotational collision number ${{Z}_{rot}}$ is set to 4.24~\cite{rui_zhang_conservative_2024}.

\begin{figure}[!h]
\centering
\subfigure[\label{Fig:case3D_cone9_Con_Rho}]{
\includegraphics[width=0.45\textwidth]{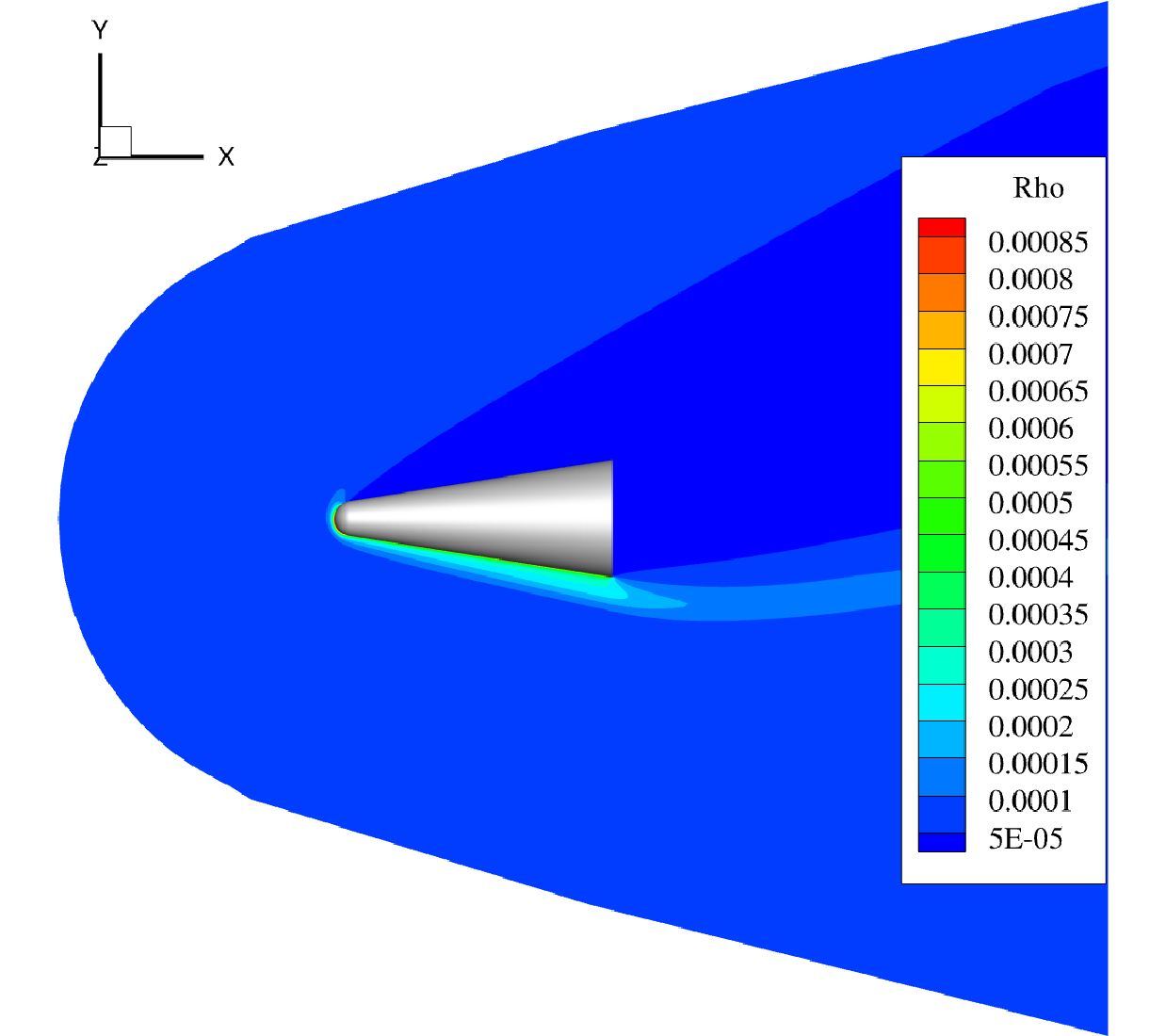}
}\hspace{0.05\textwidth}%
\subfigure[\label{Fig:case3D_cone9_Con_U}]{
\includegraphics[width=0.45\textwidth]{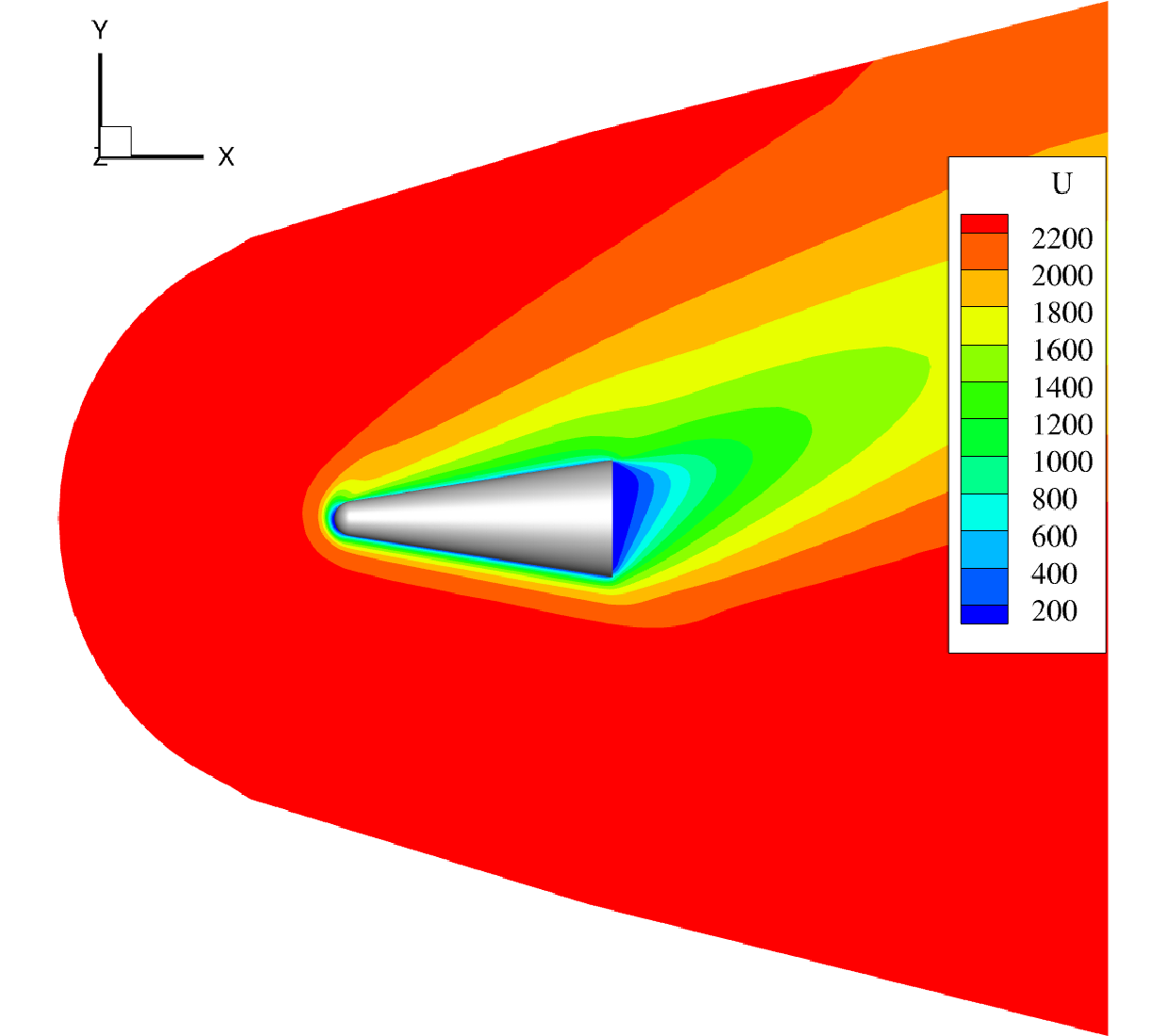}
}\\
\subfigure[\label{Fig:case3D_cone9_Con_V}]{
\includegraphics[width=0.45\textwidth]{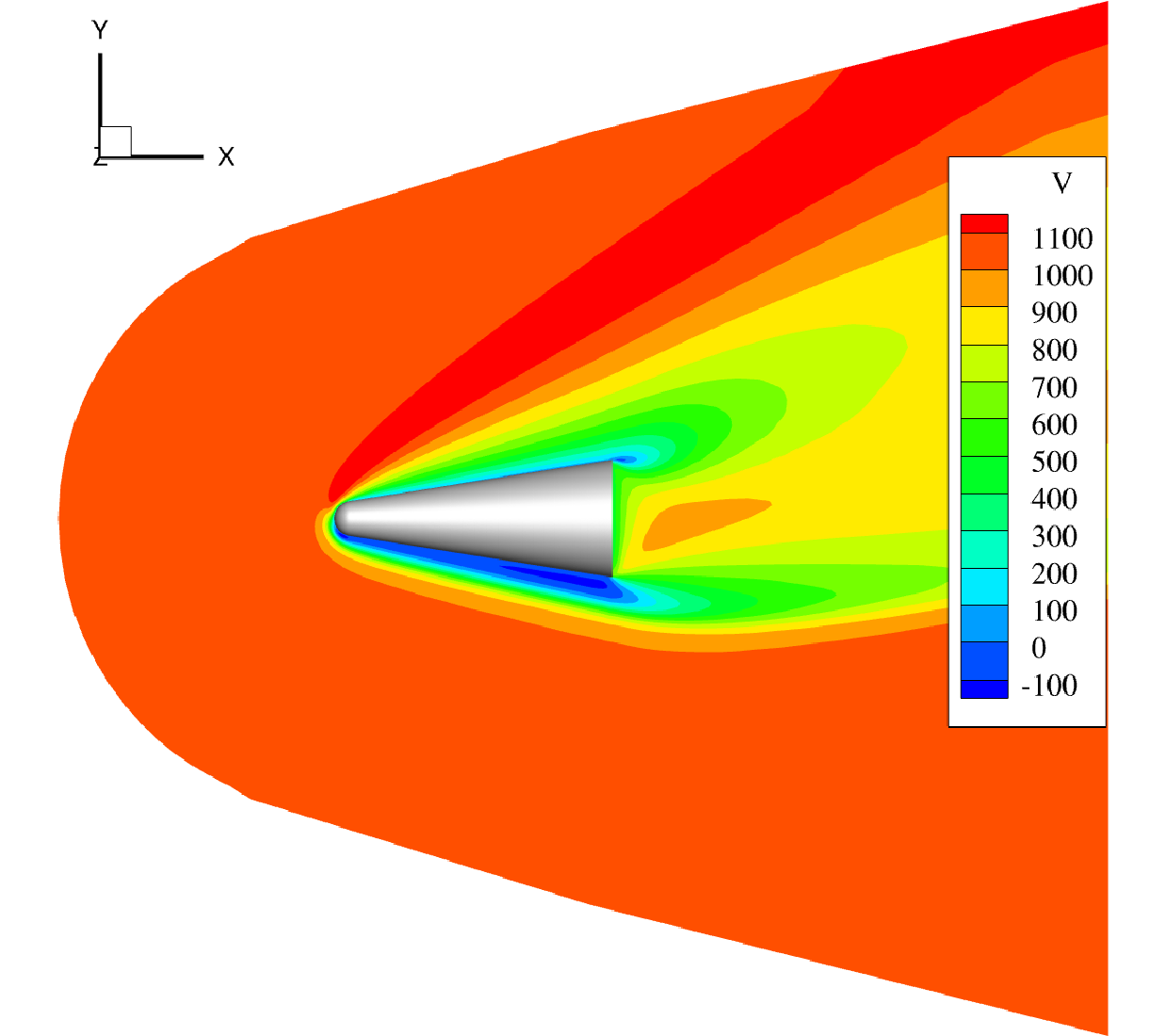}
}\hspace{0.05\textwidth}%
\subfigure[\label{Fig:case3D_cone9_Con_Teq}]{
\includegraphics[width=0.45\textwidth]{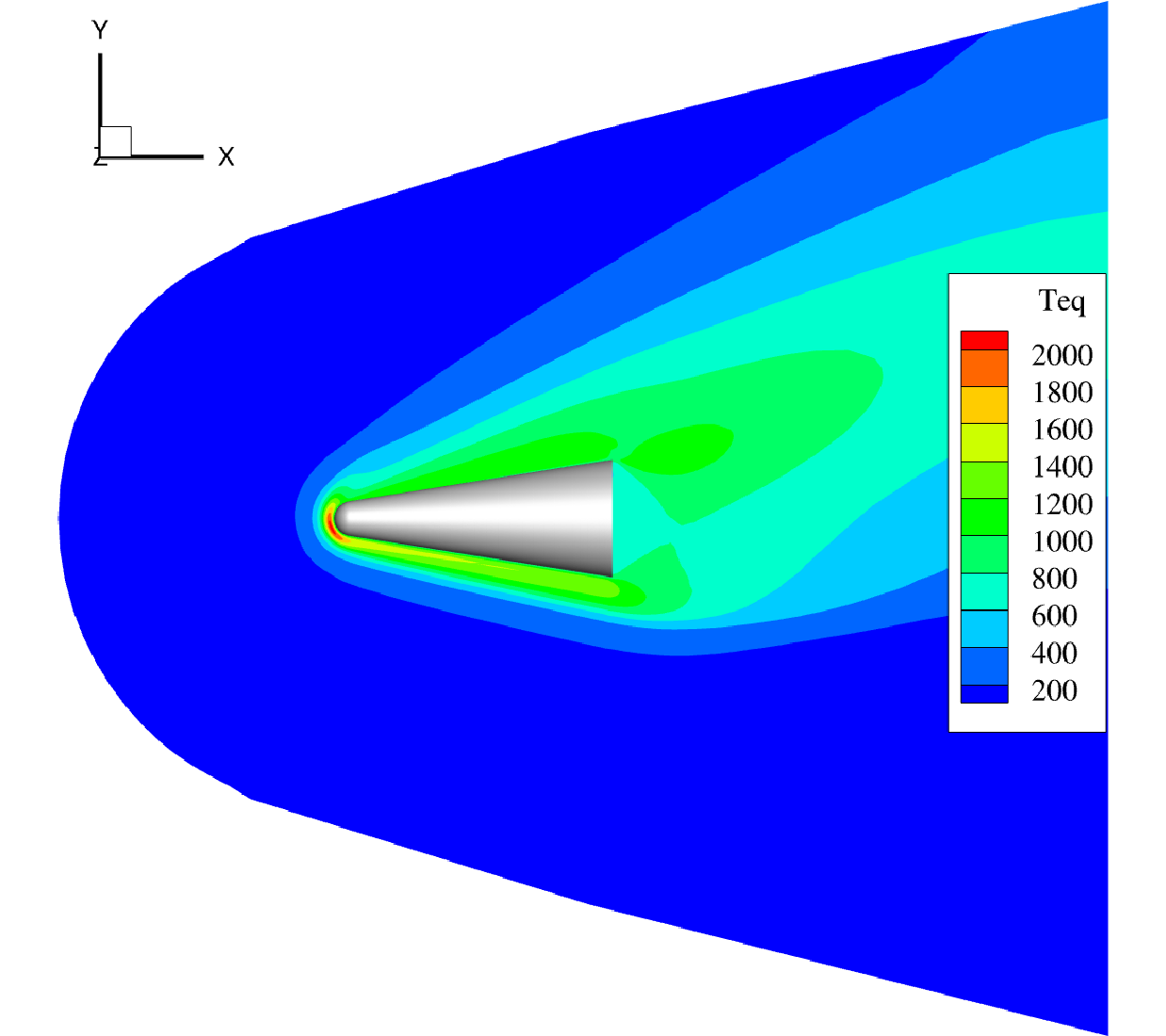}
}\\
\subfigure[\label{Fig:case3D_cone9_Con_Ttr}]{
\includegraphics[width=0.45\textwidth]{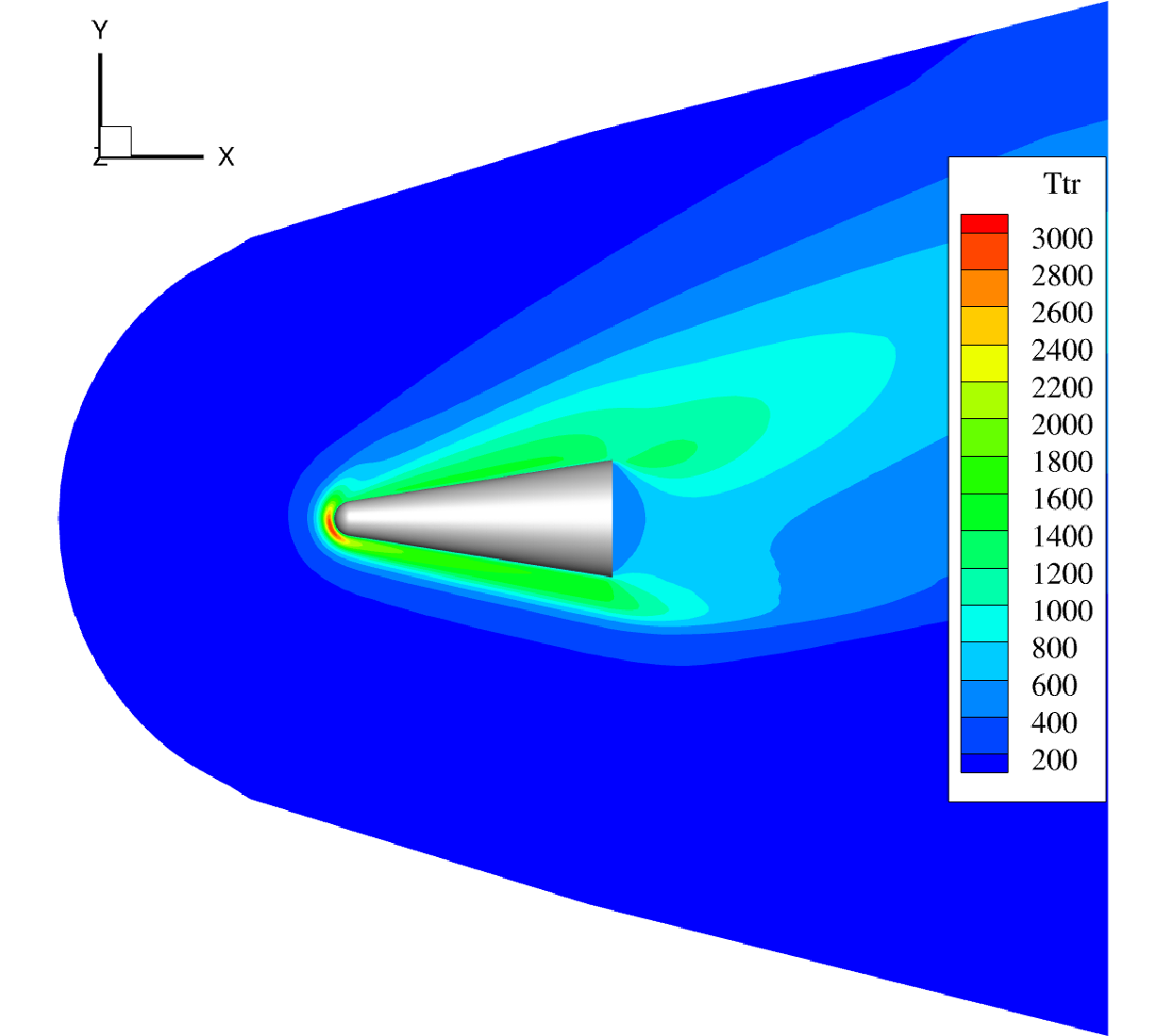}
}\hspace{0.05\textwidth}%
\subfigure[\label{Fig:case3D_cone9_Con_Trot}]{
\includegraphics[width=0.45\textwidth]{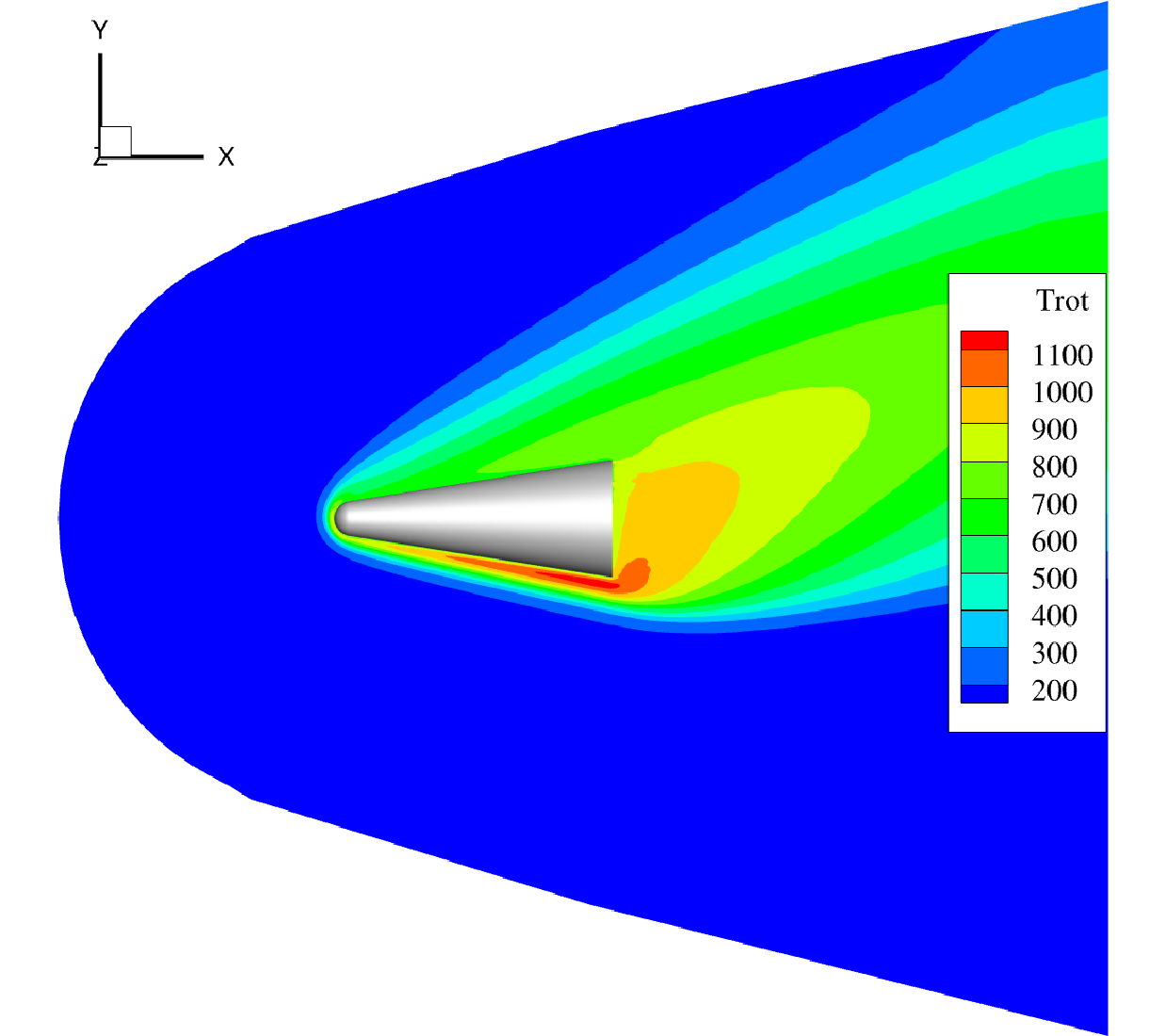}
}\\
\caption{\label{Fig:case3D_cone9_con}Contours of the blunt cone (a) density, (b) X-velocity, (c) Y-velocity, (d) equilibrium temperature, (e) translational temperature, (f) rotational temperature.}
\end{figure}

The density, velocity and temperature of the hypersonic flow around the blunt cone are shown in Fig.\ref{Fig:case3D_cone9_con}. Due to the rarefied nature of the flow, thick shock layers can be observed.

The comparison of pressure coefficient and heat flux coefficient on the surface of the blunt cone is shown in Fig.\ref{Fig:case3D_cone9_wall}. It can be observed that the results are in good agreement with the ones simulated by the DUGKS solver mentioned in Ref.\cite{rui_zhang_conservative_2024}. In addition, as can be seen in Fig.\ref{Fig:case3D_cone9_con}, the temperature contours of the case on the afterbody of the cone is not smooth enough. This situation maybe caused by that the flow field has not fully stabilized or some vortex at the location. In Ref.\cite{rui_zhang_conservative_2024}, a similar situation exists on the posterior side of the cone simulated by DUGKS, so it should not be caused by the low average step number. According to the similarity of simulation results of the SUWP and the DUGKS, it can be seen that the ability of SUWP in simulating rarefied flow is similar to UGKS methods, which is consistent with the previous derivation in Section \ref{sec::method}.

\begin{figure}[!h]
\centering
\subfigure[\label{Fig:case3D_cone9_wall_Cp}]{
\includegraphics[width=0.45\textwidth]{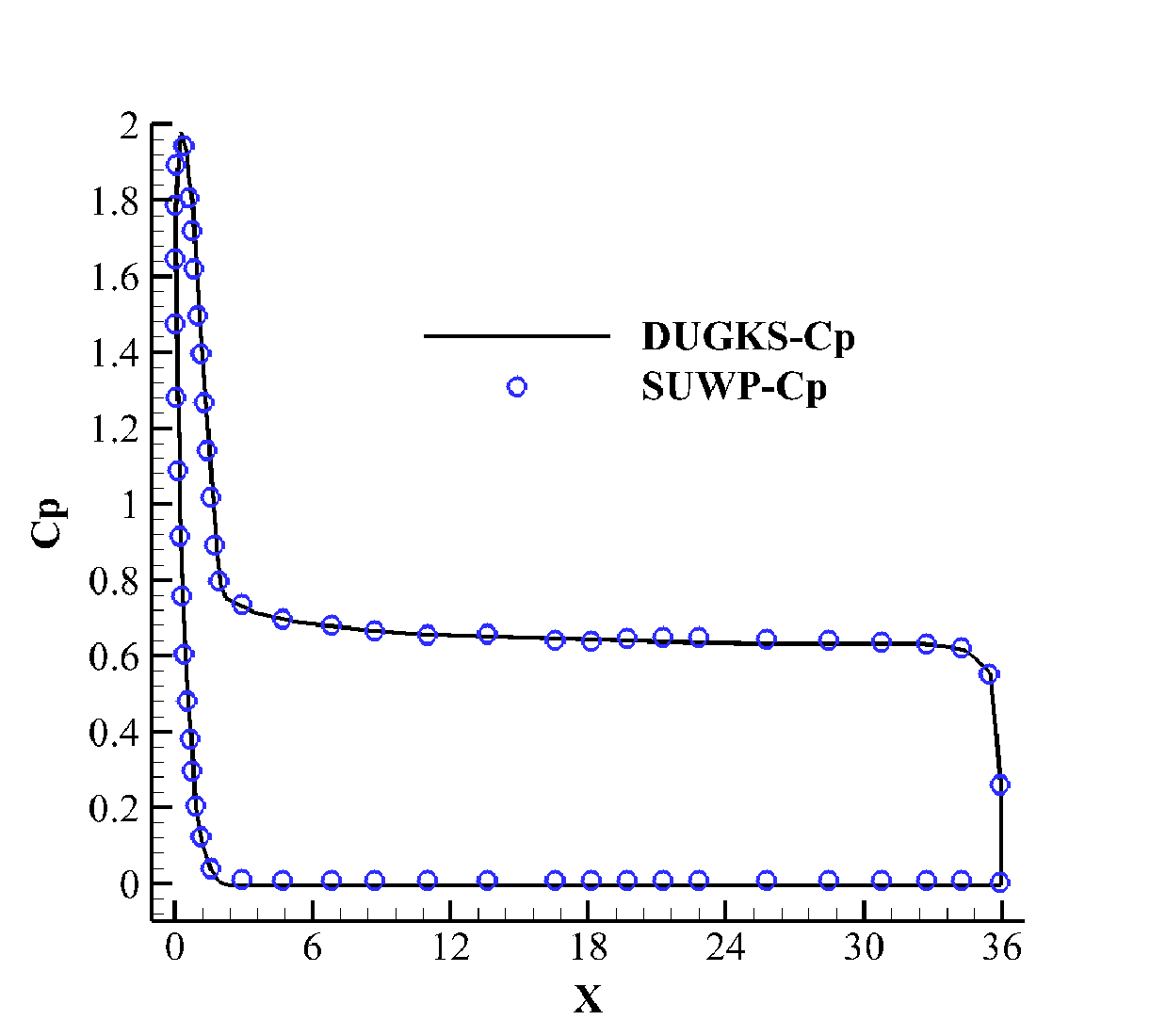}
}\hspace{0.05\textwidth}%
\subfigure[\label{Fig:case3D_cone9_wall_Ch}]{
\includegraphics[width=0.45\textwidth]{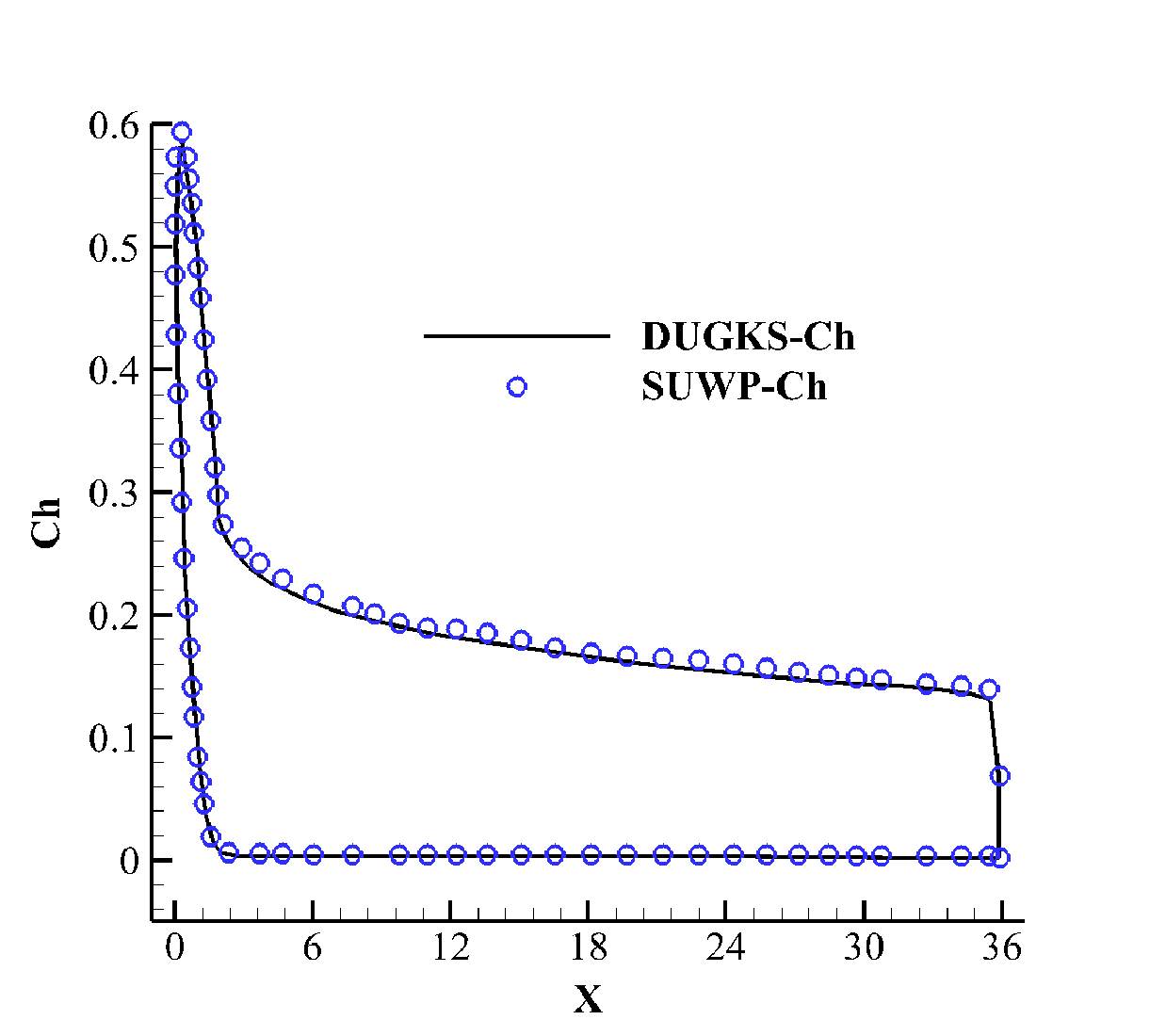}
}\\
\caption{\label{Fig:case3D_cone9_wall}The coefficient of cylinder wall (a) pressure coefficient, (b) heat flux coefficient.}
\end{figure}

\section{\label{sec::conclusions}Conclusions}

In this paper, a SUWP method for diatomic gases is developed based on the Rykov model and a two-temperature KIF scheme. The method couples the two-temperature N-S solver and the particle solver by constructing of QMC mechanism of diatomic gases. This method automatically adjusts the weights based on the flow scale, eliminating the need to partition solver regions according to the scale. Furthermore, the SUWP method can directly utilize the existing research results of N-S and DSMC solvers. This study extension applies SUWP method to diatomic gases, enabling it to simulate thermodynamic non-equilibrium problems effectively.

A series of test cases are performed to validate the SUWP-Rykov method. The accuracy of SUWP-Rykov in simulating molecular collision processes is verified by rotational relaxation in a diatomic homogeneous gas case. The normal shock structure case confirms the ability of the method to simulate molecular transport processes. The supersonic flow around a flat plate case demonstrates its ability to simulate rarefied boundary layers. The hypersonic flow past a cylinder case tests its ability to simulate hypersonic flow. The hypersonic flow past a sphere case validates its ability to simulate three-dimensional flows. The blunt cone flow case confirms its ability to simulate flow around complex geometries. 

The SUWP-Rykov method exhibits accuracy comparable to classical DSMC and other numerical methods in the simulation of rarefied flows. Future work will extend the SUWP method to include vibrational temperature and multi-species gases.

\appendix
\section{\label{sec::appendixA}Random generation of particles in three-dimensional unstructured mesh}
For the newly generated particles, their coordinates need to be randomly assigned in the present cell. For ease of handling, similar to the two-dimensional cell, when generating random particle coordinates in a three-dimensional cell, the cell is first divided into a combination of tetrahedrons. Inside each tetrahedron, a point is selected as the apex of the tetrahedron ${{\vec{P}}_{apex}}$, and the other three points form the base of the tetrahedron. For the base, one point ${{\vec{P}}_{A}}$ is chosen as the top vertex of the base, and the other two points ${{\vec{P}}_{B}}$ and ${{\vec{P}}_{C}}$ form the bottom edge of the base. Take random coordinates on the bottom edge of the base:

\begin{equation}  \label{eq:getP1D}
{{\vec{P}}_{1D}}=\left( {{{\vec{P}}}_{C}}-{{{\vec{P}}}_{B}} \right)\cdot R_n+{{\vec{P}}_{B}}.
\end{equation} 

By random coordinates of the bottom edge ${{\vec{P}}_{1D}}$ and the base vertex ${{\vec{P}}_{A}}$, take random on the bottom surface ${\vec{P}}_{2D}$:

\begin{equation}  \label{eq:getP2D}
{{\vec{P}}_{2D}}=\left( {{{\vec{P}}}_{1D}}-{{{\vec{P}}}_{A}} \right)\cdot {{\left( R_n \right)}^{\frac{1}{2}}}+{{\vec{P}}_{A}}.
\end{equation} 

By random coordinates of the base ${{\vec{P}}_{2D}}$ and the apex of tetrahedron ${{\vec{P}}_{apex}}$, take random in the tetrahedron ${\vec{P}}_{3D}$:

\begin{equation}  \label{eq:getP3D}
{{\vec{P}}_{3D}}=\left( {{{\vec{P}}}_{2D}}-{{{\vec{P}}}_{apex}} \right)\cdot {{\left( R_n \right)}^{\frac{1}{3}}}+{{\vec{P}}_{apex}}.
\end{equation} 

For each reconstructed particle, a random number is used to determine the tetrahedron it should fall into according to the tetrahedron volume, and its random coordinates in the tetrahedron are obtained by Eq.(\ref{eq:getP3D}).

\section{\label{sec::appendixB}Particle transport in three-dimensional unstructured mesh}

The way to a transport particle in unstructured meshes is to change the cell where the particle is located along the particle transport path. Compared with two-dimensional mesh, the three-dimensional mesh needs to judge the intersection of particle transport path and cell interface, which can be realized by the normal vector of cell interface. The normal vector of a cell interface is denoted as $\vec{n}$, and a point on the cell interface is represented as ${{\vec{P}}_{face}}$. It is possible to determine whether the particle crosses the plane of the cell interface by judging whether the lines to the starting point and the endpoint are consistent with the inner product of the normal vector of the cell interface. If the criterion ${{C}_{cross}}<0$, the particle passes through the plane where the interface of the cell is located. If ${{C}_{cross}}>0$, the particle does not cross the plane where the interface of the cell resides. In particular, if ${{C}_{cross}}=0$, the endpoint of the particle is located on the plane where the interface of the cell is located.

\begin{equation}  \label{eq:judgeCross}
C_{cross}=\left[ \left( {{{\vec{P}}}_{face}}-{{{\vec{P}}}_{start}} \right)\cdot \vec{n} \right]\left[ \left( {{{\vec{P}}}_{face}}-{{{\vec{P}}}_{end}} \right)\cdot \vec{n} \right].
\end{equation} 

\begin{figure}[!h]
    \centering
    \includegraphics[width=0.75\textwidth]{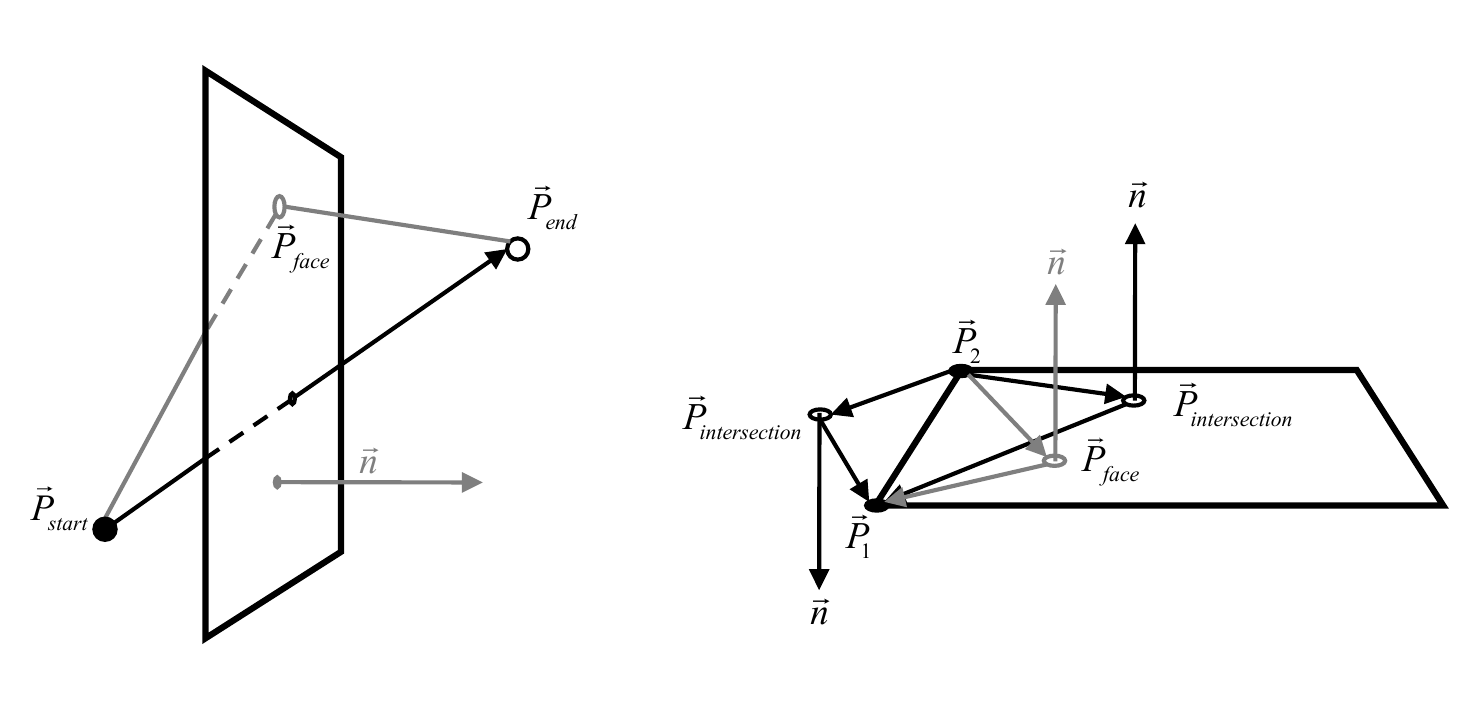}
    \setcounter{figure}{0}
    \caption{\label{Fig:crossJudge}Determine whether the particle crosses the cell interface.}
\end{figure}

According to Eq.(\ref{eq:judgeCross}), a special case may occur, although the transport path intersects the interface where the cell is located, the intersection point is outside the closed graph area of the cell interface. Therefore, further confirmation is required to ensure that the intersection of the particle and the plane ${{\vec{P}}_{inter}}$ is in the plane where the cell interface is located.

For each edge of the cell interface, the two points are denoted as ${{\vec{P}}_{1}}$ and ${{\vec{P}}_{2}}$. The normal vector of the plane is obtained by the point ${{\vec{P}}_{face}}$ on the cell interface and the intersection point ${{\vec{P}}_{inter}}$ between the particle and the plane, ${{\vec{P}}_{1}}$ and ${{\vec{P}}_{2}}$ respectively. According to the direction of the normal vector, it is judged whether the intersection point ${{\vec{P}}_{inter}}$ is on the same side of the line segment as the point ${{\vec{P}}_{face}}$ on the cell interface.

\begin{equation}  \label{eq:judgeEdge}
{{C}_{edge}}=\left[ \left( {{{\vec{P}}}_{1}}-{{{\vec{P}}}_{inter}} \right)\times \left( {{{\vec{P}}}_{inter}}-{{{\vec{P}}}_{2}} \right) \right]\cdot \left[ \left( {{{\vec{P}}}_{1}}-{{{\vec{P}}}_{face}} \right)\times \left( {{{\vec{P}}}_{face}}-{{{\vec{P}}}_{2}} \right) \right].
\end{equation} 

If for all sides, ${{\vec{P}}_{inter}}$ and ${{\vec{P}}_{face}}$ are on the same side of the line segment, i.e. $C_{edge}\ge 0$, and ${{\vec{P}}_{inter}}$ is on the cell interface. Otherwise it is outside the interface. The process is shown in Fig.\ref{Fig:crossJudge}.

\section*{Acknowledgements}
The authors thank Prof. Kun XU at the Hong Kong University of Science and Technology for discussions of the wave-particle method and Zhaoli GUO at the Huazhong University of Science and Technology for discussions of DUGKS. Sirui YANG thanks Dr. Rui ZHANG at Northwestern Polytechnical University for providing DUGKS data of blunt cone, Dr. Jianfeng CHEN at Northwestern Polytechnical University for providing the physical mesh of the flat, Dr. Hao JIN at Northwestern Polytechnical University for discussions of DSMC and Dr. Shunhao PENG at Northwestern Polytechnical University for discussions of bulk viscosities. This present work was supported by the National Natural Science Foundation of China (Grant Nos. 12172301 and 12072283) and the 111 Project of China (Grant No. B17037). This work is supported by the high performance computing power and technical support provided by Xi’an Future Artifcial Intelligence Computing Center.

\section*{Data Availability}
The data that support the findings of this study are available from the corresponding author upon reasonable request.

\bibliographystyle{elsarticle-num}
\bibliography{REF_SUWP_Rykov}
\clearpage

\end{document}